\documentclass[twocolumn]{aastex62}

\usepackage{graphicx}
\usepackage{amsmath}	
\usepackage{amssymb}
\usepackage{booktabs}
\usepackage{enumerate}
\usepackage{dirtytalk}
\usepackage{float}
\usepackage{enumitem}
\usepackage{booktabs}
\usepackage{lipsum} 
\usepackage{float}
\usepackage{xcolor}
\usepackage{multirow}
\usepackage{makecell}

\def\Rj{\,{\rm$\text{R}_{\text{J}}$}\,}
\def\Pref{\,{\rm$\text{P}_{ref}$}\,}
\def\Rp{\,{\rm$\text{R}_{\text{p}}$}\,}

\def\Xh{\,{\rm$X_{\text{H}_2\text{O}}$}}
\def\XNa{\,{\rm$X_{\text{Na}}$}}
\def\XK{\,{\rm$X_{\text{K}}$}}
\def\XNHt{\,{\rm$X_{\text{NH}_3}$}}
\def\XCO{\,{\rm$X_{\text{CO}}$}}
\def\XHCN{\,{\rm$X_{\text{HCN}}$}}
\def\XCOt{\,{\rm$X_{\text{CO}_2}$}}
\def\Rsun{\,{\rm$\text{R}_\odot\,$}}

\def\mum{\,{$\mu$m}}

\graphicspath{{./}{figures/}}

\received{2018 July 6}
\accepted{2019 March 28}

\submitjournal{ApJ}

\shorttitle{Degeneracies in transmission spectra}
\shortauthors{Welbanks \& Madhusudhan}

\begin{document}

\title{On degeneracies in retrievals of exoplanetary transmission spectra}

\author{Luis Welbanks}
\email{luis.welbanks@ast.cam.ac.uk}
\affil{Institute of Astronomy \\ University of Cambridge \\ Madingley Road Cambridge CB3 0HA, UK}

\author{Nikku Madhusudhan}
\email{nmadhu@ast.cam.ac.uk}
\affil{Institute of Astronomy \\ University of Cambridge \\ Madingley Road Cambridge CB3 0HA, UK}

\begin{abstract}
Accurate estimations of atmospheric properties of exoplanets from transmission spectra require understanding of degeneracies between model parameters and  observations that can resolve them. We conduct a systematic investigation of such degeneracies using a combination of detailed atmospheric retrievals and a range of model assumptions, focusing on H$_2$-rich atmospheres. As a case study, we consider the well-studied hot Jupiter HD~209458~b. We perform extensive retrievals with models ranging from simple isothermal and isobaric atmospheres to those with full pressure-temperature profiles, inhomogeneous cloud/haze coverage, multiple molecular species, and data in the optical-infrared wavelengths. Our study reveals four key insights. First, we find that a combination of models with minimal assumptions and broadband transmission spectra with current facilities allow precise estimates of chemical abundances. In particular, high-precision optical and infrared spectra along with models including variable cloud coverage and prominent opacity sources, Na and K being important in optical, provide joint constraints on cloud/haze properties and chemical abundances. Second, we show that the degeneracy between planetary radius and its reference pressure is well characterised and has little effect on abundance estimates, contrary to previous claims using semi-analytic models. Third, collision induced absorption due to H$_2$-H$_2$ and H$_2$-He interactions plays a critical role in correctly estimating atmospheric abundances. Finally, our results highlight the inadequacy of simplified semi-analytic models with isobaric assumptions for reliable retrievals of transmission spectra. Transmission spectra obtained with current facilities such as HST and VLT can provide strong constraints on atmospheric abundances of exoplanets.
\end{abstract}

\keywords{methods: data analysis  --- 
planets and satellites: atmospheres --- techniques: spectroscopic}

\section{Introduction}
\label{sec:intro}

Transmission spectroscopy of transiting exoplanets offers a powerful probe to study their atmospheres. Recent observational advancements have enabled high-precision transmission spectra of exoplanets over a broad spectral range. Such observations have been obtained in low resolution from space using the Hubble Space Telescope (HST) spectrographs, Space Telescope Imaging Spectrograph (STIS) in the NUV/Optical and Wide Field Camera 3 (WFC3) in the near-infrared \citep[e.g.,][]{deming_hd209,sing2016,kreidbergWASP12b}. On the other hand, spectra of comparable quality are also being obtained recently, particularly in the visible range, from large ground based facilities such as the Very Large Telescope (VLT) and the Gran Telescopio Canarias (GTC) \citep[e.g.,][]{ 2017Natur.549..238S, 2018Natur.557..526N, 2018A&A...616A.145C}.

The spectral range accessible to current facilities has the capability to constrain a wide range of atmospheric properties. While the near-infrared spectral range $(1.1-1.7\mu m)$ of the WFC3 contains strong spectral features due to H$_{2}$O \citep{deming_hd209}, the visible range probes features of several other species expected in hot Jupiters such as Na, K, TiO, VO, etc. \citep[e.g.][]{sing2016, 2016ApJ...832..191N, 2017Natur.549..238S}. In addition, optical spectra can also provide important constraints on the possibility and properties of clouds and hazes \citep[e.g.][]{brown01, 0004-637X-820-1-78, 2017ApJ...834...50B,MacDonald17}. Statistical constraints on these various properties have been reported from such datasets using rigorous atmospheric retrieval methods for various planets \citep[e.g.,][]{madhuhd209, kreidbergWASP12b, 2017ApJ...834...50B, MacDonald17}. It is clear from these studies that reliable estimates of the atmospheric properties using retrievals of transmission spectra rely heavily on a thorough understanding of the model degeneracies and the capability of the data to resolve the same. 

The role of degeneracies in interpreting transmission spectra has been investigated in some detail since the beginning of the field. Several early studies highlighted the importance of various atmospheric properties (e.g. clouds, temperature, composition) on observable spectral features \citep[e.g.][]{2000ApJ...537..916S,brown01,2005MNRAS.364..649F}. For example, \citet{brown01} alluded to possible degeneracies between chemical abundances, temperature structure and the presence of clouds. 

Later, \cite{2008A&A...481L..83L} noted the degeneracy between chemical abundance and the reference pressure in the atmosphere. Using transit spectroscopy to measure the effective radius, it was possible to derive the pressure assuming an abundance or assume a pressure to derive the abundance. 

While the above early works sought to explore the degeneracies using semi-analytic or equilibrium forward models, the advent of retrieval techniques in the last decade \citep{madhuseager} allowed investigating this problem with a rigorous statistical approach.  \citet{bennekeandseager} studied the degeneracies involved in retrieving transmission spectra of super-Earths and mini-Neptunes using synthetic spectra. They explored the interplay between chemical composition, cloud-top pressure, planetary radius, and/or a surface pressure in determining the spectral features and suggested combinations of observables that could resolve the degeneracies in different cases. \citet{2013ApJ...778..153B} comment on the degeneracy between the mean molecular mass and cloud top pressure which is present in transmission spectra especially for low-mass planets. 

\citet{2013Sci...342.1473D} showed that the slant-path optical depth at the reference radius depends on the scale height, reference pressure, temperature and the number densities of the absorbers present in the atmosphere in unique ways making their retrieval possible with high-quality data. Such constraints, in principle, also allow the determination of the planetary mass from the transmission spectrum using the retrieved gravity through the scale height \citep{2013Sci...342.1473D}, but can be challenging for low mass planets \citep{batalha2017}. 

\citet{2014RSPTA.37230086G} suggested that there can be a broad range of degenerate solutions to fit infrared data which make constraining molecular abundances challenging. Nonetheless, they suggest ways in which the degeneracy can be resolved. For example, they suggest measuring the radius of the planet at a wavelength where the atmosphere's opacity, is known, e.g., Rayleigh scattering in the optical. 

\citet{0004-637X-820-1-78} explored the influence of non uniform cloud coverage in transmission spectra. They quantitatively explore the degeneracy between clouds and mean molecular weight within an atmospheric retrieval framework. They find that partial and fully cloudy atmospheres are distinguishable and that the visible wavelengths offer an opportunity to break degeneracies between mean molecular weight and cloud coverage.

The effects of clouds and other surfaces have been studied by \citet{2016MNRAS.456.4051B, 2017MNRAS.467.2834B, 2018ApJ...865...12B}. Among their findings are the conclusions that spectral signatures in the optical encode information useful to break degeneracies between retrieved abundances and the planet's radius and that collision induced absorption potentially determines the highest pressures that can be probed in exoplanetary atmospheres in the infrared. An alternative to breaking the innate degeneracy between clouds and chemistry was offered by \citet{MacDonald17} by introducing a two-dimensional inhomogeneous cloud coverage. 

Lastly, \citet{hengkitzmann} highlighted a potential three-way degeneracy between H$_2$O abundance, reference pressure (\Pref) and planet radius (\Rp) using semi-analytic models. Their conclusions about this degeneracy were based on assumptions of isobaric and isothermal atmospheres with H$_2$O as the only molecular opacity source. Our present work investigates this further.

In the present paper we conduct a detailed analysis of the effect of model parameterisation and spectral coverage of data on atmospheric retrievals of transmission spectra. Such an analysis also helps us explore some of the key degeneracies discussed in the literature previously using semi-analytic models. Employing retrieval techniques, we test a series of atmospheric models with varying levels of complexity. In section \ref{sec:reproduction} we start by reproducing the results of previous analytic studies. We discuss the validity of their interpretations and use their assumptions as a starting point for our study. In section \ref{sec:rets} we perform a step-by-step analysis of model dependencies with retrievals using the canonical hot Jupiter HD~209458~b as our case study. 

We start with retrievals assuming a simplistic clear, isothermal and isobaric planetary atmosphere and using infrared data alone. We sequentially improve the model considerations culminating in a realistic atmospheric model with a full pressure-temperature (P-T) profile, inhomogeneous clouds, collision-induced opacities, and multiple chemical species. We also study the impact of including data in the optical wavelengths instead of using only data in the near-infrared. For each of these cases we investigate the constraints on the retrieved parameters and our ability to determine the chemical abundances, especially that of H$_2$O. In section \ref{sec:simretrieval} we asses the ability of our retrievals to constrain atmospheres with high cloud fractions. Lastly, in section \ref{sec:theory}, we revisit the notion of a three-way degeneracy between \Xh, \Rp, and \Pref. We show that the degeneracy between \Rp and \Pref is real and well characterised, but has no effect on the abundance estimates, contrary to previous assertions. We also show that the choice of a \Rp versus \Pref as a free parameter is inconsequential to constraining molecular abundances when a full retrieval study is performed. We summarise our findings in section \ref{sec:conclusions}. 

\section{The \Rp-\Pref-H$_2$O `degeneracy' }
\label{sec:reproduction}

In this section we illustrate some of the key degeneracies inherent to transmission spectra. We begin with a qualitative illustration using model spectra. We generate four forward models showing different combinations of \Rp, \Pref, and \Xh, spanning optical and infrared wavelengths. The forward models are generated using parameters for HD~209458~b with $\log_{10}$($g$)=2.963 in cgs and a stellar radius of 1.155  \Rsun \citep{Torres08}. The models shown here were chosen by inspection and use a parametric P-T profile with the parameters described by \citet{madhuseager} with values of $\log_{10}(P_1)=-1.65$, $\log_{10}(P_2)=-4.02$, $\log_{10}(P_3)=0.48$, $\alpha_1=0.67$, $\alpha_2=0.58$ and temperature of T$_0$=$1435$ K. The choice of P-T profile parameters are within 2-$\sigma$ of the best-fit values reported by \citet{MacDonald17}. 

The models have 100 pressure layers equally spaced in log-pressure between $10^{-6}$ and $10^{2}$ bar. Our prescription considers the effects of H$_2$ Rayleigh scattering and collision induced absorption due to H$_2$-H$_2$ and H$_2$-He interactions and is adapted from the recent works of \citet{2018MNRAS.480.5314P}. The only other source of opacity considered in these illustrative models is H$_2$O. The model set-up is discussed in more detail in section \ref{sec:rets}.

\begin{figure}
\centering
\includegraphics[width=0.5\textwidth]{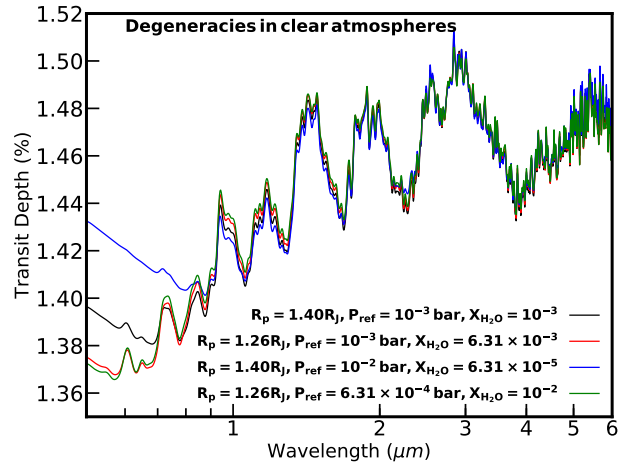}
  \caption[Degeneracies in clear atmospheres]{Degeneracies in clear atmospheres. Clear atmospheres can produce similar absorption features while having different chemical and physical properties. Spectra in red, blue and green, include variations of two or three parameters that are still capable of generating similar features as the reference spectrum shown in black. There is a clear difference in the spectra at shorter wavelengths.}
\label{fig:clear}
\end{figure}

The models are shown in Figure~\ref{fig:clear} and depict the degeneracies in cloud-free atmospheres. These models show that some spectral features in the infrared can be mimicked with different combinations of H$_2$O abundance, radius and reference pressure. While the degeneracy between radius, pressure and molecular mixing ratio allows multiple models to show similar spectral features in the infrared, there are significant differences at shorter wavelengths (i.e. below 1 $\mu$m). These differences at shorter wavelengths are the result of setting the baseline of the spectrum to different levels by changing \Rp and/or \Pref.

As alluded to in section \ref{sec:intro}, several works in the past have discussed possible degeneracies in transmission spectra \citep[e.g.][]{2008A&A...481L..83L, 2014RSPTA.37230086G, bennekeandseager, 2013Sci...342.1473D, 2017MNRAS.467.2834B}. One of the often discussed degeneracies is that between chemical abundance and reference pressure in the atmosphere for the observed radius. Such a degeneracy was formally investigated using semi-analytic models by \citet{2008A&A...481L..83L}. Their work presents the effective altitude $z$ of the atmosphere at a wavelength $\lambda$ as

\begin{equation}
    z(\lambda)=H\, \ln\left(\xi_{\text{abs}}P_{z=0}\sigma_{\text{abs}}/\tau_{\text{eq}}\times \sqrt{2\pi R_{\text{p}}/k T \mu g}\right),
\label{eq:lecavalier}
\end{equation}

\noindent where $H$ is the scale height, $\sigma_{\text{abs}}$ and $\xi_{\text{abs}}$ are the cross-section and abundance (volume mixing ratio) of the dominant absorbing species respectively. $\tau_{\text{eq}}$, also known as equivalent optical depth, is the slant optical depth at an altitude $z_{\text{eq}}$ such that the contribution of an equivalent planet completely opaque below this altitude produces the same absorption as the planet with its translucent atmosphere. $P_{z=0}$ is the reference pressure at an altitude $z=0$ corresponding to $R_{\text{p}}$, the measured radius of the planet. Additionally, $g$ is the gravity of the planet, $k$ is the Boltzmann constant, $T$ the temperature of the atmosphere and $\mu$ is the mean molecular mass of the atmosphere. This expression is one of the first indications of a degeneracy between the reference pressure and the chemical abundance. \cite{2008A&A...481L..83L} conclude that to derive an abundance, a reference pressure needs to be assumed or vice versa.

Variants of this expression have also been derived from first principles in other studies \citep{2013Sci...342.1473D, 2017MNRAS.467.2834B, 2018arXiv180407357S}. The expression was later used by \citet{hengkitzmann} (hereafter HK17) in the following form:

\begin{equation}
    R=R_0+H \left[\gamma +\, \ln \left( \frac{P_0\kappa}{g}\sqrt{\frac{2 \pi R_0}{H}} \right) \right],
\label{eq:heng}
\end{equation}

\noindent where $R_0$ is the radius of the planet at the reference pressure (\Rp in this work), $P_0$ is the reference pressure (\Pref in this work), $H$ is the scale height, $g$ is the gravity of the planet and $\kappa$ is the cross-section per unit mass. The functional form of $\kappa$ in the work of HK17 is:

\begin{equation}
\kappa=\frac{m_{\text{H$_2$O}}}{m}X_\text{H$_2$O}\kappa_{\text{H$_2$O}}+\kappa_{\text{cloud}},
\label{eq:hengkappa}
\end{equation}

\noindent with $m_{\text{H$_2$O}}$=18 amu being the molecular mass of H$_2$O, \Xh$\,$ the volume mixing ratio of H$_2$O and $\kappa_{\text{H}_2\text{O}}$ the water opacity. The additional term, $\kappa_{\text{cloud}}$, is  a constant opacity associated with clouds or aerosols. Inspecting equations \ref{eq:heng} and \ref{eq:hengkappa}, a potential degeneracy between \Xh, \Pref and \Rp becomes evident. However, the derivation of HK17 used assumptions of an isothermal and isobaric opacity along with H$_2$O as the only molecular opacity source. In what follows, we investigate these assumptions and consider other opacity sources that can be important. 

\subsection{On the $X_{\text{H$_2$O}}$-$P_{ref}$-$R_P$ degeneracy}

Here, we further investigate the three-way degeneracy claimed by HK17. The basis of the HK17 study is a semi-analytic model shown in equation~\ref{eq:heng} which was used to fit an observed transmission spectrum of a hot Jupiter WASP-12b in the near-infrared ($\sim$1.15-1.65 $\mu$m) obtained using the HST WFC3 spectrograph \citep{kreidbergWASP12b}. The model assumed an isothermal atmosphere with isobaric opacities, with H$_2$O as the only molecular opacity source. The model was fit to the near-infrared spectrum using a non-linear least-squares fitting routine to obtain best-fit values of different combinations of parameters for an assumed value of $R_p$. By repeating the fits for a range of $R_p$ values they investigate the degeneracy between \Xh-\Pref-\Rp. 

To investigate the potential three-way degeneracy reported by HK17, we follow two approaches. We first reproduce the results of HK17 using their approach, i.e., of their semi-analytic model and least-squares fit to the WFC3 transmission spectrum of WASP-12b. We then reproduce the same results using their semi-analytic model in a Bayesian retrieval approach. We later include additional opacity due to H$_2$-H$_2$ and H$_2$-He collision induced absorption (CIA) in the HK17 model to investigate the validity of their assumptions. With CIA included, we follow the same two approaches, i.e., first employing a non-linear least-squares fit and then a Bayesian retrieval. 

\begin{figure}
\includegraphics[width=0.5\textwidth]{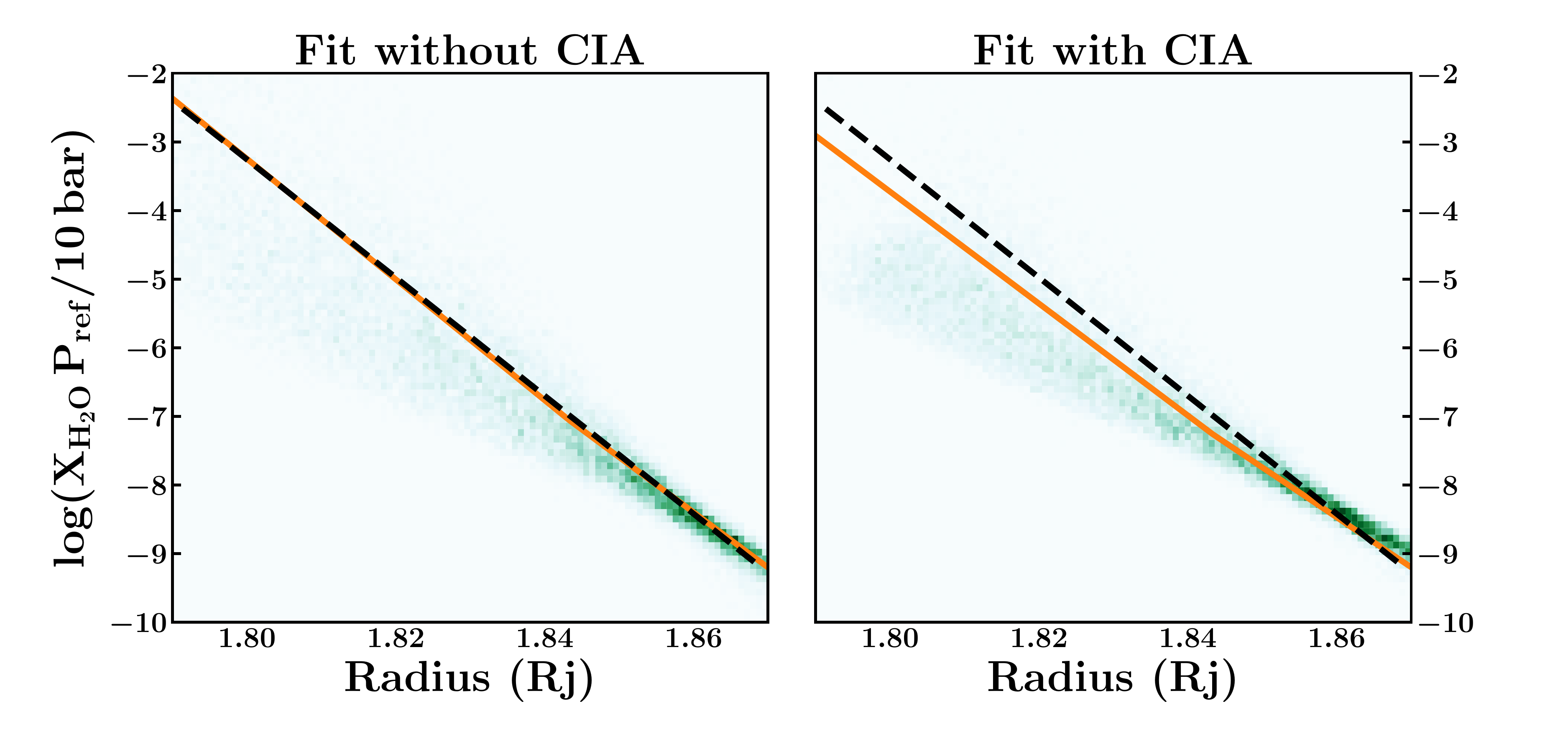}
\includegraphics[width=0.45\textwidth]{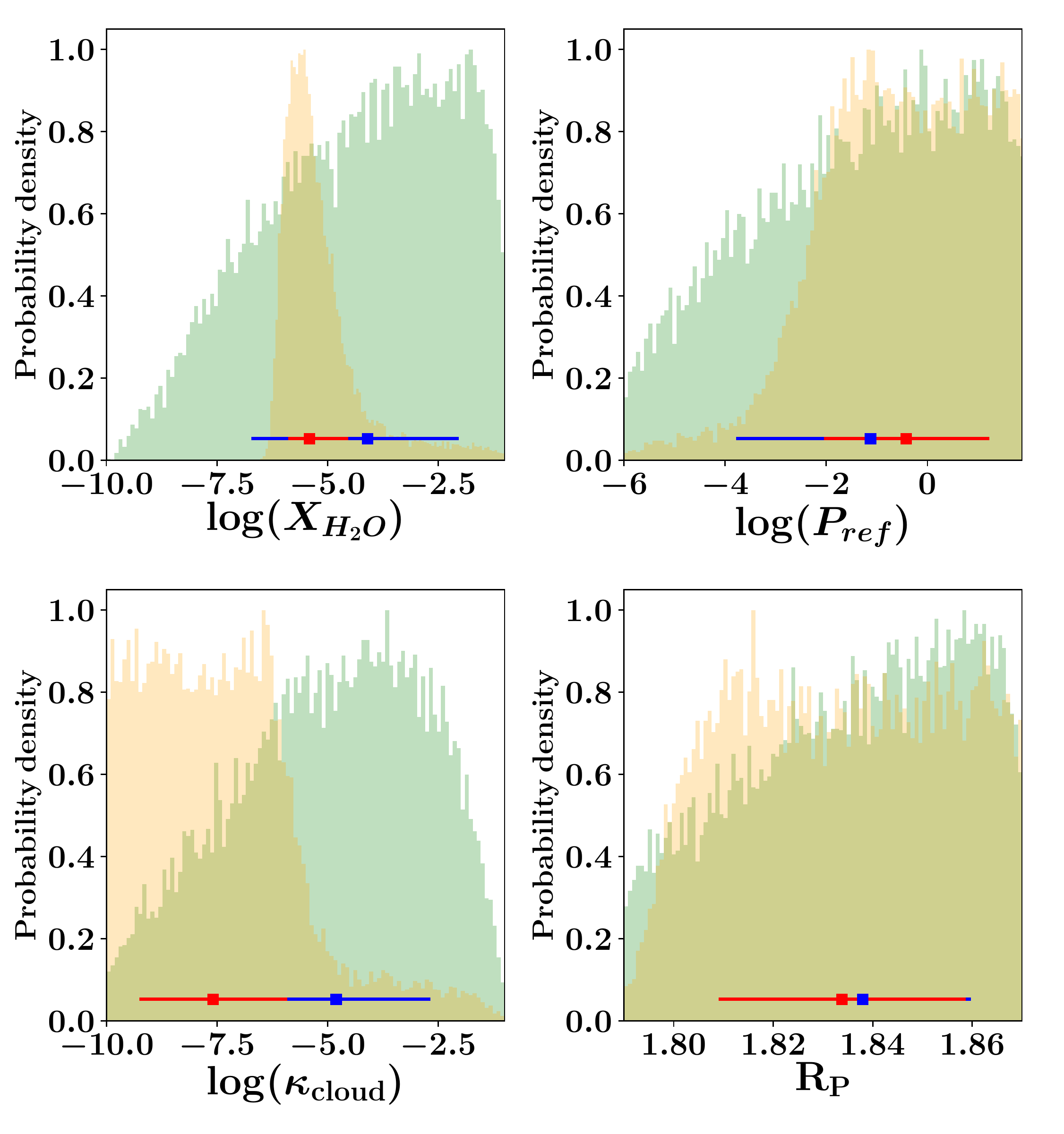}
\centering

  \caption[Results of linear fit compared to full retrieval models]{Top row: Product of H$_2$O  abundance and reference pressure (\Pref) versus the reference radius (\Rp) for simplistic model fits to WFC3 spectrum of WASP-12b. The left panel shows fits without CIA opacity and the right panel shows fits considering the effect of CIA. The black dashed line shows the result from Fig. 7 of \citet{hengkitzmann}, hereafter HK17. The orange solid line is our result using the analytic formulation and the same three parameter fit of HK17. While the orange line matches exactly with the HK17 result in the left panel, it deviates from the same in the right panel due to the inclusion of CIA opacity. The green two-dimensional histograms in the background show the same quantities using the posterior distributions from Bayesian retrievals of the same parameters.  
  Bottom rows: Posterior distributions from the retrievals corresponding to the top panels. Green (orange) shows the histograms for the retrievals without (with) CIA while the median values and 1-$\sigma$ uncertainties are shown in blue (red). The posterior distributions show the retrieved H$_2$O abundances (volume mixing ratios), \Pref (in bar), $\kappa_{\text{cloud}}$ (in m$^2$ kg$^{-1}$), and \Rp (in \Rj). The introduction of CIA in the right panel improves the constraint on the H$_2$O abundance.}
\label{fig:reproduction}
\end{figure}

We begin by following the approach of HK17 and perform a fit to the WASP-12b WFC3 data using equation~\ref{eq:heng} with a  least-squares minimisation routine (\texttt{curve\char`_fit} in Python). Our model considerations are identical to those in HK17, e.g., isothermal atmosphere and isobaric H$_2$O opacity. The top-left panel of Figure~\ref{fig:reproduction} shows our results reproducing Figure 7 in HK17. At the outset we notice two discrepancies. First, we are able to reproduce the fit in HK17 using the log of $X_{\text{H}_2\text{O}}$(P$_{ref}$/10 bar) versus \Rp. However, HK17 present their y-axis as $X_{\text{H}_2\text{O}}$(P$_{ref}$/10 bar)$^{-1}$. We interpret this as a typographic error in HK17. This is especially the case considering that equation~\ref{eq:heng} implies the product of $X_{\text{H}_2\text{O}}$  and \Pref, and also considering figures 3 to 8 of \cite{2018MNRAS.481.4698F} who use the same model and notation. Second, HK17 claim from this figure that $X_{\text{H}_2\text{O}}$ is strongly degenerate with \Rp, i.e., that the H$_2$O abundance varies by many orders of magnitude with slight changes in Rp. However, it is not possible to deduce information about the H$_2$O abundance from this figure alone given that only the product $X_{\text{H}_2\text{O}}$(P$_{ref}$/10 bar) is shown to be degenerate with \Rp and not $X_{\text{H}_2\text{O}}$ or \Pref individually.  

Next, we study this problem using a Bayesian retrieval approach. Our retrieval code is adapted from the works of \citet{2018MNRAS.480.5314P} to consider the semi-analytic model and assumptions discussed above. We replace the numerical model of \citet{2018MNRAS.480.5314P} with the semi-analytic model of HK17 while retaining the module for Bayesian parameter estimation using the Nested Sampling algorithm \citep{2009MNRAS.398.1601F, 2014A&A...564A.125B}. The model parameters remain the same as in HK17, namely $X_{\text{H}_2\text{O}}$, \Pref, $\kappa_{\text{cloud}}$, \Rp and $T$ the isothermal temperature. The prior range for the radius is \Rp=1.79 \Rj to 1.87 \Rj $\,$ to match the range shown in figure 7 of HK17 and the prior range of $\log_{10}$(\Pref) is from -6 to 2 in bar. Both the $\log_{10}(X_{\text{H}_2\text{O}})$ (volume mixing ratio) and $\log_{10}(\kappa_{\text{cloud}})$ (m$^2$ kg$^{-1}$) priors are from -1 to -10. The temperature prior is from 500K to 2000K. Similarly, we consider an isothermal atmosphere, isobaric H$_2$O opacity (at 1mbar), a fixed mean molecular weight of 2.4 amu, and a fixed gravity  of $\log_{10}(g)=2.99$ in cgs \citep{2009ApJ...693.1920H}.

The results from the retrieval are shown in green in the top left panel of Figure~\ref{fig:reproduction}. We show the posteriors from the retrieval as a two dimensional histogram of $X_{\text{H}_2\text{O}}$(P$_{ref}$/10 bar) against the retrieved \Rp. The bottom four panels of Figure~\ref{fig:reproduction} show the posterior distribution of the H$_2$O abundance, P$_{ref}$, $\kappa_{\text{cloud}}$, and R$_p$ in the green histograms. Our retrieval finds an unconstrained H$_2$O abundance with a median abundance of $\log_{10}(X_{\text{H}_2\text{O}})=-4.10 ^{+ 2.06 }_{- 2.62 }$.  We find that the posterior distributions from the retrieval closely follow the results from the linear fit (i.e. orange line) as shown in the top-most left panel. 

 We now investigate the validity of the assumptions of the semi-analytic model above by including CIA absorption as an additional source of opacity. The importance of CIA as a continuum source of opacity is highlighted in several previous studies \citep[e.g.][]{2013Sci...342.1473D,2017MNRAS.467.2834B,2018ApJ...865...12B}, which makes its inclusion imperative in model spectra of giant planets. We amended the total opacity in the formulation of HK17, shown in equation~\ref{eq:hengkappa}, to 

\begin{equation}
\kappa=\frac{m_{\text{H$_2$O}}}{m}X_\text{H$_2$O}\kappa_{\text{H$_2$O}}+\kappa_{\text{cloud}}+\kappa_{\text{CIA}},
\end{equation}

\noindent where the first two terms remain as explained above. The third term is opacity due to H$_2$-H$_2$ and H$_2$-He collision induced absorption. This and other opacity sources are discussed in section \ref{sec:rets}.

We follow the approach described above by performing a least-squares fit of the amended model to the near-infrared data. The additional opacity source (i.e. CIA) is computed following the method used to compute H$_2$O opacity and we preserve the assumption of an isobaric atmosphere by evaluating the cross-sections at 1mbar. We find that the inclusion of CIA absorption changes the slope of the resulting linear relationship between $X_{\text{H}_2\text{O}}$(P$_0$/10 bar) and \Rp. Our resulting fit is shown as an orange solid line in the top right panel of Figure~\ref{fig:reproduction} where we also show the fit of HK17 using a dashed black line. This analytic fit shows that the slope of the relation between $X_{\text{H}_2\text{O}}$(P$_0$/10 bar) and \Rp has changed. Again, it is not possible to infer from this result if \Rp is degenerate with $X_{\text{H}_2\text{O}}$ or \Pref or both. In comparison, a retrieval approach would provide the necessary insight as pursued above.  

We, therefore, now perform a retrieval using the modified model including CIA opacity. Our retrieval approach keeps the previous description although in this case we add the pressure dependent effects of CIA. This retrieval study finds a better constrained H$_2$O abundance with a median of  $\log_{10}(X_{\text{H}_2\text{O}})=-5.41 ^{+ 0.88 }_{- 0.47 }$.  Similarly to the previous retrieval, we present in the background of the top right panel of Figure~\ref{fig:reproduction} the two dimensional histogram of $X_{\text{H}_2\text{O}}$(P$_{ref}$/10 bar) against \Rp. We also show the posterior distributions of the retrieved parameters including the H$_2$O abundance for this case in the bottom orange histograms. We find that the inclusion of CIA opacity results in a better constraint on the H$_2$O abundance even within the framework of this simplistic model. 

\begin{deluxetable*}{c|ccccccccccccc}
\tablecaption{Summary of twelve cases for which retrievals were performed. Each column indicates a model assumption or a free parameter. The WFC3 column indicates the inclusion of data in the near-infrared in the retrieval. On the other hand, Optical signifies that data in the optical wavelengths were used in the retrieval. P-T means that we consider a parametric P-T profile in the retrieval. Clouds are implemented in two ways: F stands for a retrieval with full cloud cover in which the cloud fraction is fixed to $\phi$=100\%, N represents cases with non uniform clouds in which the cloud fraction $\phi$ is a free parameter in the retrieval.\label{table:cases}}
\tablecolumns{14}

\tablewidth{0pt}
\tablehead{
\colhead{} & \colhead{Isobar} & \colhead{Isotherm} & \colhead{H$_2$O} & \colhead{WFC3} & \colhead{CIA} & \colhead{P-T} & \colhead{Clouds} & \colhead{Optical} & \colhead{Na, K} & \colhead{NH$_3$} & \colhead{CO} & \colhead{HCN} & \colhead{CO$_2$}\\
\colhead{} & \colhead{} & \colhead{} & \colhead{} & \colhead{} & \colhead{} & \colhead{} & \colhead{(F/N)} & \colhead{} & \colhead{} & \colhead{} & \colhead{} & \colhead{} & \colhead{}
}
\startdata
\S \ref{sub:case1}: Case 1  & \checkmark & \checkmark   & \checkmark & \checkmark &  &  &  &  &  &  &  &  &  \\
\S \ref{sub:case2}: Case 2 & \checkmark & \checkmark & \checkmark & \checkmark & \checkmark &  &  &  &  &  &  &  &  \\
\S \ref{sub:case3}: Case 3 &  & \checkmark & \checkmark & \checkmark & \checkmark &  &  &  &  &  &  &  &  \\
\S \ref{sub:case4}: Case 4 &  &  & \checkmark & \checkmark & \checkmark & \checkmark &  &  &  &  &  &  &  \\
\S \ref{sub:case5}: Case 5 &  &  & \checkmark & \checkmark & \checkmark & \checkmark & F &  &  &  &  &  &  \\
\S \ref{sub:case6}: Case 6 &  &  & \checkmark & \checkmark & \checkmark & \checkmark & N &  &  &  &  &  &  \\
\S \ref{sub:case7}: Case 7 &  &  & \checkmark & \checkmark & \checkmark & \checkmark & N & \checkmark &  &  &  &  &  \\ 
\S \ref{sub:case8}: Case 8 &  &  & \checkmark & \checkmark & \checkmark & \checkmark & N & \checkmark & \checkmark &  &  &  &  \\ 
\S \ref{sub:case9}: Case 9  &  &  & \checkmark & \checkmark & \checkmark & \checkmark & N & \checkmark & \checkmark & \checkmark &  &  &  \\
\S \ref{sub:case10}: Case 10 &  &  & \checkmark & \checkmark & \checkmark & \checkmark & N & \checkmark & \checkmark & \checkmark & \checkmark &  &  \\
\S \ref{sub:case11}: Case 11&  &  & \checkmark & \checkmark & \checkmark & \checkmark & N & \checkmark & \checkmark & \checkmark & \checkmark & \checkmark &  \\ 
\S \ref{sec:fullret}: Case 12  &  &  & \checkmark & \checkmark & \checkmark & \checkmark & N & \checkmark & \checkmark & \checkmark & \checkmark & \checkmark & \checkmark 
\enddata

\end{deluxetable*}

Our results above demonstrate two main points. First, the retrieved molecular abundance changes with the inclusion of CIA absorption. The inclusion of CIA opacity provides a continuum to the spectrum that sets the maximum pressure probed in the atmosphere, i.e., the line of sight photosphere \citep{2013Sci...342.1473D, 0004-637X-820-1-78, 2018ApJ...865...12B}. As a result, the thickness of the atmospheric column probed by the transmission spectrum decreases compared to the non-CIA scenario thereby requiring a different H$_2$O abundance to explain the data. Second, the log-linear behaviour seen in both panels of Figure~\ref{fig:reproduction} is likely strongly influenced by a relation between \Pref and \Rp, irrespective of the H$_2$O abundance. We further discuss this relation in detail in section \ref{sec:theory}. The constraint on the H$_2$O abundance improves with the inclusion of CIA, irrespective of any degeneracy between \Pref and \Rp. Nevertheless, the H$_2$O abundance is still weakly constrained even in the CIA case. However, this is not due to a three-way degeneracy but rather a result of incomplete model assumptions and limited data. We demonstrate this in more detail in the next section. 

In summary, these results show that the conclusions of HK17 are likely due to the restricted model assumptions. The lack of consideration of CIA opacity, among other factors are likely responsible for their conclusions. We discuss this further in section \ref{sub:semianalyticlimitations}. The three-way degeneracy noted in HK17 manifests itself fully under idealised conditions encapsulated in the analytic formalism of equation~\ref{eq:heng}, namely an isothermal, isobaric, constant mean molecular weight, constant gravity, a single absorber, and a cloud-free atmosphere. In a more realistic atmosphere this degeneracy is broken in various ways. For example, for high chemical abundances the mean molecular weight becomes significant enough to affect the scale height and hence the amplitude of the spectral feature \citep[e.g.,][]{bennekeandseager, 0004-637X-820-1-78}. On the other hand, at low abundances the CIA opacity provides the continuum level for the spectrum \cite[e.g.,][]{0004-637X-820-1-78, 2013Sci...342.1473D}. Other effects influencing the spectrum include considerations of clouds, non-isothermal atmospheres, multiple-molecular absorbers, etc. Furthermore, constraining the contributions from these various effects require observed spectra in the visible in addition to the infrared spectra. The importance and effects of such considerations are studied in the rest of this work. In what follows, we perform an in-depth study of the effects of model assumptions and data coverage on atmospheric retrievals using transmission spectra.

\section{HD~209458~b: A case study.}
\label{sec:rets}

We now conduct a systematic exploration of the degeneracies in interpreting transmission spectra using fully numerical models within a rigorous retrieval framework. For this study, we choose the canonical hot Jupiter HD~209458~b which has the most data available \citep{deming_hd209,sing2016} and has been a subject of several recent retrieval studies  \citep[e.g.][]{madhuhd209,MacDonald17,2017ApJ...834...50B}. 

We used an atmospheric retrieval code for transmission spectra adapted from the recent work of \citet{2018MNRAS.480.5314P}. The code was modified to include the radius of the planet (\Rp) as one of the retrieval parameters and, unlike \cite{2018MNRAS.480.5314P}, we do not infer any stellar properties. The code computes line-by-line radiative transfer in a transmission geometry, assuming hydrostatic equilibrium. We consider a parametric P-T profile using the prescription of \cite{madhuseager}. We consider a one-dimensional model atmosphere consisting of 100 layers in pressure ranging from $10^{-6} - 10^3$ bar uniformly spaced in $\log_{10}$(P). We use the cloud/haze parametrization of \citet{MacDonald17} which allows for cloud-free to fully cloudy models, including non-homogeneous cloud cover. The haze is included as $\sigma=a\sigma_0(\lambda/\lambda_0)^\gamma$, where $\gamma$ is the scattering slope, $a$ is the Rayleigh-enhancement factor, and $\sigma_0$ is the H$_2$ Rayleigh scattering cross-section ($5.31\times10^{-31}$~m$^2$) at the reference wavelength $\lambda_0=350$~nm. Cloudy regions of the atmosphere are included as an opaque cloud deck with cloud-top pressure P$_{\text{cloud}}$. The fraction of cloud cover at the terminator is given by $\phi$.

The absorption cross-sections of the molecular and atomic species are obtained from \citet{Rothman2010} for H$_2$O, CO and CO$_2$, \citet{Yurchenko2011} for NH$_3$, \citet{Harris2006,Barber2014} for HCN, \cite{NIST_ASD} for Na and K, and \citet{RICHARD20121276} for H$_2$-H$_2$ and H$_2$-He collision induced absorption (CIA). The cross-sections are generated using the methods of \citet{Gandhi17}. Our model assumes that the atmosphere has uniform mixing ratio for each species considered and treats these mixing ratios as free parameters. Unlike the retrievals in section \ref{sec:reproduction}, these retrievals do not fix the mean molecular weight to a specific value and instead calculate it based on the retrieved molecular abundances and the assumption of a H$_2$-He dominated atmosphere with a fixed He/H$_2$ ratio of 0.17  \citep{MacDonald17}. Lastly, the reference pressure (\Pref) is a free parameter which establishes the pressure in the atmosphere at which the reference radius of the planet (\Rp) is located. In summary, our full model has 19 free parameters: \Rp, \Pref, seven chemical species (H$_2$O, CO, CO$_2$, HCN, NH$_3$, Na and K), six parameters for the P-T profile, and four parameters for clouds/hazes including the cloud deck pressure  $P_{\text{cloud}}$ and cloud fraction $\phi$.

Our goal is to investigate the effect of each model parameter and/or assumption on the retrieved parameters and their degeneracies. We start with the simplest set up and gradually increase the physical plausibility of the model and extent of the data. We start by considering an isothermal and isobaric atmosphere with only one molecule present, H$_2$O, to carry on from our reproduction of previous results in section \ref{sec:reproduction}. We later increase the number of considerations until we use a full model with a parametric P-T profile, with multiple molecules (H$_2$O, Na, K, NH$_3$, CO, HCN, and CO$_2$), and the presence of clouds/hazes. For our retrievals we use the spectrum of HD~209458~b reported in \citet{sing2016}. The spectrum has two wavelength ranges observed with HST: near-infrared (1.1-1.7 $\mu$m) obtained using WFC3 and full optical range (0.3-1.01 $\mu$m) obtained using the STIS instrument. We compare the retrieved radius values to the value reported by \citet{Torres08} of \Rp=$1.359^{+0.016}_{-0.019}$ \Rj which is consistent with the reported radius by \cite{sing2016}. 

\begin{figure*}
\centering
\includegraphics[width=\textwidth]{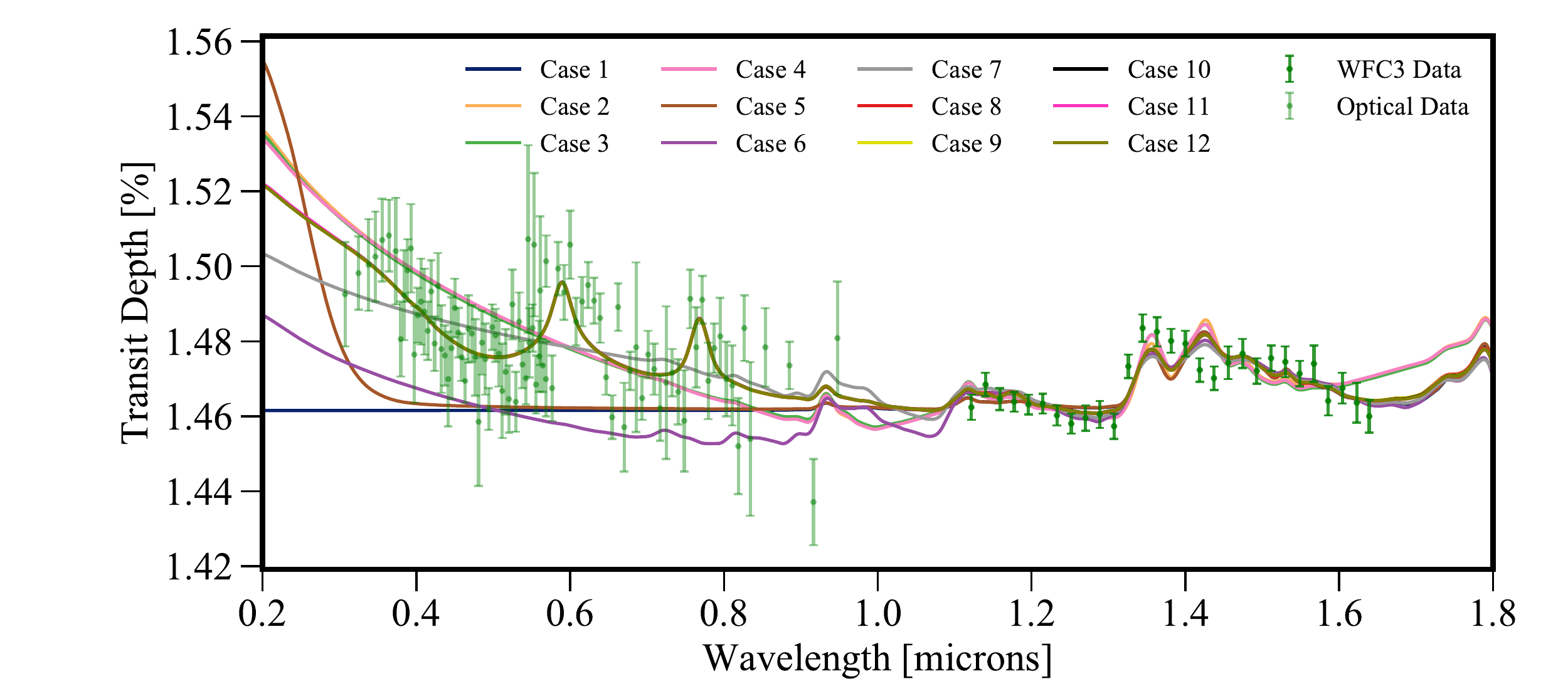}
  \caption[Spectrum of HD~209458~b and degenerate models]{Median retrieved models for each of the cases shown in Table~\ref{table:cases}. Infrared and optical data from \citet{sing2016} are shown using green markers. While all models produce some degree of fit to the data in the infrared, only cases 8-12 produce a good fit to all the data.}
\label{fig:spectramodels}
\end{figure*}

The 12 cases of our study are summarised in Table~\ref{table:cases}. The parameters, priors and results for all cases are summarised in Table~\ref{table:priors1} and Table~\ref{table:priors2} in the Appendix. The retrieved median spectra for all the cases are shown in Figure~\ref{fig:spectramodels}. The constraints on the retrieved H$_2$O abundances for the different cases are illustrated in Figure~\ref{fig:retsummary}. The posterior distributions for \Xh, \Pref and \Rp for all cases are included in the Appendix in Figure \ref{fig:collection_cases}.

\subsection{Case 0: Reproducing the semi-analytic model}
\label{sub:case0}
Before conducting our case by case study, we first consider case zero which presents a numerical analogue of semi-analytic models. Case zero has the simplest model considerations, i.e. of an isothermal and isobaric atmosphere with H$_2$O absorption as the only source of opacity. In addition, the mean molecular weight and gravity are fixed quantities. The isobaric assumption means evaluating the molecular cross section at only one pressure, in this case 1 mbar.  Following the models in section \ref{sec:reproduction}, the mean molecular weight is fixed to a value of 2.4 amu, that of an H$_2$-rich atmosphere with solar elemental abundances. The fixed value for gravity is $\log_{10}$($g$)=2.963 in cgs for HD~209458~b \citep{Torres08}. 

While generally our numerical model spans a pressure range of 10$^{-6}$ bar to 10$^{3}$ bar as discussed above, in the present case the retrieval is strongly sensitive to the edges of the pressure range due to the limited opacity sources. The deepest pressure level in the model atmosphere effectively acts as an opaque surface. In order to circumvent this edge effect, we consider a model atmosphere with an unrealistically extreme range in pressure, from $10^{-14} - 10^{14}$ bar, uniformly spaced in $\log_{10}$(P) using 400 layers. 

We use this model for a retrieval using a near-infrared WFC3 spectrum of HD~209458~b, similarly to our retrievals in section \ref{sec:reproduction}. The model parameters are \Xh, \Rp, \Pref and T, the temperature of the isotherm. The priors on the parameters are $\log_{10}(X_{\text{H}_2\text{O}})$=[-12,-1], \Rp=[1,3] \Rj, $\log_{10}$($\text{P}_{ref}$)=[-14,14] bar, and  T=[800-2710] K. The prior on $X_{\text{H}_2\text{O}}$ is chosen to be consistent with all cases investigated in this section. The prior range on $\text{P}_{ref}$, which is also the extent of the model  atmosphere, is chosen so that the edge effects are avoided as discussed above. The retrieved $X_{\text{H}_2\text{O}}$ is completely unconstrained with a retrieved value of $\log_{10}(X_{\text{H}_2\text{O}})=-6.53 ^{+ 3.86 }_{- 3.80 }$. Similarly \Pref is unconstrained with a retrieved value of $\log_{10}($\Pref$)=0.28 ^{+ 9.68 }_{- 9.88 }$ where \Pref is in bar, and R$_{\text{p}}$ is retrieved to \Rp=$1.38 ^{+ 0.10 }_{- 0.10 }$ \Rj, an unconstrained value consistent with \cite{Torres08}. Lastly, the retrieved isothermal temperature is T=$818.08 ^{+ 9.72 }_{- 8.44 }$ K. The posterior distributions for this retrieval are shown in the Appendix in Figure \ref{fig:collection_cases}.

Under these simplistic model considerations, there is a strong three-way degeneracy between \Rp, \Pref and $X_{\text{H}_2\text{O}}$ as expected \citep{hengkitzmann}. However, it is important to note that the degeneracy is a result of unrealistic model assumptions. In addition to the factors discussed in section \ref{sec:reproduction} and later in this section, several other factors deem this case unphysical. First, it is unrealistic to have an atmosphere expanding to such high pressures (e.g. $10^{14}$ bar) while maintaining the isobaric assumption for the cross sections, especially evaluating them at 1mbar. Second, such a deep atmosphere would become opaque at much lower pressures due to the effects of collision-induced absorption \citep[e.g.][]{2013Sci...342.1473D,2018ApJ...865...12B}; this is further explored in section \ref{sub:case2}. Third, assuming a fixed mean molecular weight is unrealistic at high H$_2$O abundances explored in the retrieval such as $X_{\text{H}_2\text{O}}\gtrsim 10^{-2}$ \citep[e.g.][]{bennekeandseager,0004-637X-820-1-78}. Fourth, maintaining a fixed gravity over the entire atmosphere spanning many orders of magnitude in pressure is also unrealistic. Nevertheless, the present case clearly demonstrates the three-way degeneracy between \Rp, \Pref, and $X_{\text{H}_2\text{O}}$ obtained for such a simplistic model while fitting near-infrared data alone. 

We now perform a case by case retrieval study using more realistic model atmospheres as explained at the beginning of section \ref{sec:rets}. All the cases henceforth consider models with a height-dependent $g$, a variable mean molecular weight and a pressure range of $10^{-6} - 10^3$ bar.

\subsection{Case 1: Isobar, isotherm, H$_2$O only and WFC3 data}
\label{sub:case1}

The initial model we now consider is that of an atmosphere which is best described by an isotherm at a temperature T and an isobar with only one molecule present, H$_2$O. For clarity, we specify that the isobaric assumption means evaluating the molecular cross-section at only one pressure, while density, pressure and gravity are still changing with height. Making only these assumptions in our model means neglecting CIA opacity due to  H$_2$-H$_2$ and H$_2$-He. Furthermore, we apply this model on WFC3 data only in order to test the retrievals with a limited wavelength range.

The molecular cross-sections are evaluated at 1mbar following HK17. The result of our retrieval is an unconstrained isotherm with T=$2003.65^{+248.72}_{-247.72}$ K, a H$_2$O mixing fraction of $\log_{10}$(\Xh)=$-9.54 ^{+0.15}_{-0.15}$, a retrieved \Rp=$1.49^{+0.05}_{-0.08}$ \Rj, and $\log_{10}($\Pref$)=-3.00^{+3.67}_{-2.21}$ where \Pref is in bar. The retrieved radius is consistent within 2$\sigma$ with the published photometric radius of \Rp=$1.359^{+0.016}_{-0.019}$ \Rj \citep{Torres08}. However, the reference pressure is not tightly constrained and the retrieved H$_2$O abundance is $\sim 4$ orders of magnitude smaller than that of other studies \citep{madhuhd209,MacDonald17,2017ApJ...834...50B}. The retrieved H$_2$O abundance in this case is also sensitive to the bottom of the  model atmosphere for the same reason as in section \ref{sub:case0}. In this case, the bottom of the atmosphere is at P=$10^3$ bar which limits the amplitude of the H$_2$O feature in the model spectrum, similar to the effects of an opaque surface. Changing the bottom pressure of the atmosphere can result in different H$_2$O abundance constraints. Nevertheless, for the present demonstration we have assumed a physically realistic pressure range of 10$^{3}$ to 10$^{-6}$ bar. Regardless of the pressure range, the present case is inevitably unrealistic due to the lack of various other model considerations which are incorporated in subsequent cases below. More importantly, this edge effect is not relevant once CIA opacities are considered.

\subsection{Case 2: Case 1 + H$_2$/He CIA}
\label{sub:case2}

We now consider a slightly more realistic model which includes CIA opacities due to H$_2$-H$_2$ and H$_2$-He given that the test case of HD~209458~b is a gas giant planet with an H$_2$ dominated atmosphere. All other assumptions about the isothermal and isobaric characterisation of the atmosphere in the model stay the same as in the previous retrieval. However, while we still evaluate the molecular cross-sections at 1mbar, we consider the CIA to be pressure dependent.

The inclusion of CIA decreases the retrieved isothermal temperature to T=$1070.21^{+87.56}_{-92.10}$ K, but increases the H$_2$O mixing fraction to $\log_{10}$(\Xh)=$-5.29 ^{+0.23}_{-0.20}$, a value close to that found in previous retrieval studies \citep{madhuhd209,MacDonald17,2017ApJ...834...50B}. \Rp is now retrieved to be \Rp=$1.41^{+0.02}_{-0.03}$ \Rj, and \Pref in bar to $\log_{10}($\Pref$)=-4.51^{+2.53}_{-1.11}$.

The inclusion of CIA has resulted in a value for the H$_2$O abundance that is consistent with other studies while keeping \Rp consistent with the white light radius within 2$\sigma$. This highlights the importance of considering CIA for constraints on the molecular abundances, as also discussed in section \ref{sec:reproduction}. We find that ignoring CIA leads to erroneous results. CIA opacity determines the highest pressures that can be probed and as a result provides the continuum to the spectrum \citep{2018ApJ...865...12B, 0004-637X-820-1-78}. The inclusion of CIA raises the slant photosphere of the planet to a higher altitude compared to the previous case. By decreasing the thickness of the observed slant column of the atmosphere along the line of sight, a higher abundance is required to explain the same features. In comparison, case 1, where we did not have CIA opacity, the effective column of the atmosphere is larger and hence requires less H$_2$O abundance to explain the same features. Our results show that the molecular abundance is much less biased upon the inclusion of CIA.

While the isobaric assumption makes for a simplified problem construction in analytic models, it is not necessary when numerical methods are available. It is computationally inexpensive to evaluate the molecular opacities at the corresponding pressure in the atmosphere instead of assuming a constant pressure of 1 mbar.
\newpage
\subsection{Case 3: Case 2 without an isobar}
\label{sub:case3}

\begin{figure*}
\centering
\includegraphics[width=\textwidth]{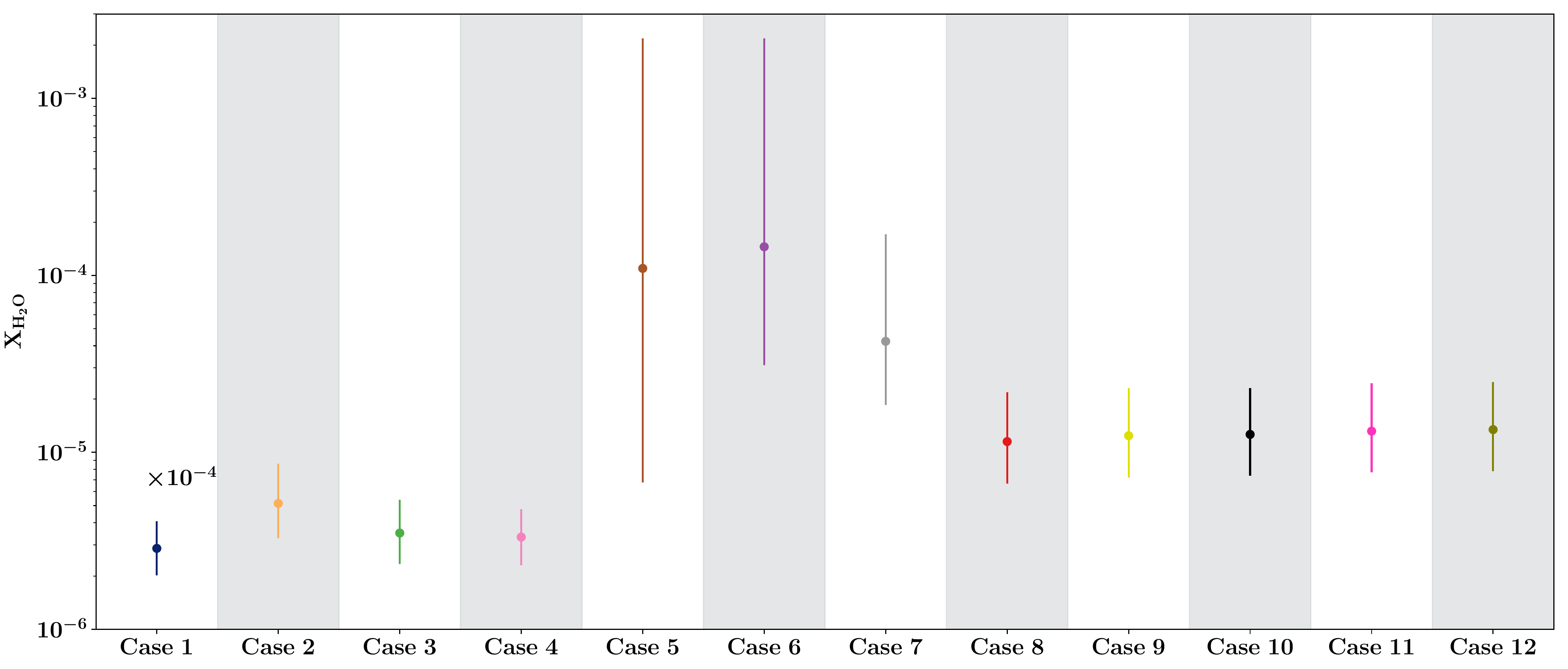}
  \caption[Retrieved H$_2$O abundances in the study of degeneracies]{Retrieved H$_2$O abundances for the different cases shown in Table~\ref{table:cases} and section \ref{sec:rets}. The abundance (i.e. mixing ratio) for case 1 has been increased by 10$^4$ to be in the same range as the abundances of other cases.}
\label{fig:retsummary}
\end{figure*}

We now remove the assumption of an isobar for the calculation of H$_2$O opacities. Instead, we calculate the molecular opacities at the corresponding pressure in the atmosphere rather than at a fixed pressure of 1 mbar.  We maintain the remaining assumption of an isotherm for the temperature profile of the atmosphere. Our retrievals obtain an isothermal profile with T=$1046.02^{+89.50}_{-95.51}$ K and $\log_{10}$(\Xh)=$-5.46 ^{+0.19}_{-0.17}$. The corresponding \Rp and \Pref in bar are \Rp=$1.41^{+0.02}_{-0.03}$ \Rj and $\log_{10}($\Pref$)=-4.35^{+2.63}_{-1.25}$. The retrieved $X_{\text{H}_2\text{O}}$ is shown in Figure~\ref{fig:retsummary}. Retrieved parameters and priors for this and other cases are shown in Tables \ref{table:priors1} and \ref{table:priors2}.

While the consideration of pressure dependent CIA is essential, assuming molecular line cross-sections to be isobaric does not make a significant difference compared to the present case given current data quality. This is because the atmosphere is mostly probed at low pressures as discussed in section \ref{sec:theory}. However, the isobaric assumption cannot be maintained when considering the effects of CIA as the CIA opacity has a stronger dependence on pressure being proportional to the pressure squared \citep{2013Sci...342.1473D}.

\subsection{Case 4: Case 3 + P-T profile}
\label{sub:case4}

We now remove the assumption of an isothermal atmosphere and consider a full P-T profile in our retrieval. We implement the parametrization used in \citet{madhuseager} which involves six parameters that capture a typical P-T profile. Along with this, we retrieve \Xh, \Pref and \Rp. This allows the atmosphere to have any pressure-temperature profile the data requires. 

With the inclusion of the parametric P-T profile, we retrieve nine parameters in total. This retrieval results in \Rp=$1.41^{+0.01}_{-0.03}$ \Rj, $\log_{10}($\Pref$)=-4.36^{+2.35}_{-1.18}$ in bar and $\log_{10}$(\Xh)=$-5.48^{+0.16}_{-0.16}$. The retrieved P-T profile has the following parameters $\log_{10}(P_1)=-0.77^{+1.88}_{-2.35}$, $\log_{10}(P_2)=-3.61^{+2.40}_{-1.62}$, $\log_{10}(P_3)=1.45^{+1.10}_{-1.76}$, $\alpha_1=0.85^{+0.11}_{-0.14}$, $\alpha_2=0.67^{+0.22}_{-0.32}$ and temperature of T$_0$=$870.11^{+82.12}_{-49.12}$ K. The retrieved values did not change significantly compared to the assumption of an isothermal atmosphere as in case 3. These numerical results agree with analytic studies that predict that while non-isothermal atmospheres distort the spectrum of an isothermal one, the effects are subtle considering present data quality with HST \citep{2018ApJ...865...12B}. The retrieved mixing fraction of H$_2$O is consistent with that of other studies \citep{madhuhd209,MacDonald17,2017ApJ...834...50B}. 

\subsection{Case 5: Case 4 + Full cloud cover}
\label{sub:case5}

We continue to remove assumptions from our model and now consider the possibility of clouds being present in the atmosphere of the planet. There is no a priori information to assume that the atmosphere of HD~209458~b is cloud-free. We consider the cloud prescription of \citet{MacDonald17} as explained at the beginning of this section. We include four parameters for clouds and hazes. For hazes we use  $a$, the Rayleigh-enhancement factor, and $\gamma$, the scattering slope. For clouds, P$_{\text{cloud}}$ and $\phi$ characterise the pressure level of the optically thick cloud deck and cloud coverage fraction respectively. In this particular case, instead of a clear atmosphere like we did in case 4, we consider the presence of a fully cloudy planet atmosphere by fixing  $\phi=100\%$.

The inclusion of a fully cloudy deck increases the number of retrieved parameters from nine to twelve. The retrieved H$_2$O abundance is $\log_{10}$(\Xh)=$-3.96^{+1.30}_{-1.21}$, while \Rp is \Rp=$1.38^{+0.04}_{-0.06}$ \Rj, and \Pref is $\log_{10}(\text{P}_{ref})=-3.84^{+2.17}_{-1.37}$ in bar. The values of the retrieved parameters for the P-T profile are T$_0$=$1940.56^{+251.87}_{-305.81}$ K, $\alpha_1=0.68^{+0.20}_{-0.23}$, $\alpha_2=0.60^{+0.24}_{-0.25}$, $\log_{10}(P_1)=-1.30^{+1.88}_{-1.78}$, $\log_{10}(P_2)=-3.86^{+1.88}_{-1.39}$, $\log_{10}(P_3)=1.19^{+1.16}_{-1.66}$. The cloud parameters are $\log_{10}(a)=0.39^{+3.81}_{-2.73}$, $\gamma=-12.14^{+7.38}_{-4.97}$ and $\log_{10}(P_{\text{cloud}})=-2.74^{+1.24}_{-1.27}$.
While the value of the retrieved planetary radius is still consistent with the observed radius, we see that the 1$\sigma$ limits have increased. Similar effects are seen with the retrieved H$_2$O abundance.  

The interesting effect of the inclusion of a fully covering cloud deck is that the H$_2$O abundance is now hardly constrained. Since the pressure at which this cloud deck could be located spans several orders of magnitude, so does the H$_2$O abundance. In this case, the cloud deck mimics a surface \citep{2017MNRAS.467.2834B} and the pressure at which the cloud is located is fully degenerate with the retrieved H$_2$O abundance. A cloud deck at a higher altitude requires higher H$_2$O abundance to account for the same features, while a lower cloud deck can explain the same features with a lower molecular abundance \citep[e.g.][]{deming_hd209, 2017ApJ...834...50B}. An alternate way to explain this is similar to what happened with the inclusion of CIA in case 2. By lowering the cloud deck altitude (i.e. increasing the cloud-top pressure), we are increasing the effective column of the observable atmosphere which requires lower abundance than a smaller observable atmosphere corresponding to a cloud deck at a higher altitude (i.e. decreasing the pressure). We also notice that the lowest H$_2$O abundance is consistent with the lowest abundance found in case 2, due to CIA providing the continuum opacity.

\subsection{Case 6: Case 4 + Non-homogeneous clouds}
\label{sub:case6}

We now consider the effects of non-homogeneous cloud coverage on constraining the H$_2$O abundance. We include the cloud fraction $\phi$ as an extra free parameter. The retrieved parameters are $\log_{10}$(\Xh)=$-3.83^{+1.17}_{-0.67}$, \Rp=$1.35^{+0.03}_{-0.05}$ \Rj, and $\log_{10}($\Pref$)=-3.61^{+2.46}_{-1.48}$ in bar. The P-T profile parameters are T$_0$=$1262.74^{+225.05}_{-230.98}$ K, $\alpha_1=0.65^{+0.21}_{-0.21}$, $\alpha_2=0.60^{+0.25}_{-0.25}$, $\log_{10}(P_1)=-1.22^{+1.86}_{-1.84}$, $\log_{10}(P_2)=-3.90^{+2.00}_{-1.34}$, $\log_{10}(P_3)=1.27^{+1.13}_{-1.65}$. The cloud parameters are $\log_{10}(a)=2.09^{+3.93}_{-3.88}$, $\gamma=-8.60^{+7.87}_{-7.40}$, $\log_{10}(P_{\text{cloud}})=-4.70^{+1.33}_{-0.85}$ and $\phi=0.68^{+0.05}_{-0.06}$.

The inclusion of non-homogeneous clouds does not change significantly the retrieved P-T profile parameters. It, however, allows to put a constraint on the cloud fraction at $\sim 68 \%$. Furthermore, \Rp and \Pref are consistent with those of cases 4 and 5.

While the median value of the retrieved H$_2$O abundance is consistent with that of case 5, the uncertainty is smaller when non-homogeneous clouds are considered. Considering a non-homogeneous cloud cover allows for a better H$_2$O constraint compared to the assumption of a fully cloudy atmosphere. It is true that the constraints in the case of a clear atmosphere are even tighter (e.g. case 4), but the validity of this assumption is not evident. Furthermore, previous studies suggest that failure to consider non-homogeneous cloud cover can bias molecular abundance findings \citep{0004-637X-820-1-78}.

We now look into other factors that could help further constrain molecular abundances. Until now we have only considered HST WFC3 data in the near-infrared for retrievals with different model assumptions. Given that the main differences in spectra with clouds manifest in the optical wavelengths, we now incorporate data in the optical.

\subsection{Case 7: Case 6 + Optical data}
\label{sub:case7}

Our seventh case considers the inclusion of an optical spectrum of HD~209458~b. We included data from 0.30 to 0.95 \mum\, from \citet{sing2016}. The addition of optical data helps constrain the Rayleigh slope and cloud properties \citep{2014RSPTA.37230086G, bennekeandseager, 0004-637X-820-1-78}. This also allows us to evaluate the effects of more data considered in our retrieval. We keep the number of parameters the same as in case 6 for a total of thirteen. We report a retrieved H$_2$O abundance of $\log_{10}$(\Xh)=$-4.37^{+0.61}_{-0.36}$. The planetary radius retrieved is \Rp=$1.35^{+0.03}_{-0.05}$ \Rj, and $\log_{10}($\Pref$)=-3.59^{+2.49}_{-1.54}$ is the retrieved reference pressure.

The retrieved values for the P-T profile parameters and cloud parameters after including optical data are T$_0$=$1306.40^{+225.24}_{-257.81}$ K, $\alpha_1=0.71^{+0.19}_{-0.22}$, $\alpha_2=0.67^{+0.21}_{-0.26}$, $\log_{10}(P_1)=-1.34^{+2.02}_{-1.84}$, $\log_{10}(P_2)=-3.79^{+1.99}_{-1.45}$, $\log_{10}(P_3)=1.23^{+1.20}_{-1.91}$. The cloud parameters are $\log_{10}(a)=7.65^{+0.24}_{-0.43}$, $\gamma=-8.97^{+1.07}_{-0.88}$, $\log_{10}(P_{\text{cloud}})=-5.29^{+0.25}_{-0.16}$ and $\phi=0.69^{+0.04}_{-0.05}$.

As expected, the parameters most affected, compared to case 6, are those responsible for clouds and hazes. The inclusion of data in the optical allows us to place tighter constraints on $a$ and $\gamma$ which characterise the slope in the optical. The cloud parameters are consistent with those of case 6 with $\phi$ mostly unchanged. However, the uncertainty in $\log_{10}(P_{\text{cloud}})$ is smaller by almost a factor of 6 compared to the values in case 6. Naturally, given that we now have information in the wavelength range where the scattering slope manifests itself, our cloud and haze prescription can fit for it, in contrast to previous cases where we fit for the slope without adequate data in the optical range.

Furthermore, by constraining the baseline of the spectrum, we are now able to place better constraints on the H$_2$O abundance. The uncertainties on the H$_2$O abundance are half as small as the ones from case 6. Thus, it is evident that the inclusion of optical data allows for better estimates of chemical abundances. Our numerical results show the importance of short wavelengths in breaking key degeneracies and in better constraining molecular abundances in agreement with previous analytic studies \citep[e.g.][]{2014RSPTA.37230086G, 0004-637X-820-1-78, bennekeandseager}.

The last step in increasing the physical reality of our model is to allow for the presence of more molecules in our atmosphere. This would prevent our models from trying to explain every spectroscopic feature with only one molecule. Furthermore, a possible way to break the degeneracy between \Rp and mixing ratios is to consider the absorption features of different absorbers \citep{bennekeandseager}. In the next cases, we incorporate several species that can be prominent in hot Jupiter atmospheres, e.g., Na, K, NH$_3$, CO, HCN and CO$_2$  \citep{madhu_molecules}.

\subsection{Case 8: Case 7 + Na and K }
\label{sub:case8}

The first species we incorporate are the alkali atomic species Na and K. Given that their spectroscopic features are present in the range covered by the additional optical data, we investigate the impact these species have on the retrieved H$_2$O abundances. The values retrieved for the species are $\log_{10}$(\Xh)=$-4.94^{+0.28}_{-0.24}$, $\log_{10}(X_{\text{Na}})=-5.55^{+0.53 }_{-0.44}$, and $\log_{10}(X_{\text{K}})=-7.17 ^{+ 0.55 }_{- 0.52 }$. The retrieved \Rp is \Rp=$1.37 ^{+ 0.02 }_{- 0.04 }$ \Rj and \Pref is $\log_{10}($\Pref$)=-3.46 ^{+ 2.33 }_{- 1.67 }$ in bar. For completion, we continue to show our retrieved values for all other parameters as in previous cases. The P-T profile parameters are  T$_0$=$1064.75 ^{+ 283.29 }_{- 185.88 }$ K, $\alpha_1=0.59 ^{+ 0.25 }_{- 0.17 }$, $\alpha_2=0.47 ^{+ 0.33 }_{- 0.21 }$, $\log_{10}(P_1)=-1.16 ^{+ 1.98 }_{- 1.79 }$, $\log_{10}(P_2)=-3.89 ^{+ 2.24 }_{- 1.43 }$, $\log_{10}(P_3)=1.22 ^{+ 1.21 }_{- 1.60 }$. The non-homogeneous cloud parameters are $\log_{10}(a)=4.42 ^{+ 0.71 }_{- 1.26 }$, $\gamma=-14.42 ^{+ 5.59 }_{- 3.76 }$, $\log_{10}(P_{\text{cloud}})=-4.69 ^{+ 0.77 }_{- 0.50 }$ and $\phi=0.49 ^{+ 0.06 }_{- 0.06 }$. Our retrieved H$_2$O abundance for this and other cases is shown in Figure~\ref{fig:retsummary} and all retrieved parameters are summarized in Tables \ref{table:priors1} and \ref{table:priors2}.

The inclusion of Na and K has further decreased the 1$\sigma$ spread of the retrieved H$_2$O abundance almost by a factor of two. While the retrieved H$_2$O abundance is consistent with that of case 7, it is important to note that the posterior distribution has shifted towards a lower H$_2$O abundance by $\sim0.5$ dex. This shift in the median value is as much as the shift between case 6 and case 7 due to the inclusion of optical data. This suggests that the Na and K, which themselves are constrained by the optical data, also strongly affect the retrieved H$_2$O abundance. This is due to better fitting the features in the optical.  An additional effect is the change in the retrieved cloud fraction from $\sim 70\%$ in case 7 to $\sim 50\%$. Evidently, these results will be sensitive to the absorption cross-sections being used. Nonetheless, it is clear that including molecules that have signatures in the optical allow us to fit the data in those wavelengths better and further constraint the H$_2$O abundance. This has little effect on the retrieved \Rp and \Pref which continue to be well constrained. 

\subsection{Case 9: Case 8 + NH$_3$}
\label{sub:case9}

Next, we include NH$_3$ as a source of opacity. The retrieval gives the following results for molecular abundances $\log_{10}$(\Xh)=$-4.91 ^{+ 0.27 }_{- 0.24 }$, $\log_{10}(X_{\text{Na}})=-5.53 ^{+ 0.51 }_{- 0.43 }$, $\log_{10}(X_{\text{K}})=-7.13 ^{+ 0.54 }_{- 0.51 }$, and $\log_{10}(X_{\text{NH}_3})=-8.02 ^{+ 1.86 }_{- 2.64 }$. The retrieved planetary radius and reference pressure are \Rp=$1.37 ^{+ 0.02 }_{- 0.04 }$ \Rj  and $\log_{10}($\Pref$)=-3.39 ^{+ 2.43 }_{- 1.69 }$ respectively. The P-T profile parameters and cloud parameters are   T$_0$=$1026.44 ^{+ 276.52 }_{- 161.11 }$ K, $\alpha_1=0.62 ^{+ 0.24 }_{- 0.18 }$, $\alpha_2= 0.49 ^{+ 0.32 }_{- 0.22 }$, $\log_{10}(P_1)=-1.18 ^{+ 1.97 }_{- 1.77 }$, $\log_{10}(P_2)=-3.95 ^{+ 2.19 }_{- 1.39 }$, $\log_{10}(P_3)=1.25 ^{+ 1.18 }_{- 1.59 }$, $\log_{10}(a)=4.38 ^{+ 0.70 }_{- 1.16 }$, $\gamma=-14.67 ^{+ 5.19 }_{- 3.57 }$, $\log_{10}(P_{\text{cloud}})=-4.57 ^{+ 0.77 }_{- 0.56 }$ and $\phi=0.47 ^{+ 0.06 }_{- 0.08 }$.

In comparison to case 8, the inclusion of NH$_3$ does not change significantly the retrieved H$_2$O abundance. Meanwhile, \Rp and \Pref also remain mostly unchanged. Although both NH$_3$ and H$_2$O have absorption features in the WFC3 spectral range, the inclusion of NH$_3$ does not affect our retrieved H$_2$O abundance. This is because H$_2$O has much stronger features than NH$_3$ in the WFC3 range; H$_2$O is also expected to be more abundant than NH$_3$ at hot Jupiter temperatures. As such, cumulative opacity of NH$_3$ is generally weaker than that of H$_2$O, as also seen in previous studies \citep{MacDonald17,2017ApJ...850L..15M}. 

\subsection{Case 10: Case 9 + CO}
\label{sub:case10}

We proceed by adding CO to our model. The retrieved molecular abundances are  $\log_{10}$(\Xh)=$-4.90 ^{+ 0.26 }_{- 0.23 }$, $\log_{10}(X_{\text{Na}})=-5.52 ^{+ 0.52 }_{- 0.43 }$, $\log_{10}(X_{\text{K}})=-7.11 ^{+ 0.54 }_{- 0.49 }$, \newline $\log_{10}(X_{\text{NH}_3})=-8.14^{+1.95}_{-2.56}$, and $\log_{10}(X_{\text{CO}})=-7.74 ^{+ 2.85 }_{- 2.72 }$. The P-T profile parameters are T$_0$=$1026.72 ^{+ 262.72 }_{- 157.60 }$ K, $\alpha_1=0.61 ^{+ 0.23 }_{- 0.18 }$, $\alpha_2= 0.49 ^{+ 0.32 }_{- 0.22 }$, $\log_{10}(P_1)=-1.14 ^{+ 1.94 }_{- 1.77 }$, $\log_{10}(P_2)=-3.87 ^{+ 2.15 }_{- 1.44 }$, and $\log_{10}(P_3)=1.29 ^{+ 1.15 }_{- 1.55 }$. The cloud parameters are $\log_{10}(a)=4.38 ^{+ 0.69 }_{- 1.14 }$, $\gamma=-14.70 ^{+ 5.17 }_{- 3.55 }$, $\log_{10}(P_{\text{cloud}})=-4.57 ^{+ 0.76 }_{- 0.54 }$ and $\phi=0.47 ^{+ 0.06 }_{- 0.08 }$. Lastly, the reference pressure and reference radius that we retrieve are \Rp=$1.37 ^{+ 0.02 }_{- 0.04 }$ \Rj and $\log_{10}($\Pref$)=-3.42 ^{+ 2.33 }_{- 1.66 }$ respectively. With all values being consistent with those presented in case 9, it is clear that the inclusion of CO did not affect the retrieved values because of the weak CO features in the WFC3 band.

\subsection{Case 11: Case 10 + HCN}
\label{sub:case11}

Second to last, we include HCN which also has some features in the WFC3 band. The resulting retrieved planetary radius and reference pressure are consistent with those of case 10 with retrieved values of \Rp=$1.37 ^{+ 0.02 }_{- 0.04 }$ \Rj and $\log_{10}($\Pref$)=-3.45 ^{+ 2.26 }_{- 1.63 }$. The retrieved molecular abundances are also consistent with values of $\log_{10}$(\Xh)=$-4.88 ^{+ 0.27 }_{- 0.23 }$, $\log_{10}(X_{\text{Na}})=-5.50 ^{+ 0.51 }_{- 0.42 }$, $\log_{10}(X_{\text{K}})=-7.09 ^{+ 0.54 }_{- 0.50 }$, $\log_{10}(X_{\text{NH}_3})=-8.11^{+ 1.90 }_{- 2.52 }$, and $\log_{10}(X_{\text{CO}})=-7.75 ^{+ 2.82 }_{- 2.73 }$. The additional molecule resulted in a retrieved abundance of   $\log_{10}(X_{\text{HCN}})=-8.60 ^{+ 2.26 }_{- 2.17 }$. Also consistent are the P-T profile parameters at T$_0$=$1013.53 ^{+ 248.73 }_{- 149.27 }$ K, $\alpha_1=0.62 ^{+ 0.23 }_{- 0.18 }$, $\alpha_2= 0.49 ^{+ 0.31 }_{- 0.22 }$, $\log_{10}(P_1)=-1.15 ^{+ 1.93 }_{- 1.78 }$, $\log_{10}(P_2)=-3.90 ^{+ 2.14 }_{- 1.42 }$, and $\log_{10}(P_3)=1.26 ^{+ 1.16 }_{- 1.55 }$. The cloud parameters are $\log_{10}(a)=4.34 ^{+ 0.69 }_{- 1.11 }$, $\gamma=-14.63 ^{+ 4.99 }_{- 3.59 }$, $\log_{10}(P_{\text{cloud}})=-4.52 ^{+ 0.73 }_{- 0.54 }$ and $\phi=0.46 ^{+ 0.06 }_{- 0.08 }$ which are also consistent. Similar to NH$_3$, our constraint on HCN is also weaker given current data. Our constraint, however, is consistent with the mixing ratio of $\sim 10^{-6}$ which was required to detect HCN on the dayside of the planet \citep{Hawker2018}.

\subsection{Case 12: Case 11 + CO$_2$}
\label{sec:fullret}

We add one last molecule, CO$_2$, in order to have what we refer to as a full retrieval. This is the equivalent to a state-of-the-art retrieval in which several molecules and atomic species are considered, a parametric P-T profile,  and non-homogeneous clouds, totalling 19 free parameters.  This retrieval gives us an atmosphere with the following molecular abundances $\log_{10}$(\Xh)=$-4.87 ^{+ 0.27 }_{- 0.24 }$, $\log_{10}(X_{\text{Na}})=-5.48 ^{+ 0.52 }_{- 0.43 }$, $\log_{10}(X_{\text{K}})=-7.07 ^{+ 0.54 }_{- 0.51 }$, $\log_{10}(X_{\text{NH}_3})=-8.09 ^{+ 1.89 }_{- 2.54 }$, $\log_{10}(X_{\text{CO}})=-7.73 ^{+ 2.79 }_{- 2.75 }$, $\log_{10}(X_{\text{HCN}})=-8.57 ^{+ 2.24 }_{- 2.22 }$, and $\log_{10}(X_{\text{CO}_2})=-8.46 ^{+ 2.43 }_{- 2.30 }$. The retrieved P-T parameters are T$_0$=$1022.15 ^{+ 246.41 }_{- 153.72 }$ K, $\alpha_1=0.60 ^{+ 0.23 }_{- 0.17 }$, $\alpha_2= 0.50 ^{+ 0.31 }_{- 0.23 }$, $\log_{10}(P_1)=-1.03 ^{+ 1.88 }_{- 1.79 }$, $\log_{10}(P_2)=-3.86 ^{+ 2.16 }_{- 1.42 }$, and $\log_{10}(P_3)=1.33 ^{+ 1.13 }_{- 1.54 }$. The retrieved cloud parameters are $\log_{10}(a)=4.37 ^{+ 0.68 }_{- 1.08 }$, $\gamma=-14.61 ^{+ 4.93 }_{- 3.58 }$, $\log_{10}(P_{\text{cloud}})=-4.52 ^{+ 0.72 }_{- 0.55 }$ and $\phi= 0.46 ^{+ 0.06 }_{- 0.08 }$. The retrieved planetary radius is \Rp=$1.37 ^{+ 0.02 }_{- 0.04 }$ \Rj, and the retrieved reference pressure is $\log_{10}($\Pref$)=-3.45 ^{+ 2.24 }_{- 1.63 }$.

Overall, with the inclusion of all the effects discussed we find that the combination of near-infrared and optical data allow strong constraints on several important parameters and in resolving key degeneracies. The H$_2$O abundance is tightly constrained and it is consistent with values of previous studies \citep[e.g][]{MacDonald17,2017ApJ...834...50B}. Other chemical species are less well constrained owing to their weaker opacities in the observed range. Nonetheless, retrieved abundance estimates are consistent with studies that investigate their presence in the planet's atmosphere, e.g. detection of HCN \citep{Hawker2018}. While the abundance of H$_2$O is retrieved, \Pref and \Rp are also retrieved with the later being consistent with the observed photometric radius of \Rp=$1.359^{+0.016}_{-0.019}$ \Rj \citep{Torres08}. The full retrieval has resolved the degeneracy between \Xh, \Rp and \Pref. Simultaneously, the cloud fraction is retrieved with tight constraints on its value indicating that the planet is not cloud free. The inclusion of multiple absorbers in our retrievals helps break key degeneracies in our results. One of the advantages of the retrieval technique is that robust models (i.e. that consider parametric P-T profiles, with many molecules, and partial clouds) can be implemented efficiently.

\subsection{Key lessons}

Here we summarise the results from our case study of HD~209458~b based on retrievals with various model assumptions. Overall, with the inclusion of all the effects, we find that the combination of near-infrared and optical data are responsible for strong constraints on several important parameters resolving key degeneracies. The combination of data and accurate models allows for high precision retrievals that impose tight constraints on the H$_2$O abundance, \Rp, \Pref, and the cloud fraction. The retrieved H$_2$O abundances under different model assumptions are shown in Figure~\ref{fig:retsummary}. The full retrieval is able to also estimate the abundance of other chemical species like HCN.

The retrieval's ability to constrain the H$_2$O abundance is not affected by \Rp and \Pref. We find that is is possible to simultaneously retrieve both \Rp and \Pref and find values for \Rp in agreement with the observed photometric radius. We also analyse the impact of the cloud fraction and the potential degeneracy between this parameter and the planetary radius and the H$_2$O abundance. We first find that there are strong differences in the retrieved H$_2$O abundances between a cloud free and fully cloudy atmosphere. Assuming a fully cloudy atmosphere introduces a degeneracy between the H$_2$O abundance and the pressure at which the cloud deck is located, since the cloud deck has the same effect on the transmission spectrum as the optically thick photosphere. An alternative to this, is to consider non-homogeneous cloud coverage in the atmosphere of the planet as there is no a priori information that favours a cloud free atmosphere or a 100\% cloudy atmosphere. On the other hand, theoretical models suggest the presence of partial clouds at the day-night terminators, i.e., the limbs, of planets \citep[e.g.][]{2016ApJ...828...22P, 2016ApJ...821....9K}. We also find that in order to better constrain the clouds and hazes, it is important to consider data points in the optical wavelength range where clouds and hazes manifest themselves. We find that there is no degeneracy between cloud fraction and radius of the planet. Furthermore, it can be seen that it is not necessary to assume a fixed cloud fraction, and instead it is better to allow for the cloud fraction to be a free parameter in the retrieval. 

A crucial lesson of our study is that CIA opacity is key in constraining molecular abundances in both clear and cloudy atmospheres. The lack of CIA due to H$_2$-H$_2$, H$_2$-He in the model skews the retrieved H$_2$O abundance by several orders of magnitude. Once CIA contribution is considered, the retrieved abundances are consistent within one order of magnitude. The CIA opacity strictly limits the location of the planetary photosphere and, hence, the column of the atmosphere above the photosphere that is probed by the observed spectrum. Without CIA the photosphere will lie deeper in the atmosphere, increasing the observable column. As such the molecular abundances will be higher when considering CIA in comparison to models without CIA. 

The inclusion of optical data in retrievals is paramount to provide highly constrained H$_2$O abundances while helping constrain the range of possible planetary radii and their associated reference pressures. In addition, we find that Na and K absorption lines in the optical significantly affect the constraints on H$_2$O abundances. The availability of a broad spectral range between optical and near-infrared helps provide joint constraints on the H$_2$O abundance and the reference pressure or the planetary radius. On the other hand, strong degeneracy still persists between the \Rp and \Pref without affecting the H$_2$O abundance. This relationship is further discussed in section \ref{sec:theory}. Optical data also allows for tight constraints on the cloud fraction of the planet, making it possible to asses whether a planet is cloud-free or not. In the next section we investigate the effectiveness of the cloud parametrization and its ability to constrain the cloud fraction in the utmost case of a fully cloudy atmosphere.

\begin{figure}
\centering
\includegraphics[width=0.5\textwidth]{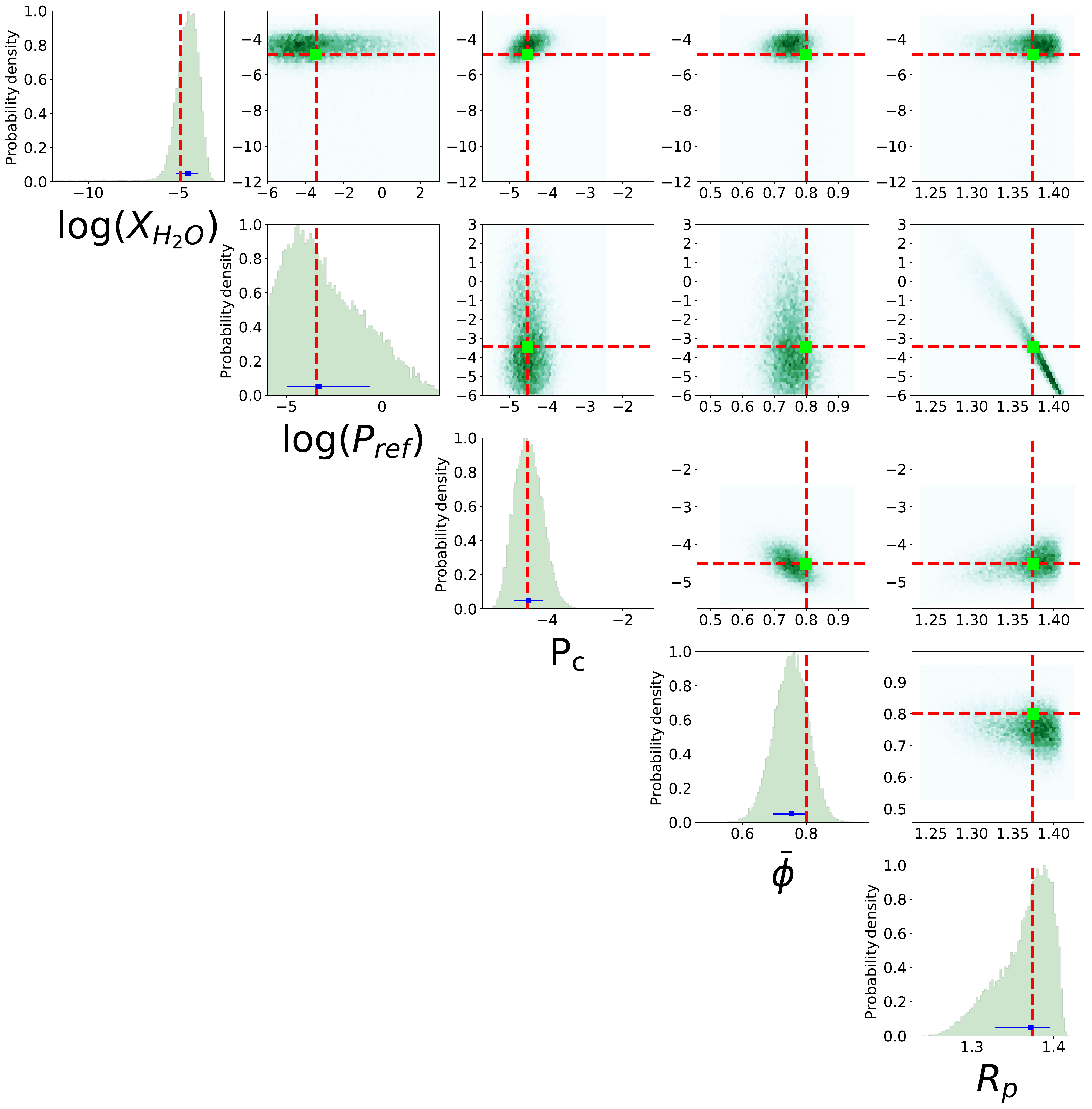}
  \caption[Simulated retrieval for assumed 80\% cloud cover]{Posterior distributions of the retrieval for simulated data of HD~209458~b with 80\% cloud coverage. The red dotted lines show the value of the simulated parameters.}
\label{fig:80sim}
\end{figure}

\begin{figure}
\centering
\includegraphics[width=0.5\textwidth]{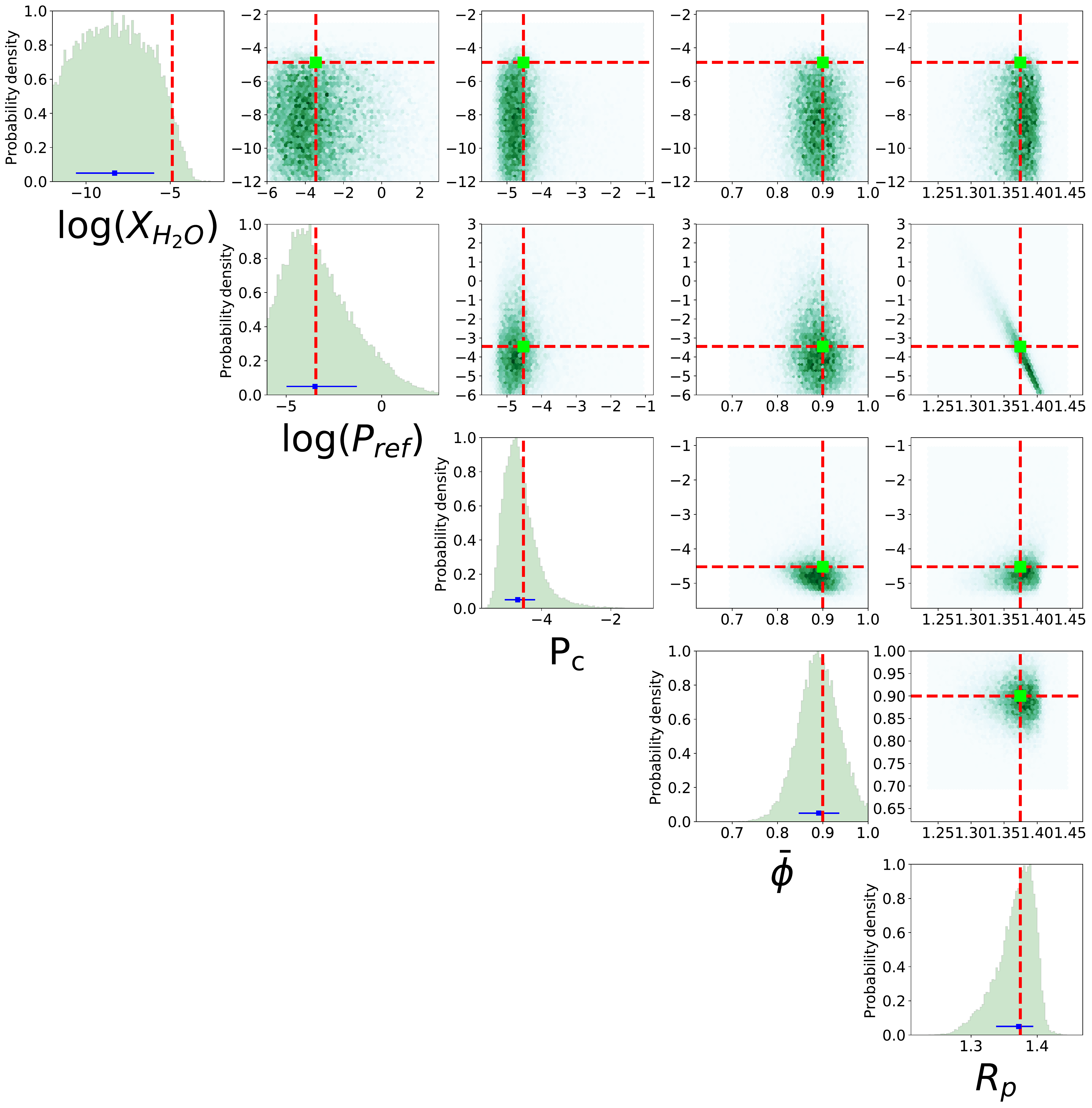}
  \caption[Simulated retrieval for assumed 90\% cloud cover]{Posterior distributions of the retrieval for simulated data of HD~209458~b with 90\% cloud coverage. The red dotted lines show the value of the simulated parameters.}
\label{fig:90sim}
\end{figure}

\begin{figure}
\centering
\includegraphics[width=0.5\textwidth]{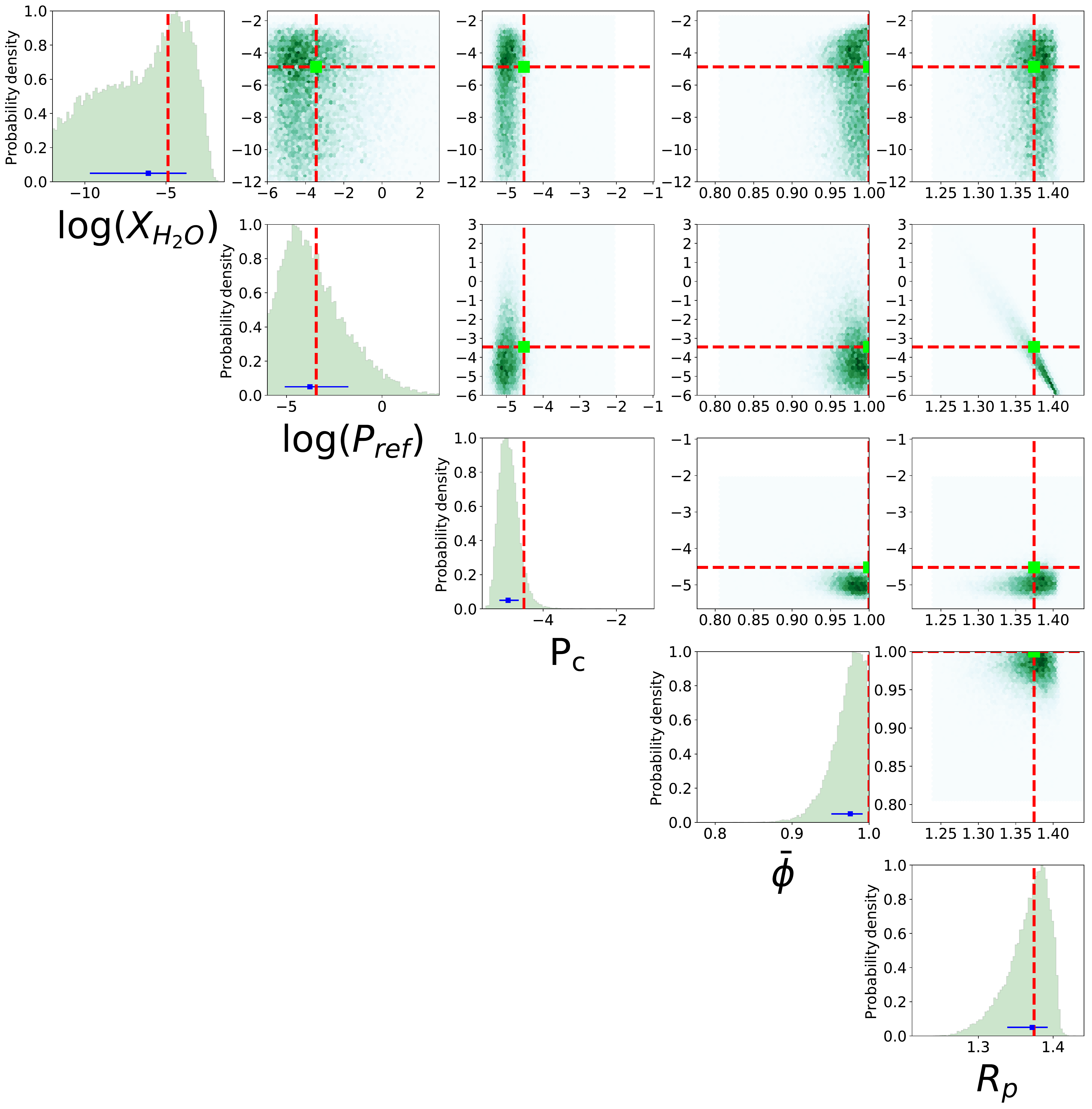}
  \caption[Simulated retrieval for assumed 100\% cloud cover]{Posterior distributions of the retrieval for simulated data of HD~209458~b with 100\% cloud coverage. The red dotted lines show the value of the simulated parameters.}
\label{fig:100sim}
\end{figure}

\section{Solutions to homogeneous cloud cover}
\label{sec:simretrieval}

Here we investigate the robustness with which clouds can be constrained. In particular, we focus on the ability to retrieve the cloud fraction of the atmosphere $\phi$ in the worst-case scenario of a fully cloudy atmosphere (i.e. $\phi=100\%$). It can be argued that a 100\% cloud deck leads to an entirely degenerate set of solutions for the H$_2$O abundances, as seen in section \ref{sub:case5}. This leads to the question of whether an inhomogeneous cloud prescription can resolve this problem. In order to answer this question, we investigate the potential of retrievals to estimate the cloud fraction covering a planet's atmosphere.  For this, we consider the median values for the full retrieval of HD~209458~b, performed in section \ref{sec:fullret}, which includes data in the near-infrared and optical ranges, multiple molecules, a parametric P-T profile and clouds. We use these values to generate three synthetic data sets with three different cloud fractions. The simulated data has the same resolution, error, and wavelength range as the data in \citet{sing2016}. In our simulated data we add random error to the binned transit depth drawn from a normal distribution. The simulated models have cloud fractions of 100\%, 90\% and 80\%. 

Figures \ref{fig:80sim}, \ref{fig:90sim}, \ref{fig:100sim} show the results of our retrievals along with the values of the parameters used in the simulated data. For all cloud fractions ($\phi$), our retrieved molecular abundances are consistent with the input value within 2$\sigma$. H$_2$O can be reliably estimated for $\phi \lesssim 80\%$. For higher $\phi$, only upper-limits are found but the $\phi$ is accurately retrieved. $\phi$ is always retrieved within $\sim 1 \sigma$. Furthermore, the retrieved $\phi$, \Rp, and \Pref are consistent with the input values in all cases. These results demonstrate that the retrieval technique can discern the cloud fraction covering the planet's atmosphere without compromising the ability to retrieve other properties.

 The worst-case scenario would be an atmosphere with 100\% cloud coverage at a very high altitude, as in the present case. Such a high-altitude cloud deck mutes almost all spectral features, resulting in a flat spectrum. Although no molecular abundances are reliably constrained for this case, the cloud fraction is still correctly retrieved to be consistent with 100\% as shown in Figure~\ref{fig:100sim}. In principle, a 100\% cloud deck at a lower altitude, e.g., at 10 mbar pressure level, would still have some spectral features. Stronger spectral features result in better constraints on the model parameters even for 100\% cloud coverage given adequate data in the optical and infrared. Lower cloud fractions are naturally retrievable in all these cases. These results are consistent with the studies of \citet{MacDonald17} and agree that non-uniform cloud coverage in models allows for a more precise determination of chemical abundances in transmission spectra in comparison to models that assume a fixed cloud fraction, effectively breaking the cloud-composition degeneracies. These results also show that non-homogeneous and homogeneous cloud scenarios are distinguishable, in agreement with \cite{0004-637X-820-1-78}.

\section{The \Rp-\Pref degeneracy}
\label{sec:theory}

As discussed in section \ref{sec:intro}, several recent studies have highlighted possible degeneracies between chemical abundances, clouds/hazes, and reference radius in interpreting transmission spectra \citep[e.g.][]{2008A&A...481L..83L, 2014RSPTA.37230086G, bennekeandseager, 2013Sci...342.1473D,0004-637X-820-1-78, hengkitzmann}. In sections \ref{sec:rets} and \ref{sec:simretrieval} we demonstrate that the combination of multi-band data and realistic models can lead to precise constraints on key chemical abundances, in this case H$_2$O, along with other properties.

Recently, HK17 inferred a three-way degeneracy between \Rp, \Pref and $X_{\text{H}_2\text{O}}$ as a fundamental hindrance for deriving chemical abundances. They argue that one way to break the three-way degeneracy is to find a functional relationship between \Rp and \Pref. In this section we interpret our results from section \ref{sec:reproduction} and \ref{sec:rets} and present an empirical relation between \Rp and \Pref.  We show that previous suggestions of a three-way degeneracy are the result of model simplifications and inadequate data, and that the primary degeneracy is between \Rp and \Pref.

The relationship between \Rp and $\text{P}_{ref}$ is explored when a fit or retrieval is performed. We briefly revisit our reproduction of previous semi-analytic results from section \ref{sec:reproduction} and our retrievals from section \ref{sec:rets}. We begin by revisiting Figure~\ref{fig:reproduction}, which shows a linear relationship between \Rp and $\log_{10}$(\Xh \Pref) obtained from fitting a near-infrared WFC3 spectrum of the hot Jupiter WASP-12b. We find that the slope can be described as $m=-1/(H\ln10)$, where $H$ is the atmospheric scale height. The slope of the linear fit obtained by HK17 and reproduced by us is m=-85.77. Using the above relationship, this slope is consistent with a scale height of 362 km, matching the estimated value for this planet reported in HK17.

\begin{figure}
\centering
\includegraphics[width=0.5\textwidth]{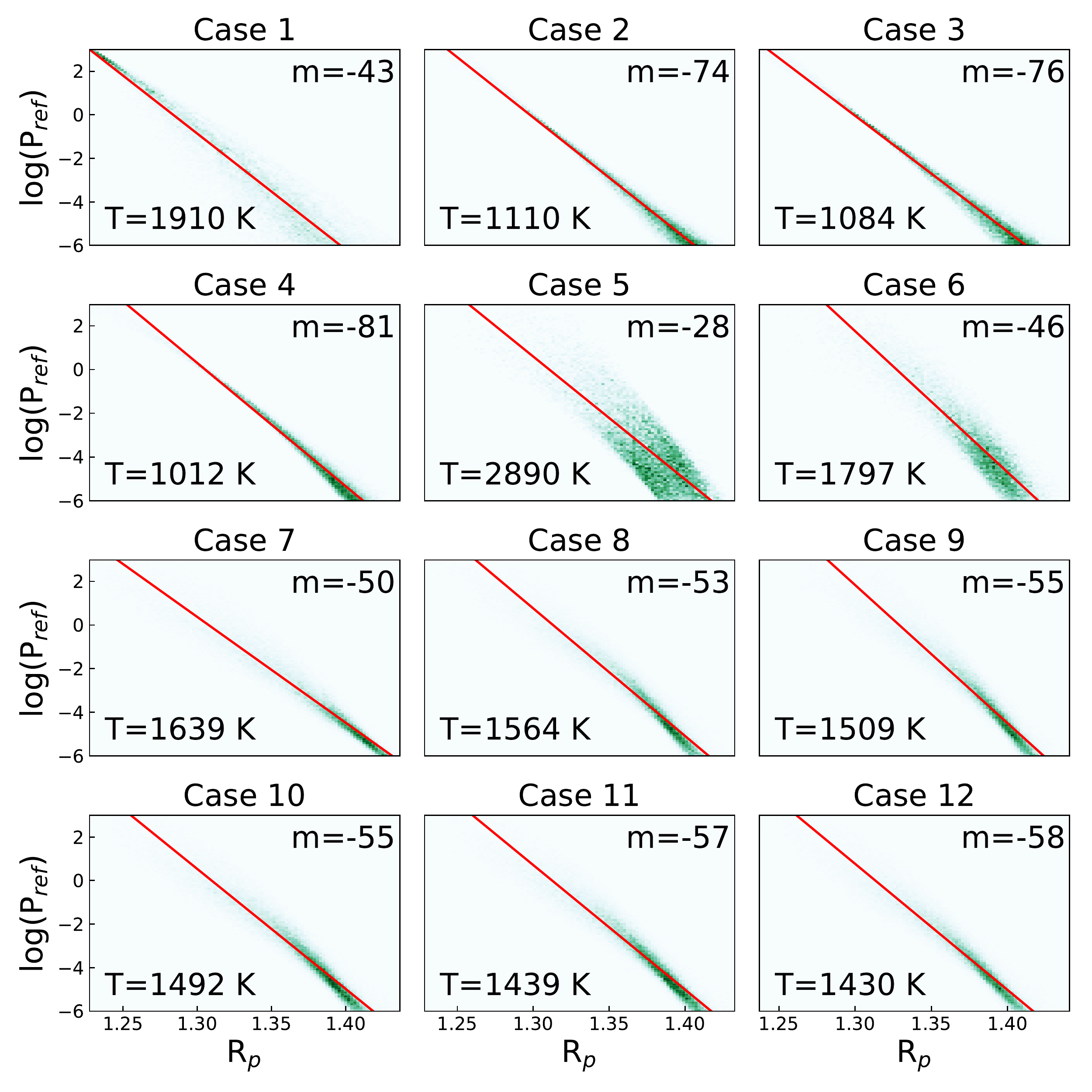}
\caption[Linear fit to the H$_2$O abundance vs planetary radius relationship]{Relationship between the retrieved planetary radius (\Rp in units of \Rj) and the associated reference pressure (\Pref in units of bar). A linear fit is shown in red along with the slope in each panel. In the top right corner of each panel, the slope of a linear fit is presented. On the bottom left corner we present the temperature (in units of K) derived from the best fit slope assuming it is defined as $m=-1/(H\ln10)$. The twelve panels correspond to the cases explained in section \ref{sec:rets}.}
\label{fig:radpref}
\end{figure}

\begin{figure}
\centering
\includegraphics[width=0.5\textwidth]{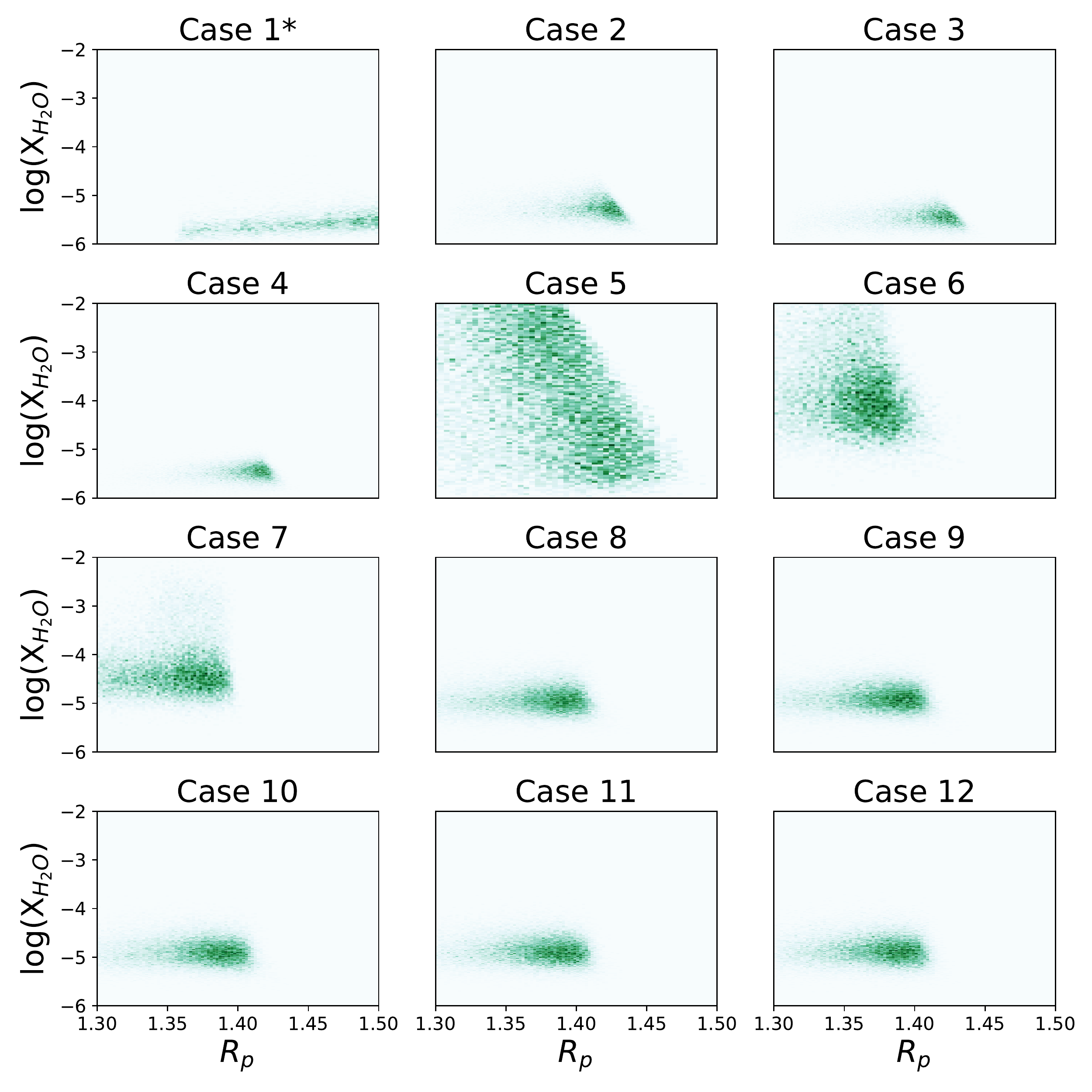}
\caption[Retrieved planetary radius vs retrieved H$_2$O mixing fraction]{Correlation between the retrieved planetary radius (\Rp in units of \Rj) and H$_2$O mixing ratio ($X_{\text{H}_2\text{O}}$). The twelve panels correspond to the cases explained in section \ref{sec:rets}. The spread in the retrieved values changes under different assumptions with case 12 being the most general case. The H$_2$O abundance (i.e. mixing ratio) in case 1 has been multiplied by $10^{4}$ to be in the same range as the H$_2$O abundance of other cases.}
\label{fig:abundances}
\end{figure}

We now investigate this empirical finding using the retrievals from section \ref{sec:rets}, under different model assumptions. The correlations between $\log_{10}$(\Pref) and \Rp for each of our retrievals of section \ref{sec:rets} are shown in Figure~\ref{fig:radpref}, along with a linear fit and the corresponding slope. The fit is obtained using \texttt{polyfit} included in \texttt{NumPy} \citep{numpy}. Accompanying this figure we have Figure~\ref{fig:abundances} where we show $\log_{10}$(\Xh) as a function of \Rp for the same cases. Figure~\ref{fig:abundances} shows the posterior distributions of $X_{\text{H}_2\text{O}}$ which become more localised as different assumptions are removed from the retrievals. It is clear from Figures \ref{fig:radpref} and \ref{fig:abundances} that while $\log_{10}$(\Pref) and \Rp are strongly degenerate, there is almost no degeneracy between $\log_{10}$(\Xh) and \Rp in most of the cases. The only exception is case 5 with an assumed cloud fraction of 100\%. This assumption introduces a degeneracy between the cloud level (i.e. P$_{\text{cloud}}$) and \Rp. The different combinations of \Rp and P$_{\text{cloud}}$ that explain the spectrum for an assumed $\phi=100\%$ result in the wide spread of H$_2$O abundances observed in Figure~\ref{fig:abundances} case 5.

From Figure~\ref{fig:radpref} it can be observed that there is a log-linear relation between \Pref and \Rp. The superimposed linear fit gives us an idea of what the scale height for each model is, i.e., $m=-1/(H\ln10)$ as we did above with our analysis of Figure~\ref{fig:reproduction}. While the slopes vary between cases ($m=-28$ to $-81$), they converge to a value of $m=-58$ as the model and data in our retrieval become more robust (i.e. case 7 and above). Figure~\ref{fig:radpref} also shows a temperature estimate for the photosphere of the planet, which we obtain using the slope of the linear fit and assuming a mean molecular weight of 2.4 amu and a planet gravity of $\log_{10}$($g$)=2.963 in cgs. We find that these temperature estimates range between $\sim1012$ K and $\sim1910$ K for all the cases except case 5; we discussed the exception of case 5 previously. The temperature values converge in case 12 to 1430 K which is consistent with the equilibrium temperature of the planet as well as the photospheric temperature estimated in previous studies \citep[e.g.][]{MacDonald17}. These findings suggest that the relationship between \Rp and \Pref is indeed governed by the atmospheric scale height.

As our retrieval cases build towards full model considerations and adequate data, the estimated slope and the scale height converge. This is to be expected as data at short wavelengths help constrain the continuum and, hence, the molecular abundances, the mean molecular mass and the scale height \citep{bennekeandseager, 2013Sci...342.1473D}. As such, the spread in H$_2$O abundances seen in Figure~\ref{fig:abundances} is not a result of the \Rp-\Pref degeneracy but a result of data quality and model assumptions. The better the data and model, the better constraints we can impose on the molecular abundances.

Here, we investigate a possible justification for the log-linear relationship we empirically observe between \Pref and \Rp. Generally, the pressure and distance in a planetary atmosphere are related by the consideration of hydrostatic equilibrium. We explore whether the same can explain the observed \Pref-\Rp relation. 

An observed transmission spectrum consists of transit depths, i.e. $(r/R_s)^2$, as a function of wavelength. By knowing the radius of the star, we know the observed radius (or effective radius) of the planet as a function of wavelength. An observed effective radius should correspond to an effective height in the atmosphere, and the corresponding pressure level, where the atmosphere has a slant optical depth of $\tau_\lambda \sim \tau_{eq}$ \citep{2008A&A...481L..83L}. The equivalent slant optical depth ($\tau_{eq}$) corresponding to observed spectral features is discussed in more detail in section \ref{sub:probingbelow}.

The pressure (P) and distance (r) in the atmosphere are related by hydrostatic equilibrium as

\begin{equation}
\ln\left(\frac{P}{P_{ref}}\right)=-\frac{\mu g }{k_B} \int _{R_p}^{r}\frac{1}{T}dr'.
\end{equation}

\noindent Here, \Rp and $\text{P}_{ref}$ are a reference planet radius and the corresponding pressure, respectively. This equation can be solved if the temperature profile with distance is known. Assuming an isotherm, P and r are related by
\begin{equation}
\label{eq:hydro}
\ln\left(P\right)=-\frac{r}{H}+\ln(P_{ref})+\frac{R_p}{H},
\end{equation}
where $H=k_BT$($\mu $g)$^{-1}$ is the scale height.

This relation is linear in $\ln(P)$ and $r$ with a slope of $-1/H$ and an intercept of $\ln(P_{ref})+R_p/H$. We rewrite equation~\ref{eq:hydro} as 
\begin{equation}
\label{eq:simplified}
\ln\left(P\right)=-a\,r+b.
\end{equation}

\begin{figure*}
\centering
\includegraphics[width=1.1\textwidth]{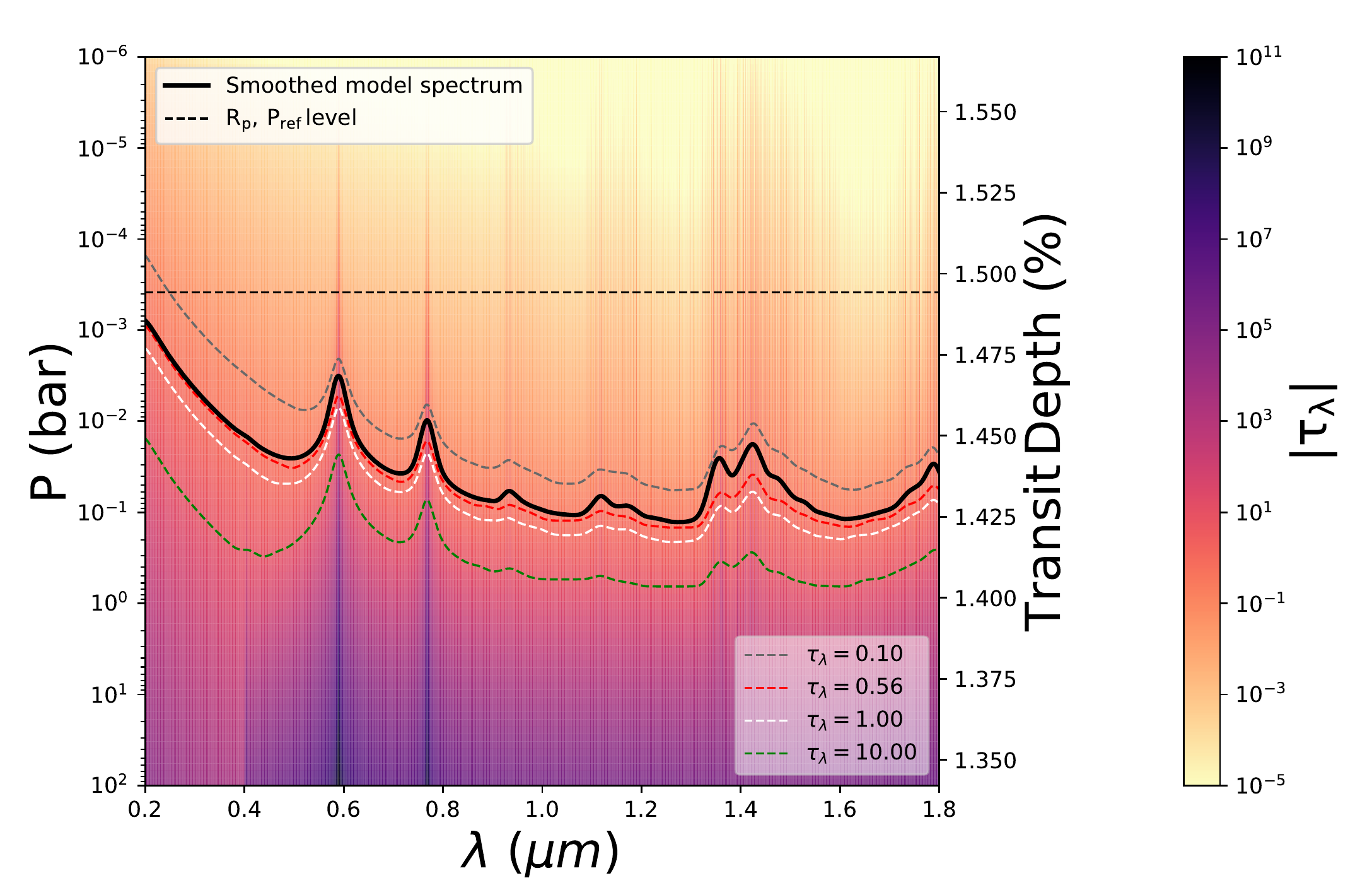}
  \caption[Optical depth colour map]{Photospheric level of spectral features. The black solid line shows the cloud free model spectrum using the median values retrieved for HD~209458~b as explained in section \ref{sub:probingbelow}. The colour map represents the slant optical depth ($\tau_\lambda$) and the $\tau_\lambda=0.10,\,0.56,\,1.00,\,10.00$ levels are shown in grey, red, white and green respectively. The black dashed line shows the reference pressure level and its corresponding transit depth, which in turn corresponds to the transit depth of the retrieved \Rp. The transit depth and radius are related by the expression $\Delta=\text{R}_{\text{p}}^2/\text{R}_{\text{star}}^2$.}
\label{fig:depth}
\end{figure*}

The observed radii also provide another constraint. Given a set of observations $r_{\lambda,i}$'s, the corresponding $P_{\lambda,i}$'s are those for which the slant optical depths satisfy $\tau_{\lambda,i} \sim \tau_{eq}$. From a procedural point of view, a retrieval tries to find the best fitting model parameters for which the distances in the model atmosphere at $r = r_{\lambda,i}$ satisfy $\tau_\lambda \sim \tau_{eq}$. The atmospheric model consists of a fixed pressure grid, as discussed in section~\ref{sec:rets}. For a given \Rp and \Pref, among other parameters drawn in a model fit, the pressure grid is related to a distance grid using hydrostatic equilibrium as shown in equation~\ref{eq:hydro}. These properties in turn are used to create a grid of slant optical depths corresponding to the altitude, or pressure, as a function of wavelength. The differential optical depth along the line of sight is given by $d\tau_\lambda=n\sigma_\lambda \,ds$ where $\sigma_\lambda$ is the absorption cross-section, $n$ is the number density, and $s$ is the distance along the line of sight. Thus, the model has a distance grid on a one-to-one correspondence with the pressure grid and an associated $\tau_\lambda$ map. In a retrieval, the acceptable fit parameters are those for which the locations of the observed $r_{\lambda,i}$ in the model distance grid have $\tau_\lambda \sim \tau_{eq}$.

Thus, from equation~\ref{eq:simplified}, given a set of observations $r_{\lambda,i}$'s the values of $a$ and $b$ can be  constrained. $a$ independently constraints the scale height of the planet and the slope of hydrostatic equilibrium since $a=1/H$. Similarly, $b$ helps determine a unique relationship by $b=R_p/H+\ln(P_{ref})$. Rearranging for $\ln(P_{ref})$ we obtain $\ln(P_{ref})=-R_p/H+b$, where it is evident that \Rp and \Pref, by construction, will also need to satisfy hydrostatic equilibrium with the same slope determined by $a$. 

We can thus conclude that there is indeed a degeneracy between \Pref and \Rp but it is well defined and it does not affect the retrieved molecular abundance. It is the functional form of $b$ in equation~\ref{eq:simplified} that seems to explain the $-1/H$ behaviour seen in Figures \ref{fig:reproduction} and \ref{fig:radpref}, and what defines the relationship between \Pref and \Rp. Now, we inspect more closely the requirement imposed for the observed radii to correspond to a constant slant optical depth $\tau_\lambda \sim \tau_{eq}$.

\newpage
\subsection{The slant photosphere.}
\label{sub:probingbelow}

Following the previous section, we investigate how the observed radius at a given wavelength corresponds to a pressure through $\tau_\lambda$. The one-to-one correspondence between a set of observations $r_{\lambda,i}$'s and their associated pressures is determined by the slant optical depth $\tau$ of the photosphere. In this section we explore further this notion of the equivalent slant optical depth and how this helps constrain \Rp and \Pref. For illustration, we use the retrieved values of HD~209458~b for case 12 in section~\ref{sec:rets} and generate a model spectrum for a cloud-free and isothermal atmosphere with temperature set to the retrieved $T_0$. For each wavelength in our model we obtain the slant optical depth as a function of the pressure in the atmosphere corresponding to the impact parameter. We show a contour of $\tau_\lambda$ in the $P-\lambda$ space in Figure~\ref{fig:depth}.

Figure~\ref{fig:depth} shows both a pressure axis and a transit depth axis that are related by our selection of \Rp and \Pref and hydrostatic equilibrium. The colour map shows that the equivalent slant photosphere appears at pressures between 0.1 and 0.01 bar for most wavelengths. Furthermore, it is clear from the model spectrum that the slant optical depth at the apparent radius is $\sim$0.5. This $\tau_\lambda$ surface is close to $\tau_\lambda=0.56$; a value first encountered numerically by \cite{2008A&A...481L..83L} and later shown extensively by \cite{2013Sci...342.1473D}. The cumulative contribution of the atmosphere to the spectrum is consistent with an opaque planet below the $\tau_{eq}$ surface. This factor provides an additional constrain when fitting equation~\ref{eq:simplified}. Following Figure~\ref{fig:depth} we find that $\tau \gtrsim 0.5$ generally determines the equivalent radius and motivates the condition $\tau_\lambda \sim \tau_{eq}$ discussed in the previous section. This condition is true for hot Jupiters and for most planetary atmospheres as long as \Rp/H is between $\sim 300$ and $\sim 3000$ \citep{2008A&A...481L..83L}.

\subsection{Retrieving \Rp vs. \Pref}
\label{sec:rp_or_pref}

So far, the retrievals presented here have both \Rp and \Pref as parameters in the retrieval. We have shown above that the degeneracy between these variables can be characterised through an empirical relationship. Several retrieval analyses use only one of \Rp or $\text{P}_{ref}$ as a free parameter and assume a fixed value for the other \citep[e.g][]{bennekeandseager, kreidbergWASP12b, 0004-637X-820-1-78, 2017Natur.549..238S, 2018AJ....155...29W, 2018A&A...616A.145C, 2018arXiv181102573V}. Here, we conduct retrievals that assume \Pref and retrieve \Rp and vice versa, in addition to case 12 in section~\ref{sec:fullret} where both were considered to be free parameters. We compare the results and discuss whether the retrievals are sensitive to these assumptions.

We start by assuming a reference pressure and retrieving a planetary radius. Our retrieval is set up in the same way as in section \ref{sec:fullret} for case 12: the model includes volatiles, a parametric P-T profile, inhomogeneous cloud cover, and uses data in the near-infrared and optical. The retrieved planetary radius is \Rp=$1.37^{+0.01}_{-0.01}$ \Rj at an assumed pressure in bar of $\log_{10}($\Pref$)=-3.45$. The assumed reference pressure was chosen to match the retrieved value in section \ref{sec:fullret}. The H$_2$O abundance is retrieved to a value of $\log_{10}$(\Xh)=$-4.87 ^{+ 0.30 }_{- 0.25 }$. The other retrieved parameters are $\log_{10}(X_{\text{Na}})=-5.44 ^{+ 0.59 }_{- 0.45 }$, $\log_{10}(X_{\text{K}})=-7.04 ^{+ 0.60 }_{- 0.54 }$, $\log_{10}(X_{\text{NH}_3})=-8.34 ^{+ 2.06 }_{- 2.43 }$, $\log_{10}(X_{\text{CO}})=-7.72 ^{+ 2.88 }_{- 2.81 }$, $\log_{10}(X_{\text{HCN}})=-8.61 ^{+ 2.28 }_{- 2.23 }$, and $\log_{10}(X_{\text{CO}_2})=-8.41 ^{+ 2.47 }_{- 2.35 }$. The retrieved P-T parameters are T$_0$=$1049.35 ^{+ 280.69 }_{- 173.03 }$ K, $\alpha_1=0.59 ^{+ 0.24 }_{- 0.16 }$, $\alpha_2= 0.54 ^{+ 0.29 }_{- 0.24 }$, $\log_{10}(P_1)=-0.20 ^{+ 2.15 }_{- 2.21 }$, $\log_{10}(P_2)=-3.44 ^{+ 2.57 }_{- 1.74 }$, and $\log_{10}(P_3)=1.01 ^{+ 1.31 }_{- 1.71 }$. The retrieved cloud parameters are $\log_{10}(a)=44.42 ^{+ 0.68 }_{- 0.98 }$, $\gamma=-14.57 ^{+ 5.30 }_{- 3.68 }$, $\log_{10}(P_{\text{cloud}})=-4.56 ^{+ 0.70 }_{- 0.53 }$ and $\phi= 0.47 ^{+ 0.06 }_{- 0.08 }$. All the retrieved values are consistent within 1-$\sigma$ with the obtained values when retrieving both \Rp and $\text{P}_{ref}$ in case 12. 

Then, we perform the retrieval in which we assume a planetary radius and retrieve the reference pressure.  Here we assume a radius of \Rp=1.359 \Rj, using the value reported by \citet{Torres08} and retrieve a H$_2$O abundance of $\log_{10}$(\Xh)=$-4.84 ^{+ 0.28 }_{- 0.25 }$ and a reference pressure in bar of $\log_{10}($\Pref$)=-2.48 ^{+ 0.46 }_{- 0.45 }$. The retrieved P-T profile parameters are T$_0$=$1000.05 ^{+ 264.42 }_{- 143.28 }$ K, $\alpha_1=0.62 ^{+ 0.24 }_{- 0.18 }$, $\alpha_2=0.48 ^{+ 0.33 }_{- 0.22 }$, $\log_{10}(P_1)=-1.16 ^{+ 1.98 }_{- 1.77 }$, $\log_{10}(P_2)=-3.92 ^{+ 2.19 }_{- 1.43 }$, and $\log_{10}(P_3)=1.25 ^{+ 1.20 }_{- 1.56 }$. The retrieved cloud parameters are $\log_{10}(a)= 4.40 ^{+ 0.68 }_{- 0.99 }$, $\gamma=-14.65 ^{+ 5.08 }_{- 3.63 }$, $\log_{10}(P_{\text{cloud}})=-4.54 ^{+ 0.69 }_{- 0.53 }$ and $\phi= 0.47 ^{+ 0.06 }_{- 0.08 }$. Lastly, the additional retrieved molecular abundances are $\log_{10}(X_{\text{Na}})=-5.44 ^{+ 0.55 }_{- 0.45 }$, $\log_{10}(X_{\text{K}})=-7.04 ^{+ 0.58 }_{- 0.53 }$, $\log_{10}(X_{\text{NH}_3})=-8.11 ^{+ 1.94 }_{- 2.57 }$, $\log_{10}(X_{\text{CO}})=-7.75 ^{+ 2.88 }_{- 2.77 }$, $\log_{10}(X_{\text{HCN}})=-8.58 ^{+ 2.34 }_{- 2.24 }$, and $\log_{10}(X_{\text{CO}_2})=-8.37 ^{+ 2.43 }_{- 2.35 }$. Again, the retrieved values are consistent with those of section \ref{sec:fullret} to within  1-$\sigma$.

\begin{figure}
\centering
\includegraphics[width=0.5\textwidth]{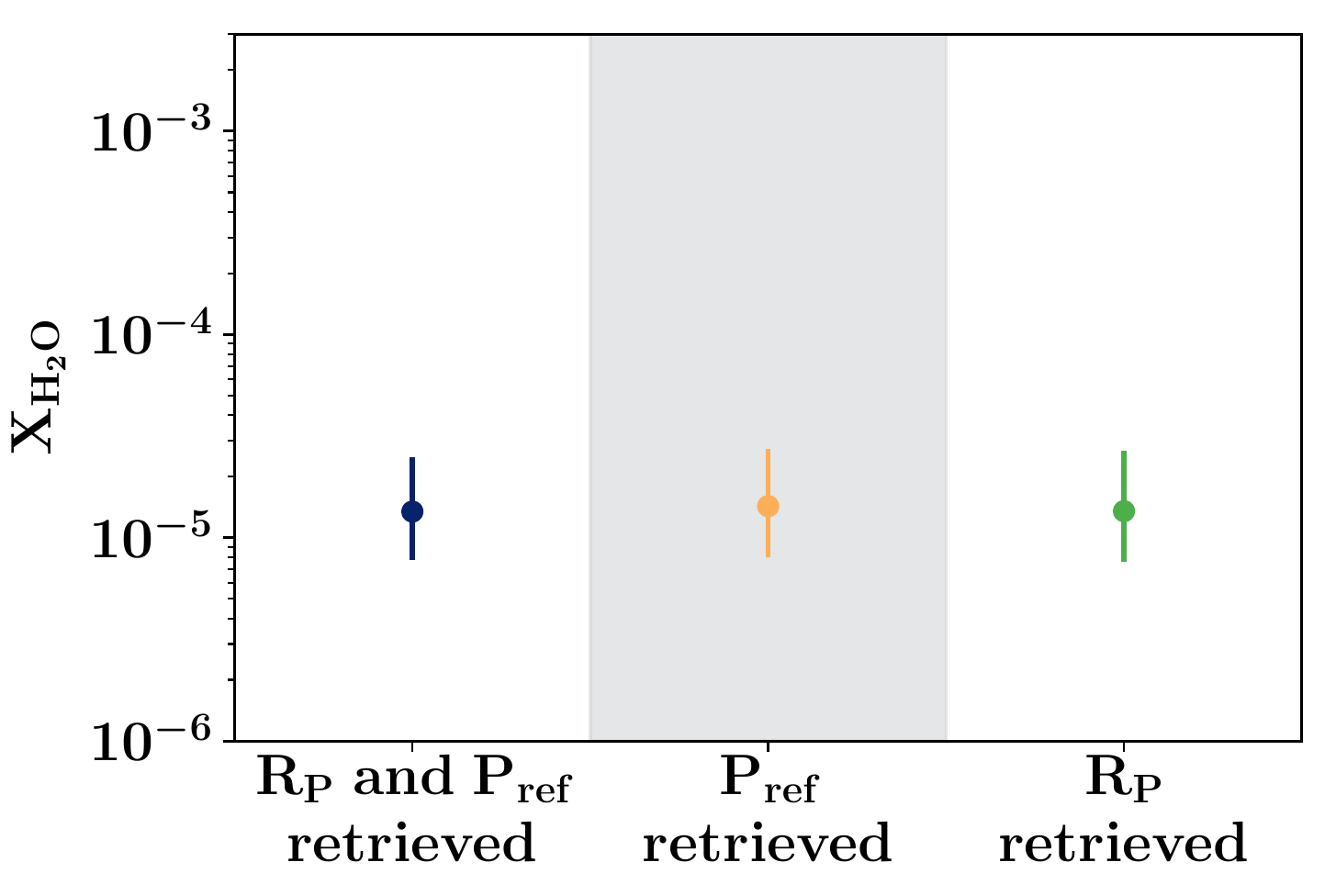}
  \caption[Retrieved H$_2$O abundances for different retrieval parameters]{Retrieved H$_2$O abundances for three model considerations, median value and 1-$\sigma$ error bars. In blue we show the case were both \Rp and \Pref are retrieved. In yellow, only \Pref is retrieved for an assumed \Rp, and in green the opposite is shown. All three retrievals provide a consistent H$_2$O abundance showing that the assumption of a radius or reference pressure is inconsequential in the retrieval of H$_2$O abundance. }
\label{fig:threeway}
\end{figure}

We show the retrieved H$_2$O abundances and their error bars in Figure~\ref{fig:threeway} for the three cases. First we show the retrieval in section \ref{sec:fullret} where we retrieved both \Rp and \Pref. Second we show the retrieval assuming a \Rp and retrieving \Pref. The third panel shows the remaining permutation where we assume a \Pref and retrieve \Rp.  All three retrieved H$_2$O mixing fractions are consistent with each other. These results confirm that it is not necessary to retrieve both \Rp and \Pref. Assuming one will retrieve the other and both will be used to determine the atmospheric structure as discussed in section \ref{sec:theory}. While our paper was in review a follow up paper to HK17 was published \citep{2018MNRAS.481.4698F} which retracts some of the claims of HK17. They suggest to break the three-way degeneracy of HK17 for cloud-free atmospheres by deriving a $\text{P}_{ref}$ for an \Rp assuming it is associated to a part of the atmosphere opaque to optical and infrared radiation, a similar procedure to that suggested in previous studies \citep[e.g.][]{2014RSPTA.37230086G}. They come to a similar conclusion that it is not necessary to retrieve both \Rp and $\text{P}_{ref}$ and it is possible to assume the value of one or the other, as is common practice in the literature \citep[e.g.][]{bennekeandseager, kreidbergWASP12b, 0004-637X-820-1-78, 2017Natur.549..238S, 2018AJ....155...29W, 2018A&A...616A.145C, 2018arXiv181102573V}. However, here we show that no such assumption of wavelength is necessary. The \Rp can be fixed to any measured value and the retrieval will automatically derive the corresponding \Pref.

\subsection{Limitations of semi-analytic analysis}
\label{sub:semianalyticlimitations}

Based on our above results, here we summarise some key factors that may have limited some previous studies using semi-analytic models for constraining chemical abundances \citep[e.g.][]{hengkitzmann, 2018MNRAS.481.4698F}. These key factors include ignoring the effects of CIA opacity, incorrect inferences from least-square fits and generalised conclusions drawn from inadequate data.

As we show in section \ref{sec:rets}, ignoring the effects of CIA leads to an incorrect estimate of molecular abundances by several orders of magnitude. In the work of HK17 CIA effects were not considered, thereby rendering their analytic solution incomplete. While the assumption of isobaric opacities for molecular line absorption does not affect the retrieved molecular abundances substantially, the inclusion of CIA is incompatible with an isobaric assumption. Molecular features in current data are less strongly affected by the pressure dependence because the spectrum probes lower pressures (P $\lesssim$ 0.1 bar) for these features. On the other hand, CIA opacities have strong dependence on the pressure. In other words, CIA absorption is proportional to P$^2$ \citep{2013Sci...342.1473D}, and its impact on the spectrum is underestimated if evaluated at only one pressure \citep[e.g.][]{2018MNRAS.481.4698F} or completely ignored \citep[e.g.][]{hengkitzmann}.

The more problematic assumption in the work of HK17 comes in their inferred H$_2$O abundances obtained from best fits to the observed WFC3 spectrum of WASP-12b as shown in their Figure 7. In that figure, they claim to be showing values of $X_{\text{H}_2\text{O}}$(P$_0$/10 bar)$^{-1}$ as a function of the assumed planetary radius R$_0$. A close inspection of this graph suggests that they are instead showing $X_{\text{H}_2\text{O}}$(P$_0$/10 bar) as a function of the assumed planetary radius. The linear trend they obtain in their figure is likely a manifestation of the relationship between \Pref and \Rp, and not \Xh. They claim that a small change in the assumed planetary radius leads to a large change in the ordinate; and hence the H$_2$O abundance. While this claim may be partly true, their inference on the H$_2$O abundance is manifestly incorrect. A large change in the product of $X_{\text{H}_2\text{O}}$(P$_0$/10 bar) is not because of a change in H$_2$O abundance but a change in the reference pressure. Changing the assumed planetary radius will change the associated reference pressure. As shown in sections \ref{sec:rets} and \ref{sec:theory}, the inferred H$_2$O abundance is largely unaffected by the \Pref-\Rp degeneracy.


Lastly, their work considers only the interpretation of WFC3 data and ignores the effects of optical data. On the other hand, as we have shown here, it is the inclusion of optical data that helps constrain molecular abundances the most. The inclusion of optical data helps constrain the effects of clouds, the reference pressure, and the scale height. These in turn improve the constraint on the H$_2$O abundance. Thus, retrievals which do not take these factors into account are inherently biased towards incorrect chemical abundances \citep[e.g.][]{2018MNRAS.481.4698F}. As such, their abundance estimates (e.g. of HD~209458~b) do not agree with retrievals that use cloudy models and optical data (e.g. \citet{MacDonald17, 2017ApJ...834...50B}; and the present study).

In this work we have shown an empirical relationship between \Rp and $\text{P}_{ref}$ that seems to be related to hydrostatic equilibrium. Furthermore, we show that this relationship is independent of \Xh, effectively breaking the three-way degeneracy. Complimentary to this is the importance of choosing the right models and assumptions in the model atmospheres. Models ignoring CIA, considering only H$_2$O opacity, along with constant gravity and mean molecular weight lead to poor constraints in the retrieved H$_2$O abundance. Equally important are the consideration of inhomogeneous cloud coverage and inclusion of optical data in constraining molecular abundances. 

\section{Summary and Discussion}
\label{sec:conclusions}

We conduct a detailed analysis of degeneracies in transmission spectra of transiting exoplanets. We investigate the effect of various model assumptions and spectral coverage of data on our ability to constrain molecular abundances. Utilising atmospheric retrievals we test simple isobaric and isothermal atmospheric models for their ability to constrain molecular abundances using infrared spectra alone. We later remove one by one these assumptions until resulting in a realistic atmospheric model composed of H$_2$/He collision induced absorption, multiple molecular species, a full P-T profile, inhomogeneous cloud coverage and the inclusion of broadband data spanning infrared to optical. We conduct this investigation using the canonical example of HD~209458~b, a hot Jupiter which has the best data currently available. 

We identify several key properties that need to be accounted for in models for reliable estimates of chemical abundances, in particular H$_2$/He CIA opacities, a full P-T profile, and possible inhomogeneities in cloud cover. The inclusion of CIA has the most effect in accurately constraining molecular abundances as it provides a natural continuum in the model spectrum. 

We find that the degeneracies between molecular abundances and cloud properties can be alleviated by the inclusion of optical data. Optical data provides constraints on the scattering slope in the optical as well as a continuum for the full spectrum. As such, optical data are key to constrain molecular abundances using transmission spectra. When optical data is included, considering Na and K absorption significantly influences the retrieved H$_2$O abundances, making them more precise. We also find that assuming a cloud fraction a priori (e.g. 100\% cloud cover) leads to erroneous estimates on molecular abundances. Leaving the cloud fraction as a free parameter allows for more accurate molecular estimates than those obtained when assuming a fixed cloud cover fraction. 

We show, using simulated data, that even in the case of an atmosphere with 100\% cloud cover the retrievals are able to closely retrieve abundances and other properties. In principle, \Rp is degenerate with the pressure level of the cloud top. However, the range of altitudes, and hence pressures, of the cloud top where it affects the spectrum is limited. A 100\% cloud deck must be at an altitude higher than the line-of-sight photosphere and lower than the level where the atmosphere is optically thin. In the former case the cloud deck does not contribute significantly and in the latter case the cloud deck causes a featureless spectrum, contrary to observed features. On the other hand, we show that an inhomogeneous cloud model accurately retrieves the cloud fraction even for a 100\% cloudy case. Among the considerations that we leave unexplored in this set of retrievals are the effect of stellar activity on the transmission spectra and its consequence in resolving degeneracies, and the presence of shifts or offsets between data taken from different instruments. Other aspects to consider in the future include inhomogeneities across the limb due to the 3D structure of an atmosphere \citep[e.g.][]{2019arXiv190109932C}, refraction in the atmosphere \citep[e.g.][]{2016MNRAS.456.4051B}, height-varying chemical abundances \citep[e.g.][]{2018A&A...617A.110P}, and various cloud properties \citep[e.g.][]{2014ApJ...789L..11V, 2016Icar..271..400B}. Future work and retrieval frameworks like that of \cite{2018MNRAS.480.5314P} could help elucidate on these aspects. Overall, the quality of the data and the wavelengths they span are fundamental in breaking degeneracies and retrieving molecular abundances.

We also discuss the limitations of semi-analytic studies in fully assessing degeneracies in transmission spectra. One important finding is that the degeneracy between \Pref and \Rp does not lead to an inability to determine molecular abundances in transit spectroscopy, contrary to previous suggestion. We show an empirical relationship between the planetary radius and the reference pressure that characterises their degeneracy. We find that $\ln$(\Pref) and \Rp have a linear relationship with a slope of $-1/H$ and suggest that this behaviour is rooted in hydrostatic equilibrium. For each \Rp there is an associated \Pref and vice versa. As such, it is redundant to perform retrievals that consider both quantities as free parameters. Instead, we demonstrate that it is justified to assume a value for one quantity and retrieve the other. Current studies usually assume an \Rp and retrieve a \Pref; here we demonstrate that the inverse is also consistent, i.e., that it is possible to assume a \Pref and retrieve the radius of the planet corresponding to that \Pref. 

We investigate the origins of spectral features in transmission spectra by following the line-of-sight opacity of the planet as a function of the vertical pressure level and wavelength. This allows us to calculate the height and pressure levels in the atmosphere at which the observed features are generated and compare it to the white light radius. We show that the effective radius corresponding to the observed transit depth at a given wavelength corresponds to a level in the atmosphere with a slant optical depth of $\tau \gtrsim 0.5$, as also suggested by previous studies. 

Overall, our study demonstrates the effectiveness of high-precision spectra and realistic models to retrieve atmospheric abundances. Data with current facilities such as HST and VLT over the visible and near-infrared can already provide valuable constraints on abundances of key species such as H$_2$O, Na, K, etc. The upcoming JWST and ground-based facilities therefore hold great promise for characterising exoplanetary atmospheres using transmission spectra. 

\acknowledgments
LW is grateful for research funding from the Gates Cambridge Trust. We thank the anonymous reviewer for their thoughtful comments on the manuscript. LW thanks Siddharth Gandhi for helpful discussions and Anjali Piette for inputs on Figure~\ref{fig:depth}. 



\appendix

\section{Extended results from retrievals.}

\begin{figure}
\gridline{\fig{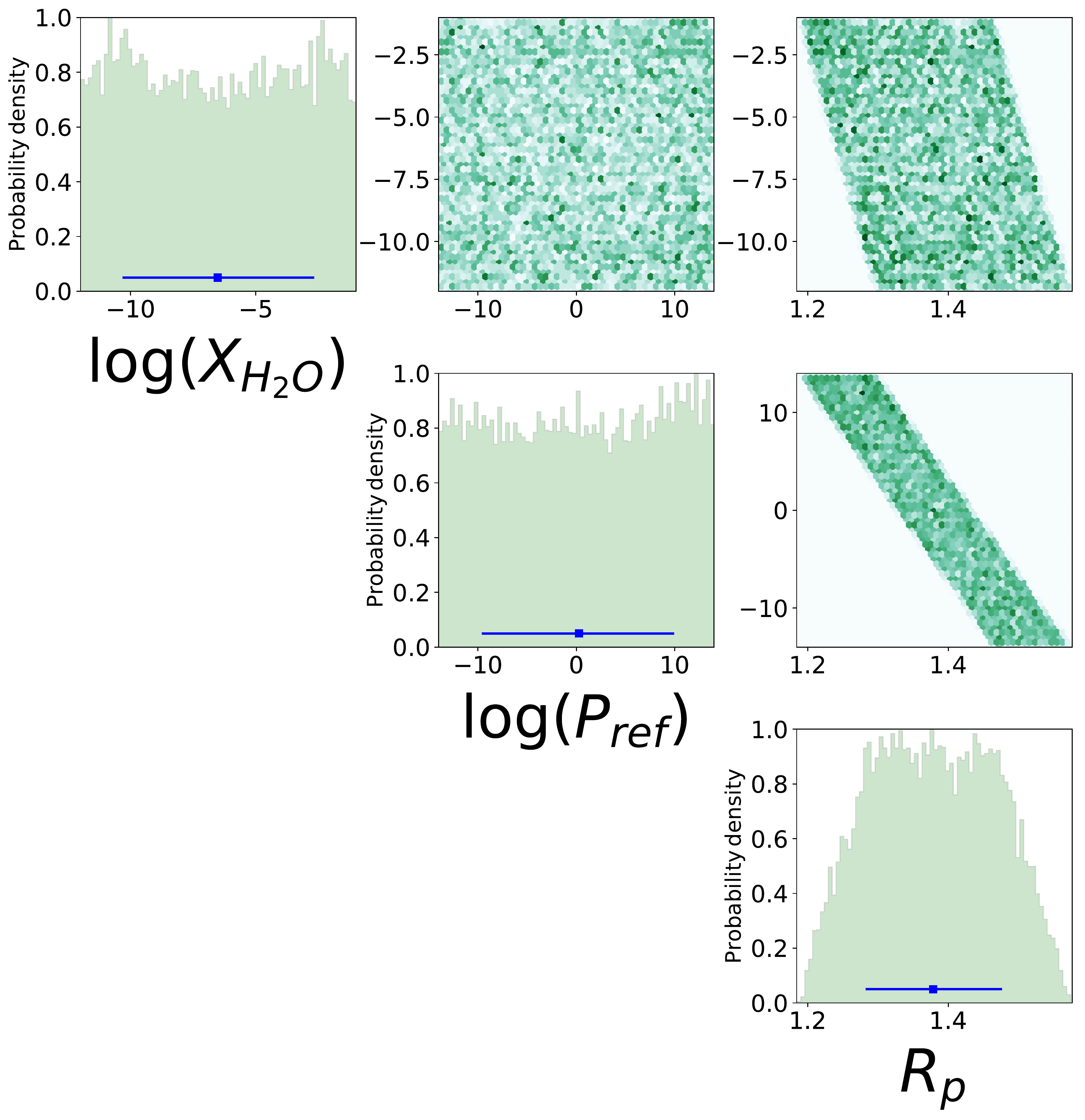}{0.25\textwidth}{(Case 0)}
          \fig{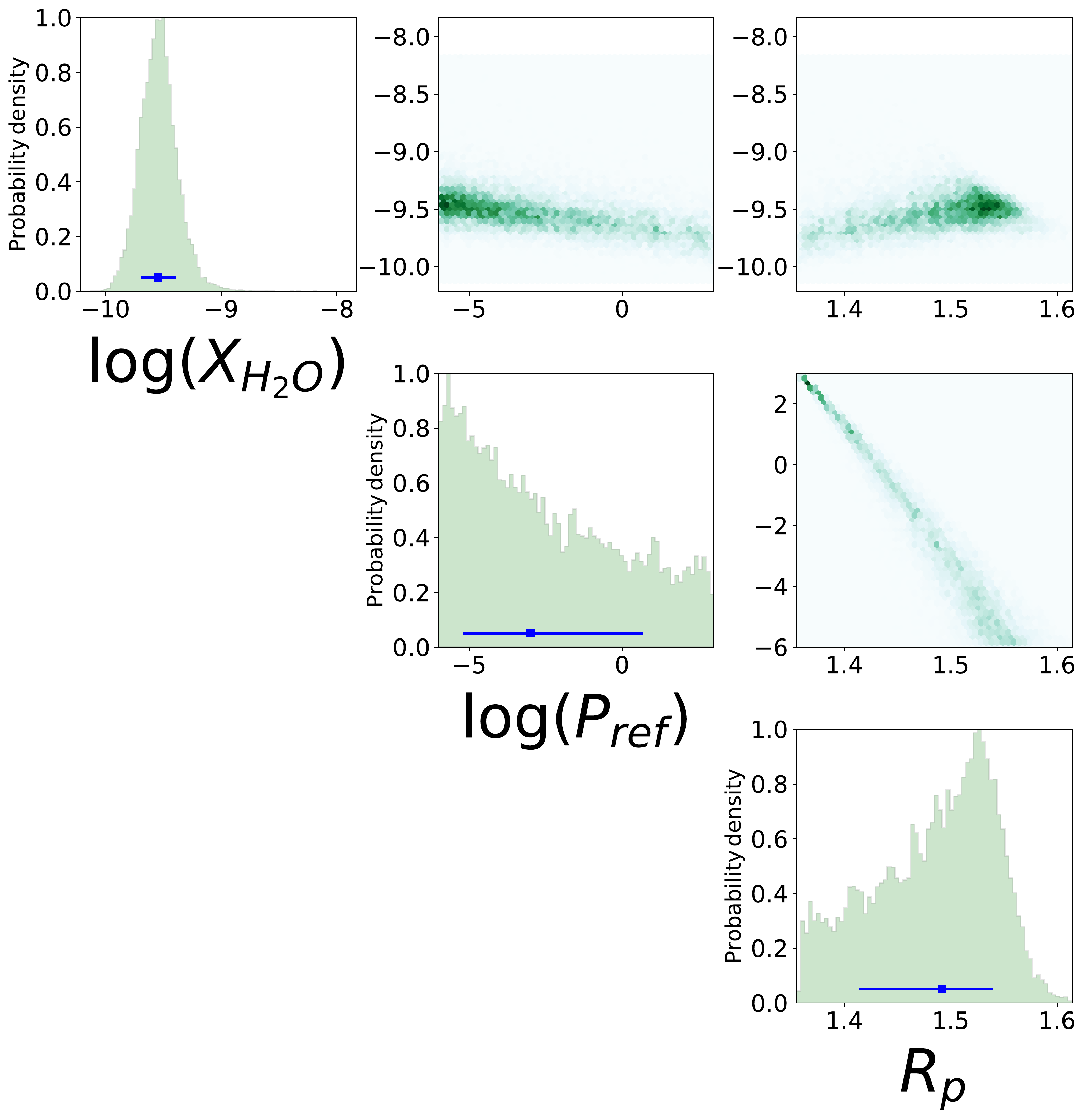}{0.25\textwidth}{(Case 1)}
          \fig{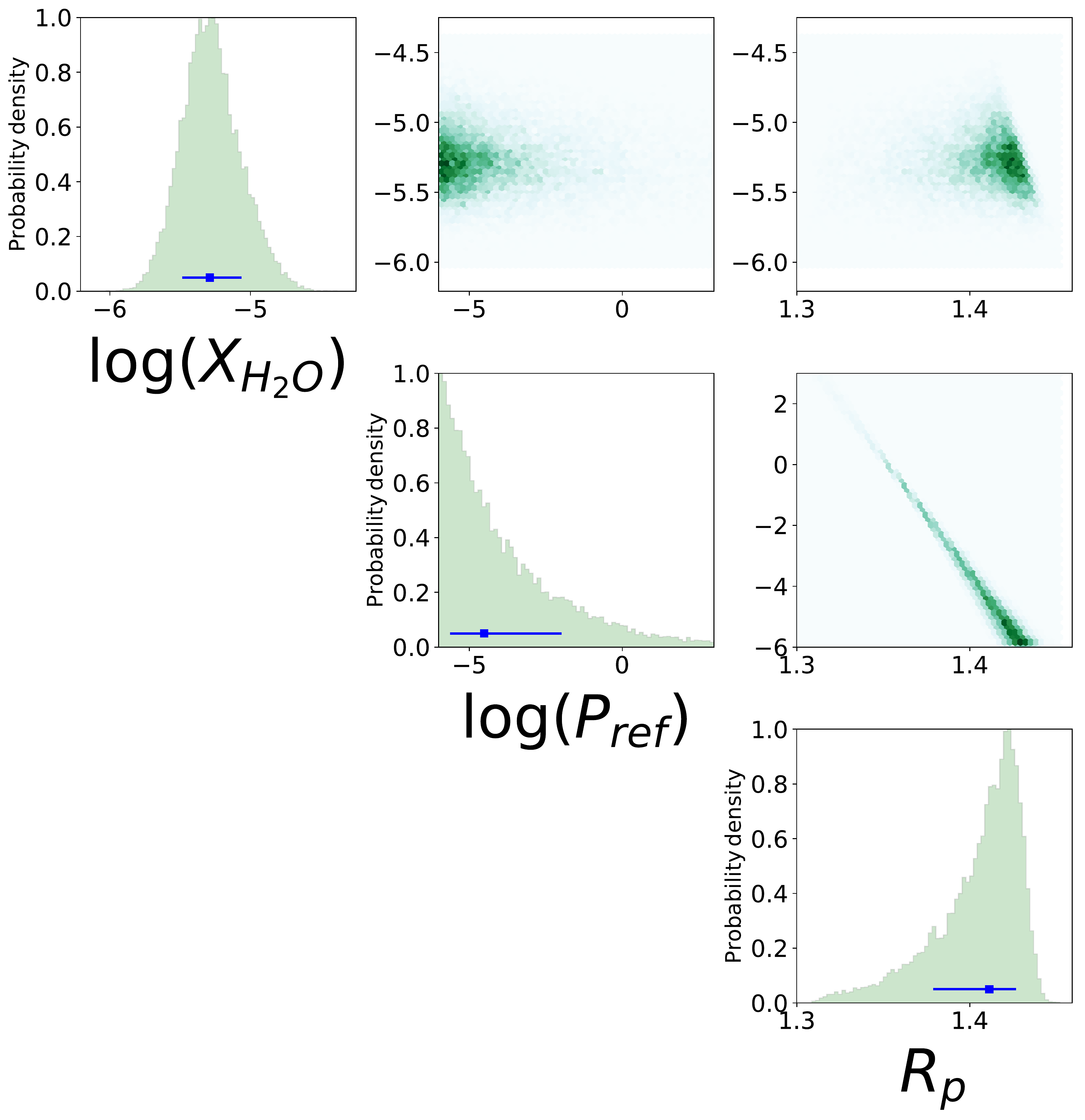}{0.25\textwidth}{(Case 2)}
          \fig{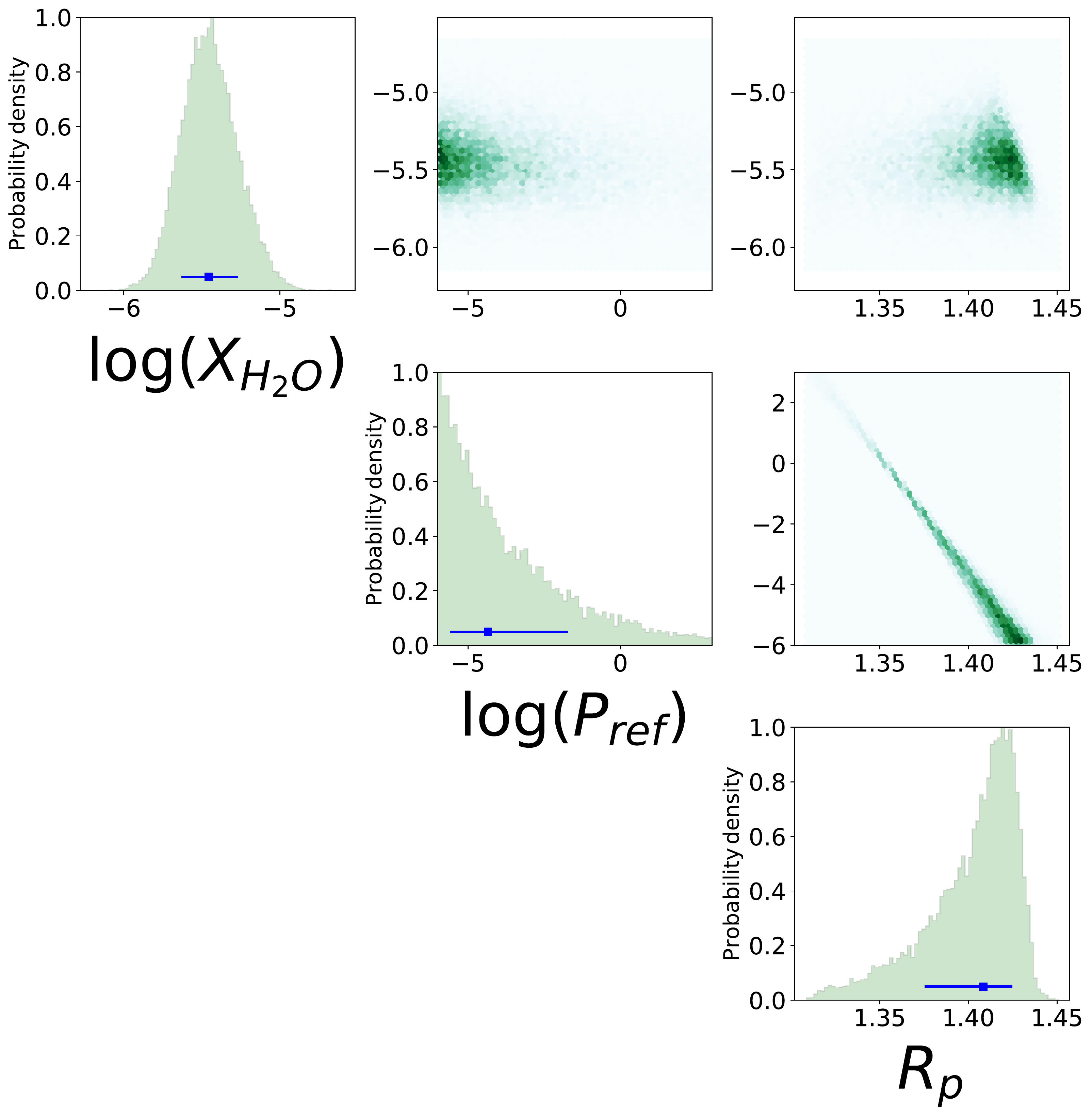}{0.25\textwidth}{(Case 3)}}
\gridline{\fig{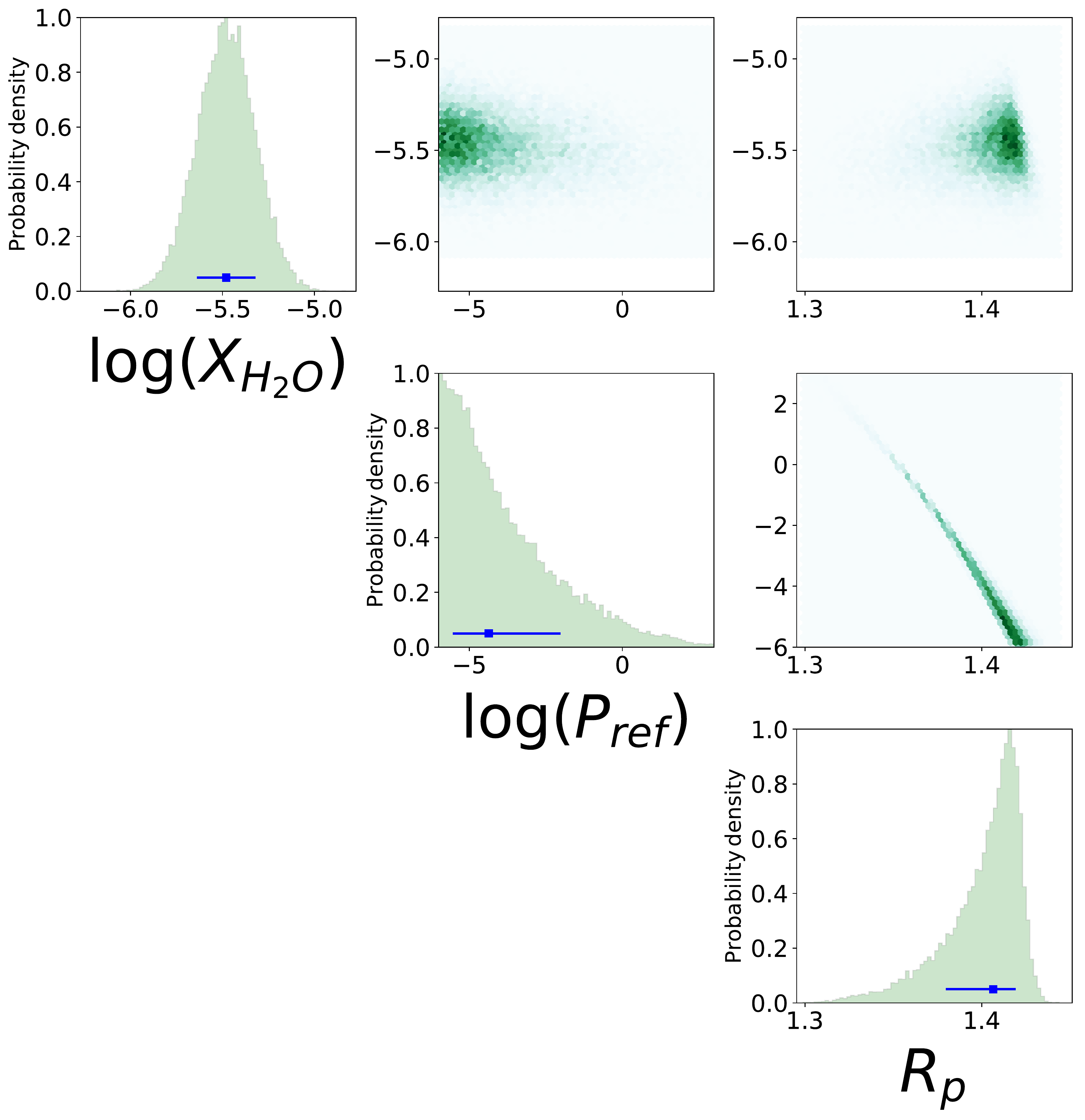}{0.25\textwidth}{(Case 4)}
          \fig{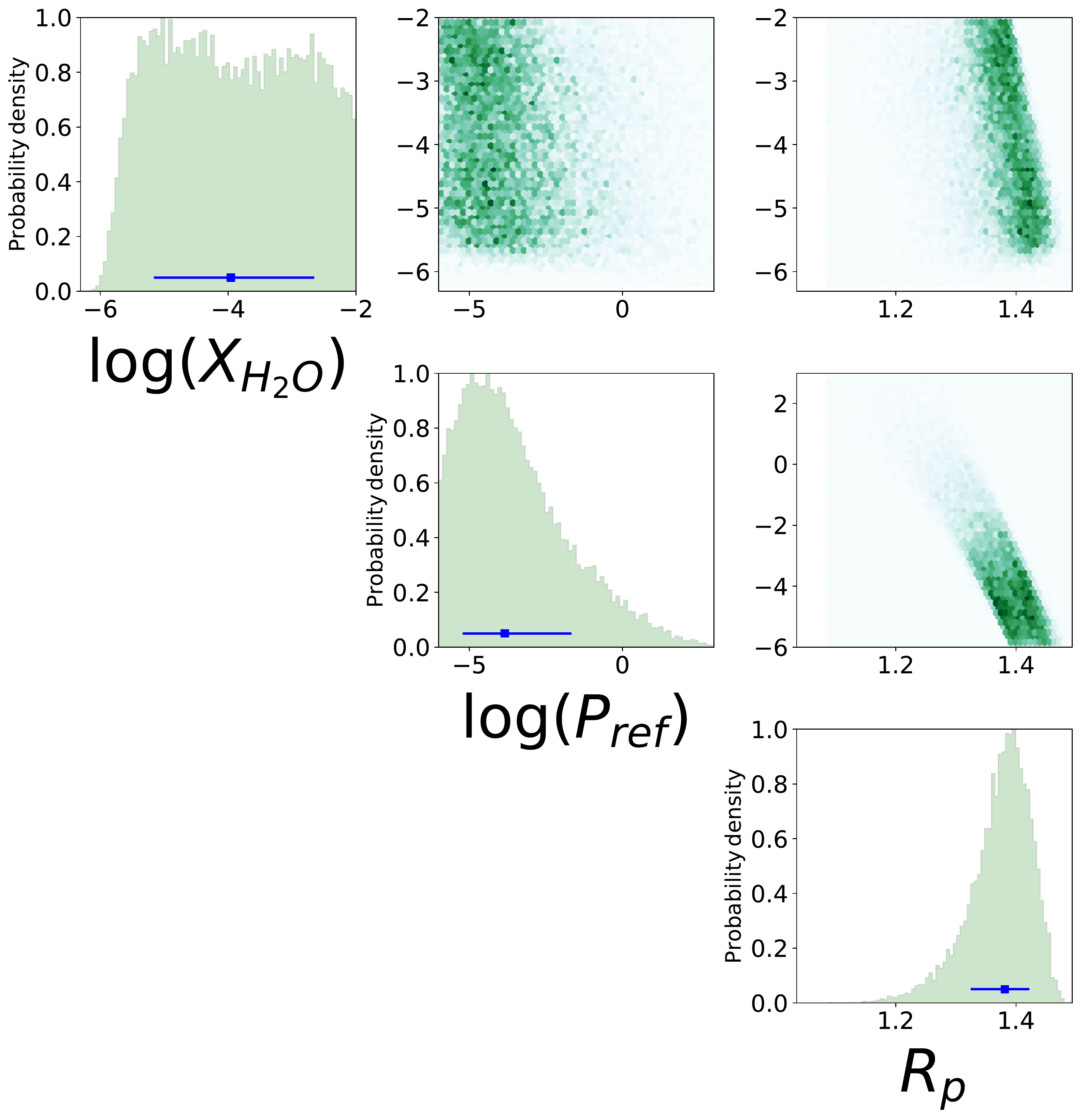}{0.25\textwidth}{(Case 5)}
          \fig{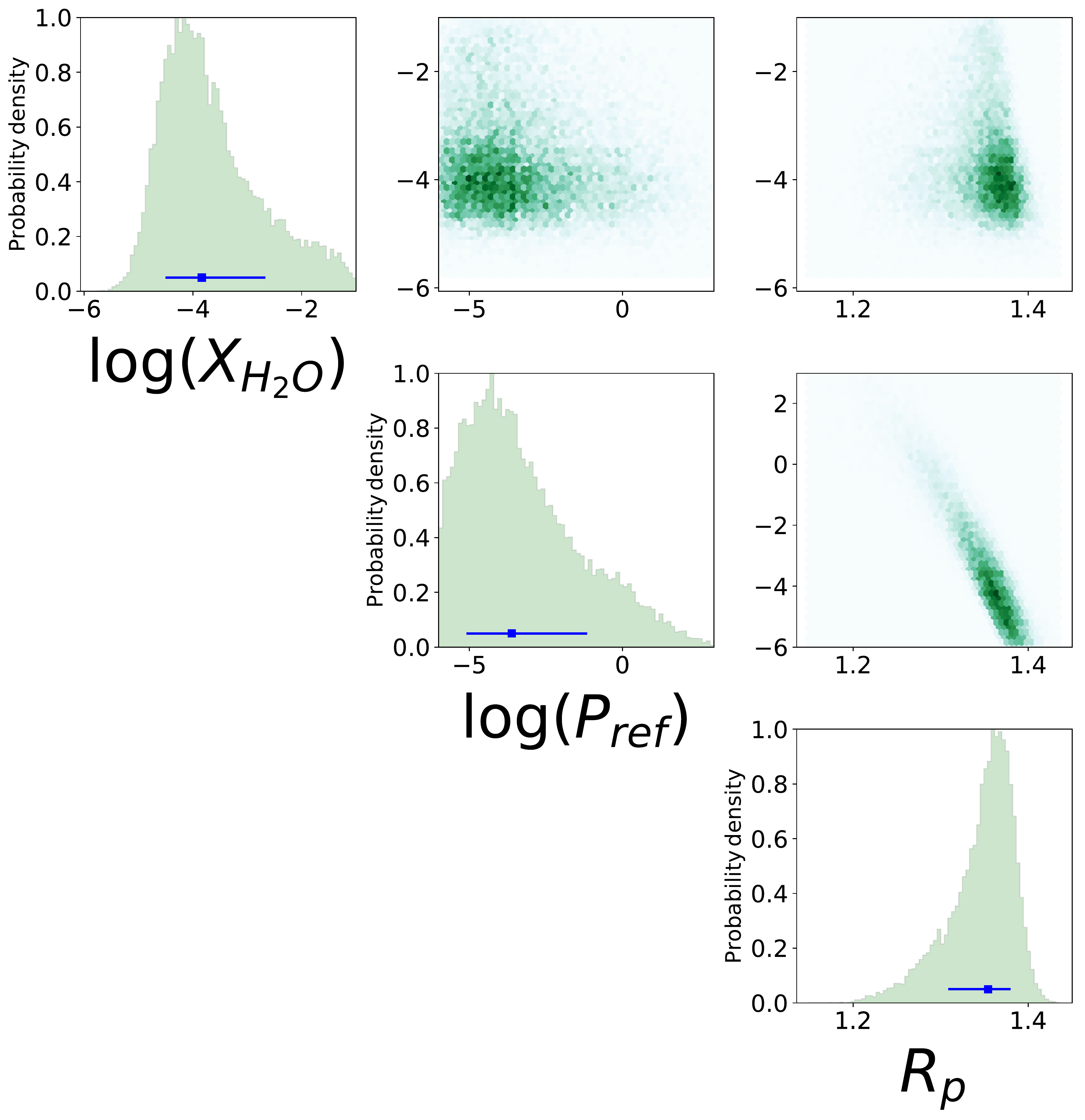}{0.25\textwidth}{(Case 6)}
          \fig{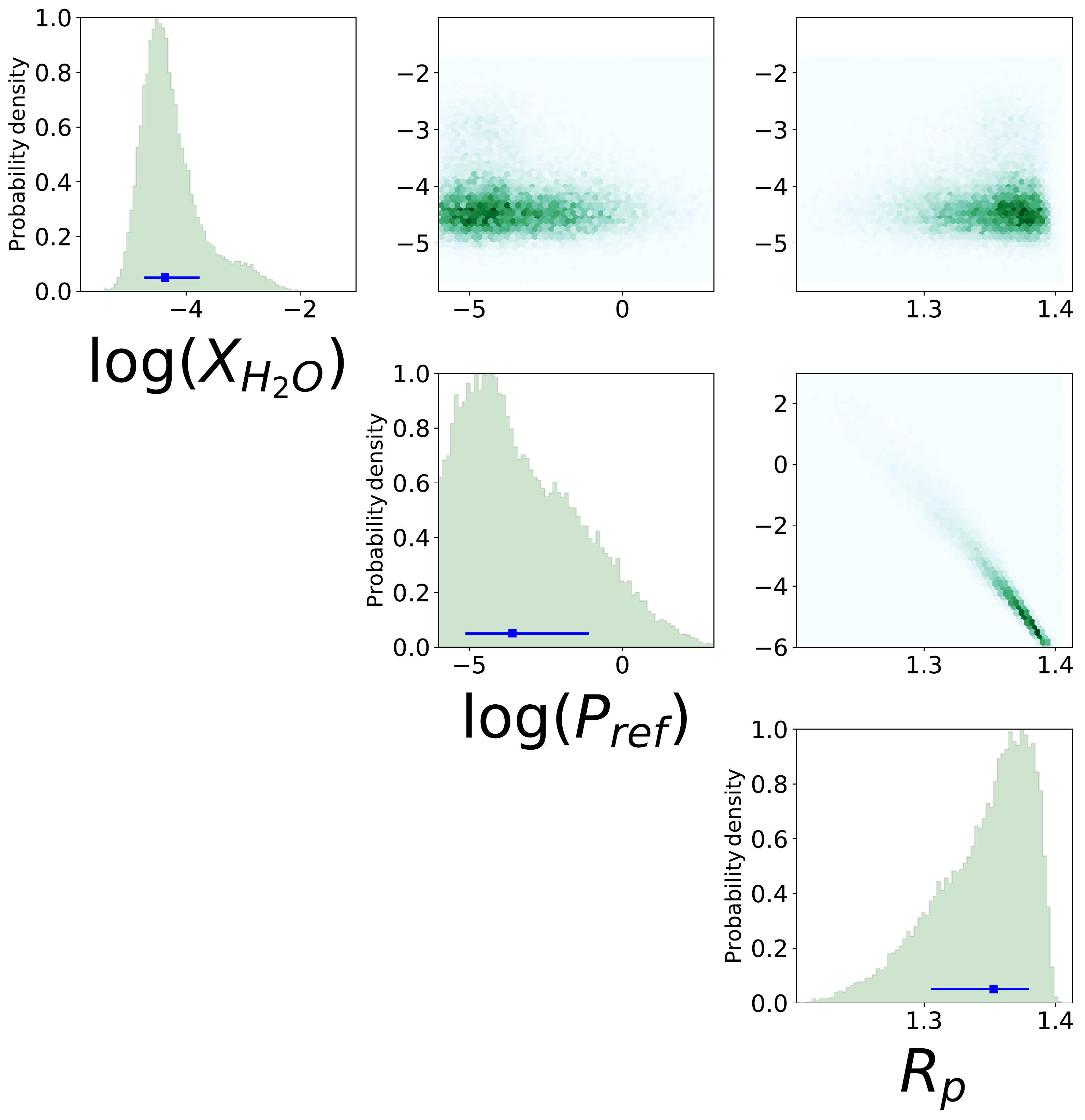}{0.25\textwidth}{(Case 7)}}
\gridline{\fig{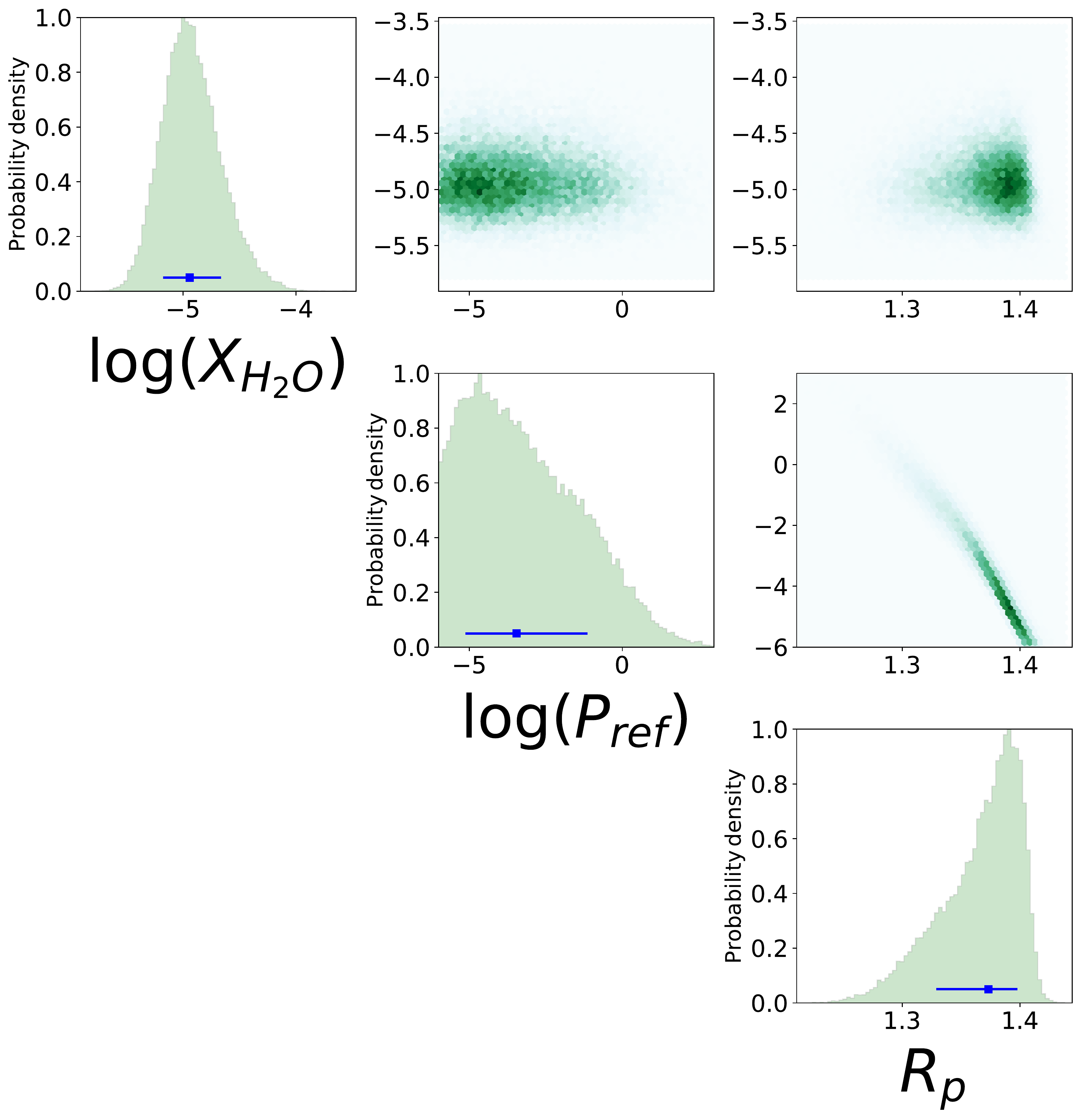}{0.25\textwidth}{(Case 8)}
          \fig{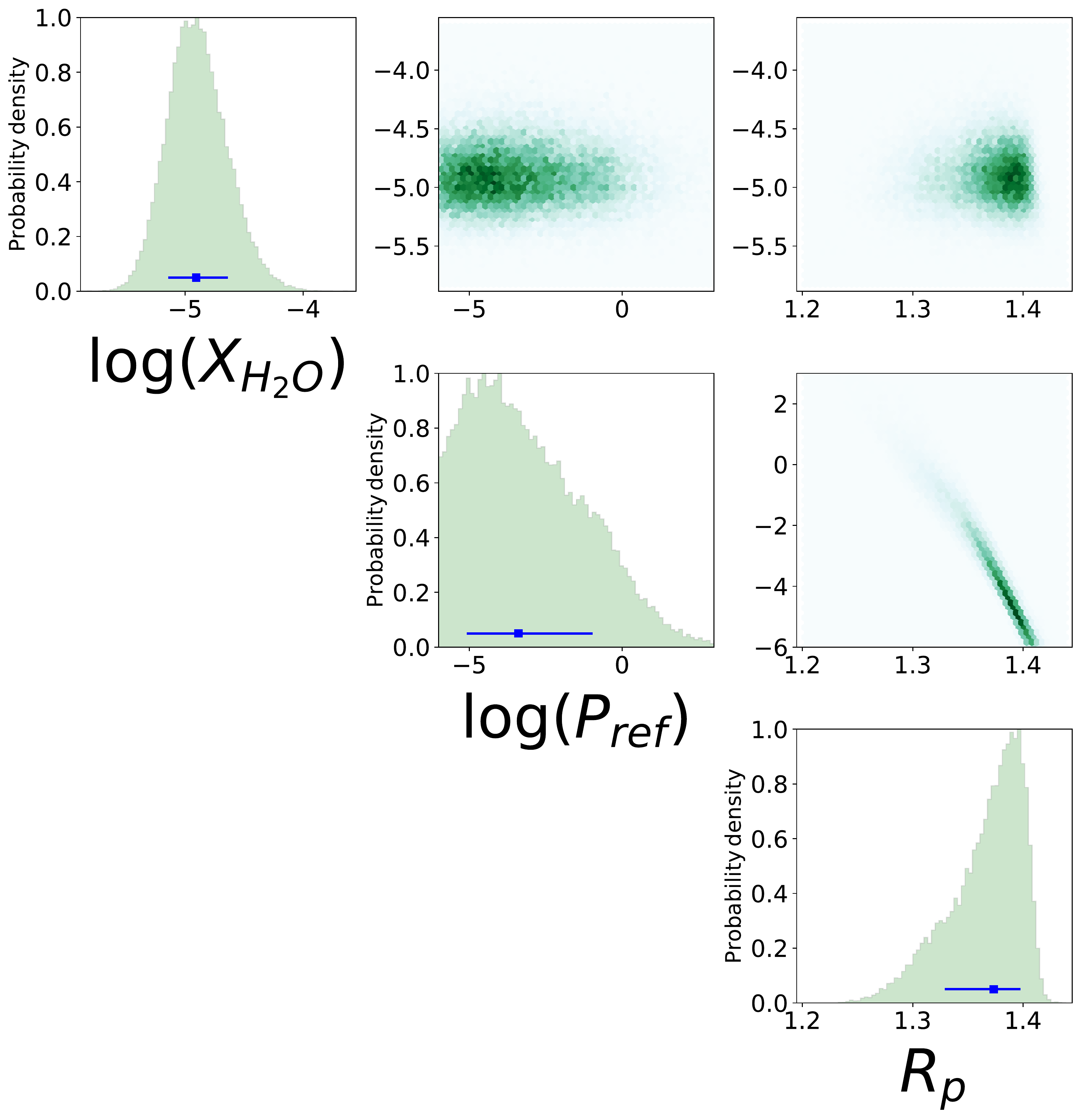}{0.25\textwidth}{(Case 9)}
          \fig{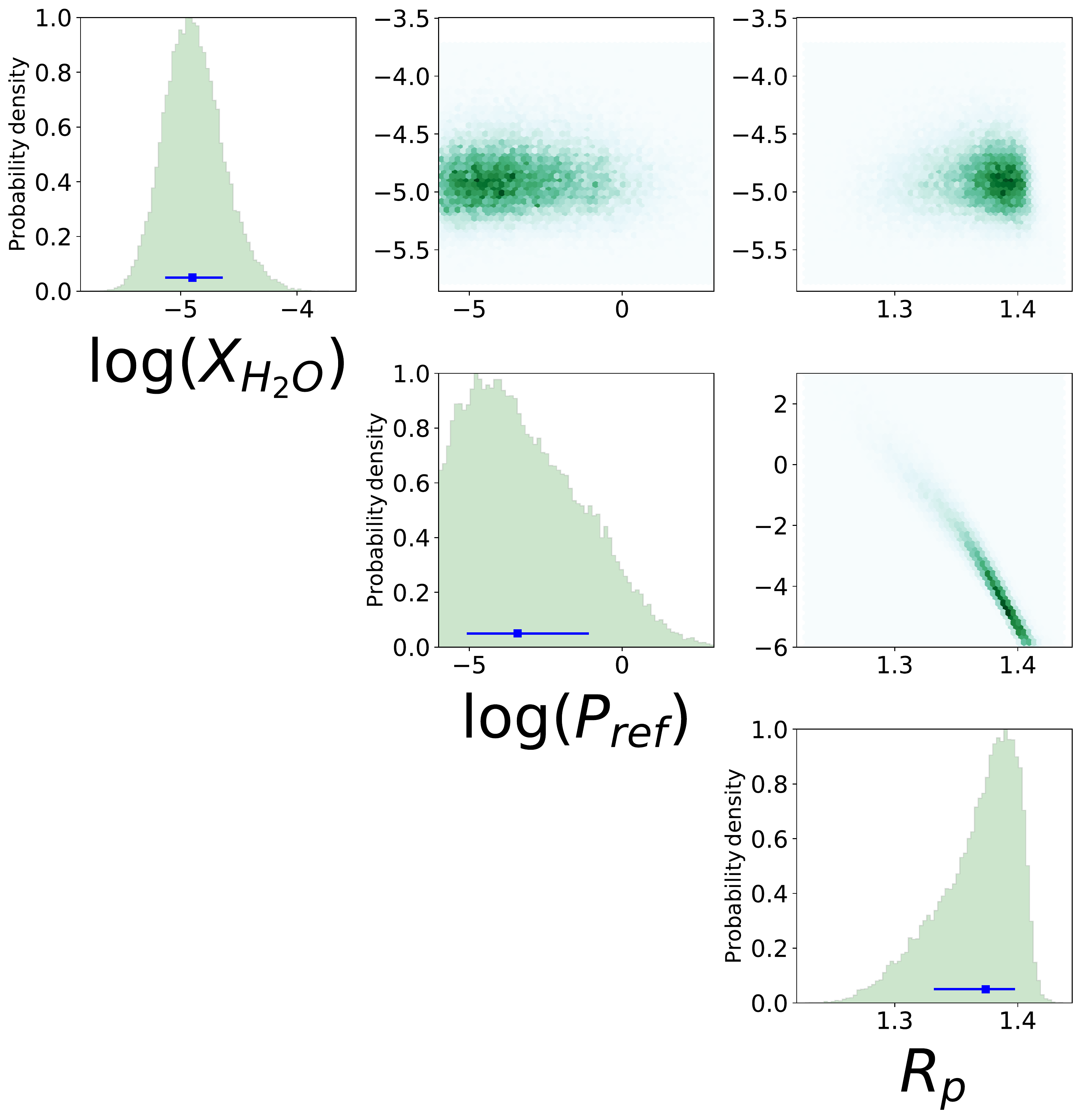}{0.25\textwidth}{(Case 10)}
          \fig{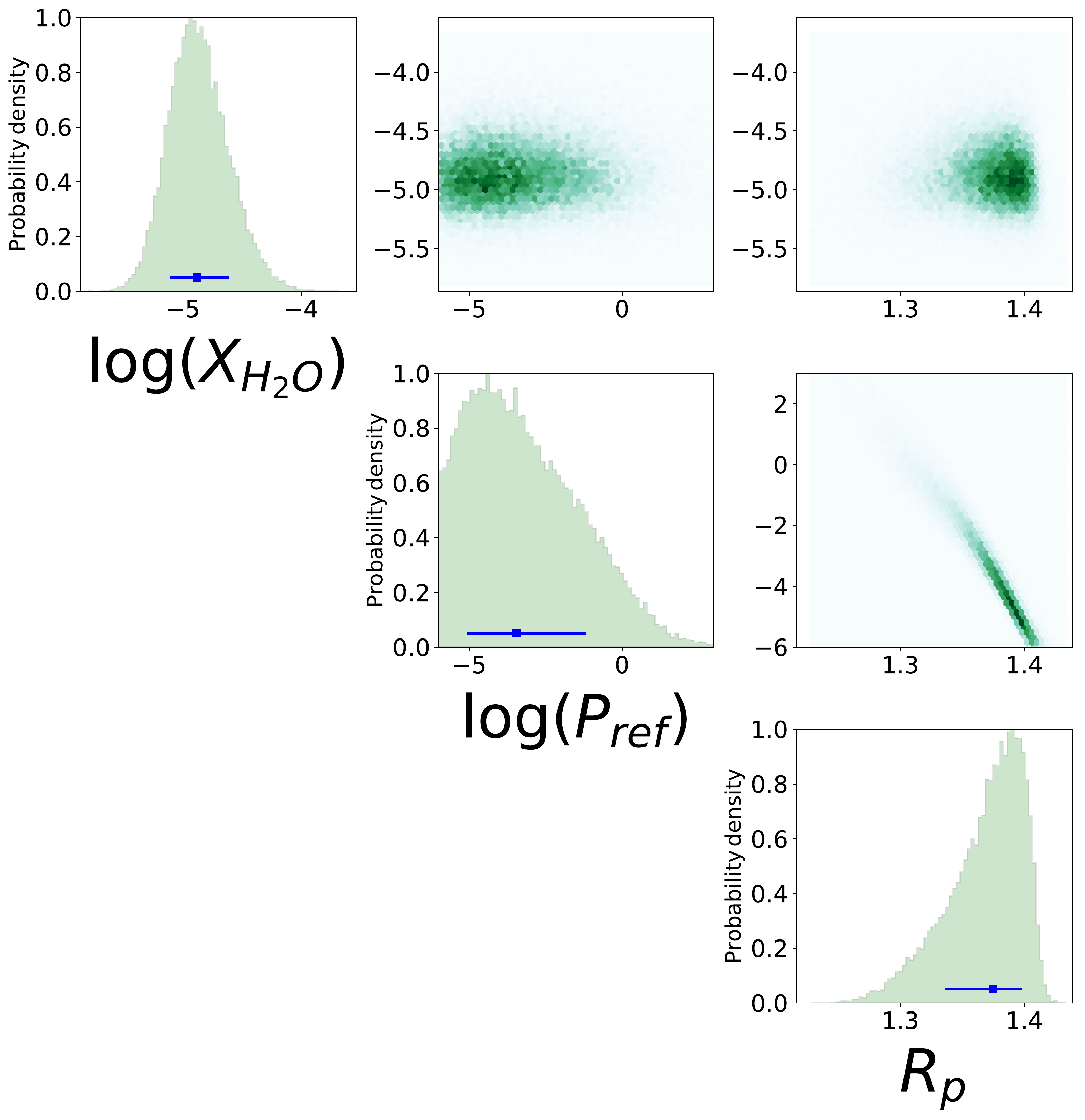}{0.25\textwidth}{(Case 11)}}
\gridline{\fig{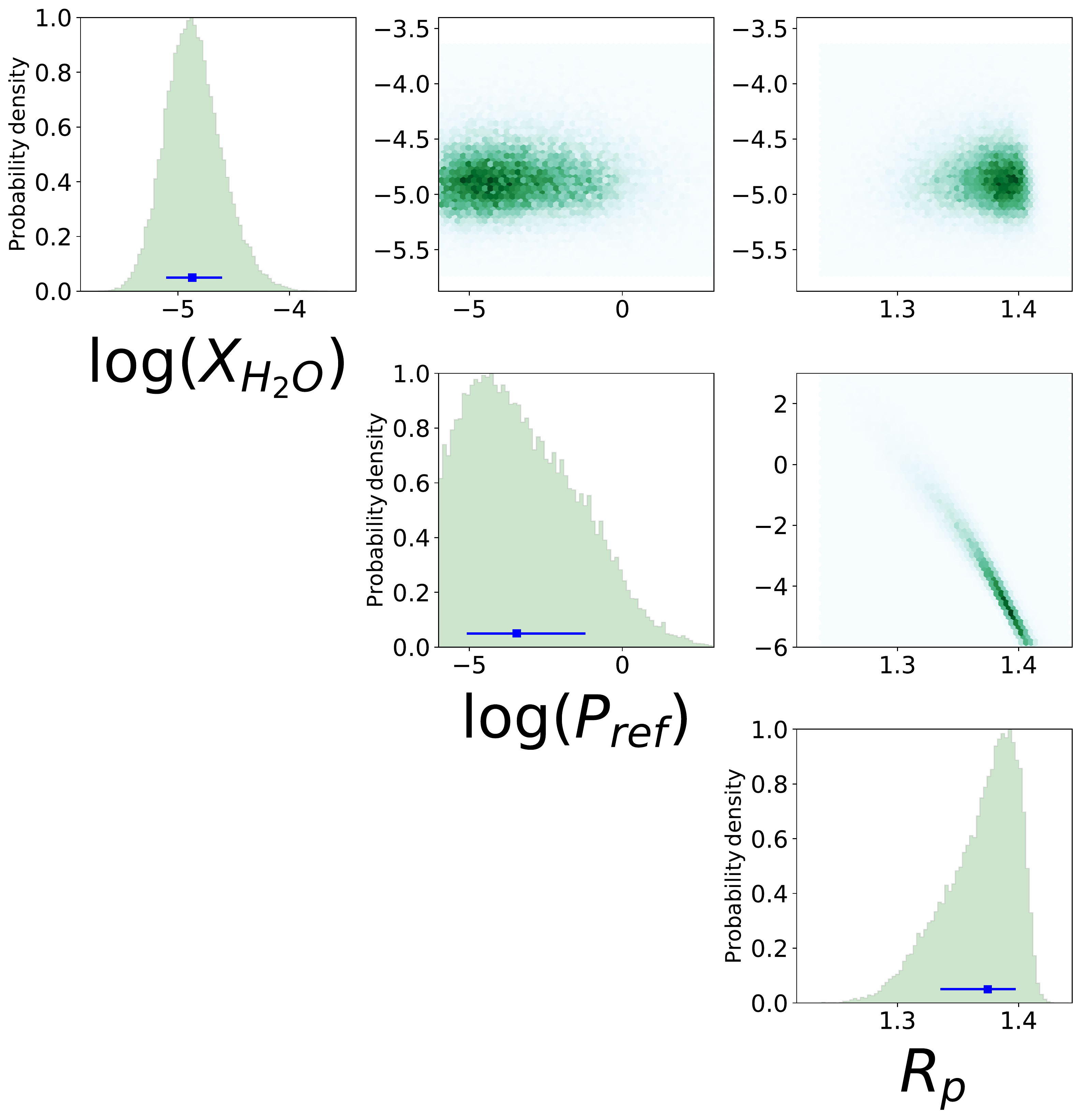}{0.25\textwidth}{(Case 12)}}
\caption{Posterior distributions for cases 0-12 from section \ref{sec:rets}.}
\label{fig:collection_cases}
\end{figure}

\begin{longrotatetable}
\begin{deluxetable*}{cl|llllllll}
\tablecaption{Results from retrievals of cases 1-12 as discussed in section \ref{sec:rets}. The uniform priors and the parameter estimates are shown for cases 1-7; remaining cases are shown in Table~\ref{table:priors2}. $^{*}$The $\log_{10}$(H$_2$O) prior for cases 4 and 5 is -12, -2. \label{table:priors1}}
\tablecolumns{15}
\tabletypesize{\small}
\tablehead{
 \colhead{}& \colhead{Parameter} & \colhead{Priors} & \colhead{Case 1} & \colhead{Case 2} & \colhead{Case 3} & \colhead{Case 4$^{*}$} & \colhead{Case 5$^{*}$} & \colhead{Case 6} & \colhead{Case 7} \\}
\startdata
 &$\text{R}_{\text{p}}$ ($\text{R}_\text{J}$) & 0.1,3.0 & $1.49^{+0.05}_{-0.08}$ & $1.41^{+0.02}_{-0.03}$ & $1.41^{+0.02}_{-0.03}$  & $1.41^{+0.01}_{-0.03}$  & $1.38^{+0.04}_{-0.06}$  & $1.35^{+0.03}_{-0.05}$ & $1.35^{+0.03}_{-0.05}$  \\
 &$\log_{10}$($\text{P}_{ref}$) (bar)& -6,3  & $-3.00^{+3.67}_{-2.21}$  & $-4.51^{+2.53}_{-1.11}$ & $-4.35^{+2.63}_{-1.25}$   & $-4.36^{+2.35}_{-1.18}$  & $-3.84^{+2.17}_{-1.37}$ & $-3.61^{+2.46}_{-1.48}$ & $-3.59^{+2.49}_{-1.54}$\\
 &$\log_{10}$(\Xh)& -12,-1  & $-9.54 ^{+0.15}_{-0.15}$   &  $-5.29 ^{+0.23}_{-0.20}$  & $-5.46 ^{+0.19}_{-0.17}$  & $-5.48^{+0.16}_{-0.16}$ & $-3.96^{+1.30}_{-1.21}$ & $-3.83^{+1.17}_{-0.67}$ & $-4.37^{+0.61}_{-0.36}$  \\
 \hline
\multirow{6}{*}{\parbox[t]{6mm}{{\rotatebox[origin=c]{90}{P-T Parameters }}}} &T$_0$ (K) & 800,2710  & $2003.65^{+248.72}_{-247.72}$  & $1070.21^{+87.56}_{-92.10}$   & $1046.02^{+89.50}_{-95.51}$  & $870.11^{+82.12}_{-49.12}$ &  $1940.56^{+251.87}_{-305.81}$ & $1262.74^{+225.05}_{-230.98}$  & $1306.40^{+225.24}_{-257.81}$\\
 & $\alpha_1$ & 0.02,1.00  &   &   &   & $0.85^{+0.11}_{-0.14}$  & $0.68^{+0.20}_{-0.23}$ & $0.65^{+0.21}_{-0.21}$ & $0.71^{+0.19}_{-0.22}$  \\
 &$\alpha_2$ & 0.02,1.00  &   &   &   & $0.67^{+0.22}_{-0.32}$ & $0.60^{+0.24}_{-0.25}$ & $0.60^{+0.25}_{-0.25}$  & $0.67^{+0.21}_{-0.26}$ \\
 &$\log_{10}$($\text{P}_{1}$) (bar)& -6,3  &   &   &   & $-0.77^{+1.88}_{-2.35}$  & $-1.30^{+1.88}_{-1.78}$ & $-1.22^{+1.86}_{-1.84}$  & $-1.34^{+2.02}_{-1.84}$\\
 &$\log_{10}$($\text{P}_{2}$) (bar)& -6,3  &   &   &   &  $-3.61^{+2.40}_{-1.62}$ & $-3.86^{+1.88}_{-1.39}$  & $-3.90^{+2.00}_{-1.34}$ & $-3.79^{+1.99}_{-1.45}$ \\
 &$\log_{10}$($\text{P}_{3}$) (bar) & -2,3  &   &   &   & $1.45^{+1.10}_{-1.76}$  & $1.19^{+1.16}_{-1.66}$ & $1.27^{+1.13}_{-1.65}$ & $1.23^{+1.20}_{-1.91}$  \\
 \hline
\multirow{4}{*}{\parbox[t]{6mm}{{\rotatebox[origin=c]{90}{\makecell{Cloud \\ \,Parameters \,}}}}} &$\log_{10}$(a)& -4,8  &   &   &   &  & $0.39^{+3.81}_{-2.73}$ & $2.09^{+3.93}_{-3.88}$ &  $7.65^{+0.24}_{-0.43}$ \\
 &$\gamma$ & -20,2  &   &   &   &  & $-12.14^{+7.38}_{-4.97}$ &$-8.60^{+7.87}_{-7.40}$ & $-8.97^{+1.07}_{-0.88}$ \\
 &$\log_{10}$($\text{P}_{\text{cloud}}$) (bar) & -6,2  &   &   &   &  & $-2.74^{+1.24}_{-1.27}$  & $-4.70^{+1.33}_{-0.85}$ & $-5.29^{+0.25}_{-0.16}$\\
 &$\phi$  & 0,1  &   &   &   &  &  & $0.68^{+0.05}_{-0.06}$ & $0.69^{+0.04}_{-0.05}$  \\
\enddata
\end{deluxetable*}
\end{longrotatetable}

\begin{longrotatetable}
\begin{deluxetable*}{cl|llllll}
\tablecaption{Results from retrievals of cases 1-12 as discussed in section \ref{sec:rets}. The uniform priors and the parameter estimates are shown for cases 8-12; remaining cases are shown in Table~\ref{table:priors1}. 
\label{table:priors2}
}
\tablecolumns{15}
\tabletypesize{\small}
\tablehead{
 \colhead{}& \colhead{Parameter} & \colhead{Priors} & \colhead{Case 8} & \colhead{Case 9} & \colhead{Case 10} & \colhead{Case 11} & \colhead{Case 12}\\}
\startdata
 &$\text{R}_{\text{p}}$ ($\text{R}_\text{J}$)& 0.1,3.0 & $1.37 ^{+ 0.02 }_{- 0.04 }$ & $1.37 ^{+ 0.02 }_{- 0.04 }$  & $1.37 ^{+ 0.02 }_{- 0.04 }$ & $1.37 ^{+ 0.02 }_{- 0.04 }$ & $1.37 ^{+ 0.02 }_{- 0.04 }$ \\
 &$\log_{10}$($\text{P}_{ref}$) (bar)& -6,3  & $-3.46 ^{+ 2.33 }_{- 1.67 }$  & $-3.39 ^{+ 2.43 }_{- 1.69 }$ & $-3.42 ^{+ 2.33 }_{- 1.66 }$ & $-3.45 ^{+ 2.26 }_{- 1.63 }$ &  $-3.45 ^{+ 2.24 }_{- 1.63 }$\\
 &$\log_{10}$(\Xh)& -12,-1  & $-4.94^{+0.28}_{-0.24}$ & $-4.91 ^{+ 0.27 }_{- 0.24 }$ & $-4.90 ^{+ 0.26 }_{- 0.23 }$ & $-4.88 ^{+ 0.27 }_{- 0.23 }$  & $-4.87 ^{+ 0.27 }_{- 0.24 }$  \\
 \hline
\multirow{6}{*}{\parbox[t]{6mm}{{\rotatebox[origin=c]{90}{P-T Parameters }}}} &T$_0$ (K) & 800,2710  & $1064.75 ^{+ 283.29 }_{- 185.88 }$ & $1026.44 ^{+ 276.52 }_{- 161.11 }$ & $1026.72 ^{+ 262.72 }_{- 157.60 }$ & $1013.53 ^{+ 248.73 }_{- 149.27 }$ &  $1022.15 ^{+ 246.41 }_{- 153.72 }$\\
 & $\alpha_1$ & 0.02,1.00 & $0.59 ^{+ 0.25 }_{- 0.17 }$ & $0.62 ^{+ 0.24 }_{- 0.18 }$  & $0.61 ^{+ 0.23 }_{- 0.18 }$ & $0.62 ^{+ 0.23 }_{- 0.18 }$ &  $0.60 ^{+ 0.23 }_{- 0.17 }$\\
 &$\alpha_2$ & 0.02,1.00  & $0.47 ^{+ 0.33 }_{- 0.21 }$ & $0.49 ^{+ 0.32 }_{- 0.22 }$ & $0.49 ^{+ 0.32 }_{- 0.22 }$  & $0.49 ^{+ 0.31 }_{- 0.22 }$ & $0.50 ^{+ 0.31 }_{- 0.23 }$ \\
 &$\log_{10}$($\text{P}_{1}$) (bar) & -6,3 & $-1.16 ^{+ 1.98 }_{- 1.79 }$ & $-1.18 ^{+ 1.97 }_{- 1.77 }$ & $-1.14 ^{+ 1.94 }_{- 1.77 }$ & $-1.15 ^{+ 1.93 }_{- 1.78 }$ & $-1.03 ^{+ 1.88 }_{- 1.79 }$ \\
 &$\log_{10}$($\text{P}_{2}$) (bar)& -6,3  & $-3.89 ^{+ 2.24 }_{- 1.43 }$  & $-3.95 ^{+ 2.19 }_{- 1.39 }$  & $-3.87 ^{+ 2.15 }_{- 1.44 }$ & $-3.90 ^{+ 2.14 }_{- 1.42 }$ & $-3.86 ^{+ 2.16 }_{- 1.42 }$ \\
 &$\log_{10}$($\text{P}_{3}$) (bar) & -2,3  & $1.22 ^{+ 1.21 }_{- 1.60 }$ & $1.25 ^{+ 1.18 }_{- 1.59 }$ & $1.29 ^{+ 1.15 }_{- 1.55 }$ & $1.26 ^{+ 1.16 }_{- 1.55 }$ & $1.33 ^{+ 1.13 }_{- 1.54 }$  \\
 \hline
\multirow{4}{*}{\parbox[t]{6mm}{{\rotatebox[origin=c]{90}{\makecell{Cloud \\ \,Parameters \,}}}}} &$\log_{10}$(a)& -4,8  & $4.42 ^{+ 0.71 }_{- 1.26 }$ & $4.38 ^{+ 0.70 }_{- 1.16 }$ & $4.38 ^{+ 0.69 }_{- 1.14 }$ & $4.34 ^{+ 0.69 }_{- 1.11 }$  & $4.37 ^{+ 0.68 }_{- 1.08 }$ \\
 &$\gamma$ & -20,2 & $-14.42 ^{+ 5.59 }_{- 3.76 }$ & $-14.67 ^{+ 5.19 }_{- 3.57 }$ &   $-14.70 ^{+ 5.17 }_{- 3.55 }$  & $-14.63 ^{+ 4.99 }_{- 3.59 }$ &  $-14.61 ^{+ 4.93 }_{- 3.58 }$ \\
 &$\log_{10}$($\text{P}_{\text{cloud}}$) (bar) & -6,2  & $-4.69 ^{+ 0.77 }_{- 0.50 }$ & $-4.57 ^{+ 0.77 }_{- 0.56 }$ & $-4.57 ^{+ 0.76 }_{- 0.54 }$  & $-4.52 ^{+ 0.73 }_{- 0.54 }$ &  $-4.52 ^{+ 0.72 }_{- 0.55 }$\\
 &$\phi$  & 0,1 & $0.49 ^{+ 0.06 }_{- 0.06 }$  & $0.47 ^{+ 0.06 }_{- 0.08 }$ & $0.47 ^{+ 0.06 }_{- 0.08 }$  & $0.46 ^{+ 0.06 }_{- 0.08 }$ & $0.46 ^{+ 0.06 }_{- 0.08 }$\\
 \hline
&$\log_{10}$(\XNa)& -12,-2  &  $-5.55^{+0.53 }_{-0.44}$ & $-5.53 ^{+ 0.51 }_{- 0.43 }$ & $-5.52 ^{+ 0.52 }_{- 0.43 }$ & $-5.50 ^{+ 0.51 }_{- 0.42 }$ & $-5.48 ^{+ 0.52 }_{- 0.43 }$ \\
  \multirow{5}{*}{\parbox[t]{6mm}{{\rotatebox[origin=c]{90}{\makecell{Chemical\\ Species} }}}} &$\log_{10}$(\XK)& -12,-2 &   $-7.17 ^{+ 0.55 }_{- 0.52 }$& $-7.13 ^{+ 0.54 }_{- 0.51 }$ & $-7.11 ^{+ 0.54 }_{- 0.49 }$  & $-7.09 ^{+ 0.54 }_{- 0.50 }$ & $-7.07 ^{+ 0.54 }_{- 0.51 }$ \\
&$\log_{10}$(\XNHt) & -12,-2  &  &  $-8.02 ^{+ 1.86 }_{- 2.64 }$ & $-8.14^{+1.95}_{-2.56}$ & $-8.11 ^{+ 1.90 }_{- 2.52 }$  & $-8.09 ^{+ 1.89 }_{- 2.54 }$ \\
 &$\log_{10}$(\XCO)& -12,-2 & &    & $-7.74 ^{+ 2.85 }_{- 2.72 }$ & $-7.75 ^{+ 2.82 }_{- 2.73 }$ & $-7.73 ^{+ 2.79 }_{- 2.75 }$  \\
 &$\log_{10}$(\XHCN) & -12,-2   &  & &  & $-8.60 ^{+ 2.26 }_{- 2.17 }$  & $-8.57 ^{+ 2.24 }_{- 2.22 }$ \\
 & $\log_{10}$(\XCOt) & -12,-2 &  &  & & & $-8.46 ^{+ 2.43 }_{- 2.30 }$\\
\enddata
\end{deluxetable*}
\end{longrotatetable}

\begin{figure}
\centering
\includegraphics[width=1.0\textwidth]{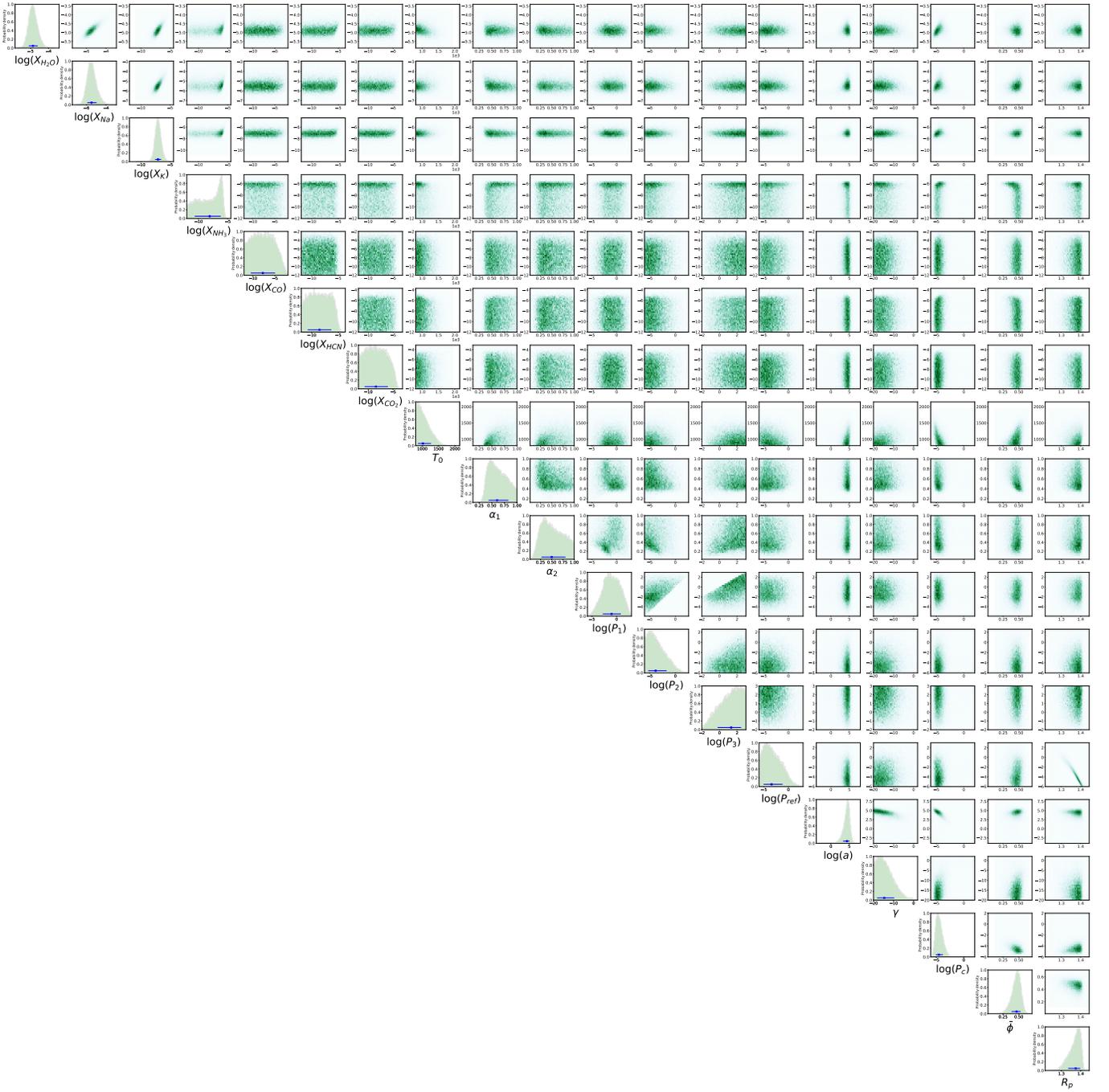}
  \caption[Full posterior distribution 12]{Full posterior distribution for case 12 as explained in section \ref{sec:rets} and section \ref{sec:fullret}. This is a full retrieval of HD~209458~b  using data in the near-infrared and optical wavelengths from \citet{sing2016}. The model includes the effects H$_2$-H$_2$ and H$_2$-He CIA opacity, absorption from H$_2$O, Na, K, NH$_3$, CO, HCN, and CO$_2$, a parametric P-T profile and the presence of clouds/hazes. Both \Rp and \Pref are simultaneously retrieved.}
\label{fig:case12fullpost}
\end{figure} 

\bibliographystyle{aasjournal}
\bibliography{biblio}

\end{document}